\documentclass[preprint,floats,aps,epsfig,nofootinbib,amssymb]{revtex4}
\pdfoutput=1
\usepackage[pdftex]{graphicx}
\usepackage{mathrsfs}
\usepackage{slashed}
\usepackage{graphicx,color}
\usepackage{epsfig}
\usepackage{subfigure}
\usepackage{epsfig}
\usepackage{dcolumn}
\usepackage{bm}

\begin{document}

\title{Study on perturbation schemes for achieving the real PMNS matrix from various symmetric textures}

\author{\bf Bin Wang$^1$, Jian Tang$^2$, Xue-Qian Li$^1$}

\affiliation{1. School of Physics, Nankai University, Tianjin, 300071, China\\
2. Center for Particle Physics, University of Alberta, Canada}

\begin{abstract}
The PMNS matrix displays an obvious symmetry, but not exact. There are several textures proposed in literature, which possess various symmetry patterns and seem to originate from different physics scenarios at high energy scales. To be consistent with the experimental measurement, all of the regularities slightly decline, i.e. the symmetry must be broken. Following the schemes given in literature, we modify the matrices (9 in total) to gain the real PMNS matrix by perturbative rotations.  The transformations may provide hints about the underlying physics at high energies and the breaking mechanisms which apply during the evolution to the low energy scale, especially the results may be useful for the future model builders.


\end{abstract}

\draft

\maketitle

\section{Introduction}

The mixing among fermions is one of the most mysterious aspects in particle physics. Unlike the mixing matrix for quarks, the mixing among leptons displays an obvious regularity which is manifested in the lepton mixing matrix. It is well known that the mixing among fermions originates from the fact that the weak eigenstates of fermions (quarks and leptons) are not that of the mass Hamiltonian, and the rotation from the weak basis to the mass basis results in the mixing matrix\cite{Fritzsch}. As observed, the structures of the quark and lepton mixing matrices are so different, and it implies the mechanisms which determine their mass eigenstates would be different. Lam suggests that a higher horizontal symmetry $U(1)\times SO(3)$ is broken into
the tetrahedral $A_4$ and nematic $Z_2\times Z_2\times Z_2$ sectors
which correspond to the lepton and quark mixing respectively \cite{Lam}. Definitely, it is only one of the possible structures which were discussed in literature. It is believed that there must be a higher symmetry at high energy scales and later it is broken during the evolution from high energy to the weak energy scale. It is worth pointing out that the Lam's mechanism which determines an $A_4$ symmetry for the lepton mixing demands $\theta_{13}$ in the mixing matrix to be zero. And most of the proposed symmetries would result in the same zero-$\theta_{13}$. However, the recent experiments of T2K\cite{T2K}, Double-Chooz \cite{DoubleChooz1, DoubleChooz2}, the Daya-Bay \cite{DayaBay1, DayaBay2, DayaBay3} and RENO\cite{RENO} collaborations all confirm that $\theta_{13}$ is not zero, but sizable as near $9^{\circ}$. This implies that even though the lepton mixing matrix displays an approximate symmetric form, its original symmetry must be broken.

The most plausible way to break the symmetry is to perturb the matrix to realize a practical lepton mixing matrix which is obtained by fitting the data while the unitarity of the matrix must be retained. The form of the perturbation may hint us the breaking mechanism which is important for understanding the nature. Moreover, in the process of perturbing the matrix and comparing with data, we notice that several textures of the matrix are disfavored or marginally favored, even though a perturbative rotation would make them to be in marginal agreement with data (see the text). That is the breaking mechanism. A careful analysis of the breaking (indeed the perturbation) indicates that one may have an opportunity to realize what original symmetric texture(s) is more realistic, so would be able to trace back to high energy scale physics where the mixing originates. In particular, such a study about the patterns of perturbation may be useful for the future model builders.

As well known, non-zero neutrino masses; neutrino or lepton mixing and relatively small splitting among neutrino masses are the three conditions leading to the quantum mechanical phenomena: observable neutrino oscillations \cite{PDG, XZhBook}.  The mixing matrix in the lepton sector $U_l^\dag U_\nu$ are named as the Pontecorvo \cite{PMNS1}-Maki-Nakawaga-Sakata \cite{PMNS2} (PMNS) matrix
\begin{eqnarray}
U_{\rm{PMNS}}=U_l^\dag U_\nu.
\end{eqnarray}
which is a $3\times 3$ unitary matrix and can be parameterized via mixing angles $\theta_{12}$, $\theta_{23}$, $\theta_{13}$ and one CP phase $\delta$ \cite{PDG}
\begin{eqnarray}
U_{\rm{PMNS}}=\left(
\begin{array}{ccc}
c_{12}c_{13}&s_{12}c_{13}&s_{13}e^{-i\delta}\\
-s_{12}c_{23}-c_{12}s_{23}s_{13}e^{i\delta}&
c_{12}c_{23}-s_{12}s_{23}s_{13}e^{i\delta}&s_{23}c_{13}\\
s_{12}s_{23}-c_{12}c_{23}s_{13}e^{i\delta}&
-c_{12}s_{23}-s_{12}c_{23}s_{13}e^{i\delta}&c_{23}c_{13}
\end{array}
\right), \label{PMNS2}
\end{eqnarray}
where $c_{ij}\equiv \cos\theta_{ij}$, $s_{ij}\equiv \sin\theta_{ij}$. If neutrinos are Majorana particles, there could be one additional matrix diag($e^{\alpha_1/2}, e^{\alpha_2/2}, 1$), and since it is not revelent to neutrino oscillations at all, we ignore it in this work.
The mixing angles and Jarlskog invariant $J_{\rm{CP}}$, which determines the magnitude of CP violation in neutrino oscillation \cite{Jarlskog1, Jarlskog2, PDG} are
\begin{eqnarray}
T_{12}&\equiv&\tan\theta_{12}=\frac{|U_{e2}|}{|U_{e1}|},\label{T12}\\
T_{23}&\equiv&\tan\theta_{23}=\frac{|U_{\mu 3}|}{|U_{\tau 3}|},\label{T23}\\
S_{13}&\equiv&\sin\theta_{13}=|U_{e3}|,\label{S13}\\
J_{\rm{CP}}&\equiv& {\rm{Im}}(U_{\mu 3}U_{e3}^*U_{e2}U_{\mu 2}^*). \label{JcpD}
\end{eqnarray}

The recent data indicate that the angle $\theta_{13}$ is sizable:
\begin{itemize}

\item{\bf KamLAND} Global $\theta_{13}$ analysis incorporating CHOOZ, atmospheric, and long-baseline accelerator experiments indicates $\sin^2\theta_{13}=0.009^{+0.013}_{-0.007}$ ( i.e. $\theta_{13}={5.444^{+3.086}_{-2.881}}^\circ$ ) and non-zero $\theta_{13}$ at 79$\%$ C.L. \cite{KamLAND}.

\item{\bf T2K} At 90$\%$ C.L. and for $\delta_{CP}=0$, $4.99^\circ(5.77^\circ)<\theta_{13}<15.97^\circ(17.83^\circ)$ for normal (inverted) hierarchy \cite{T2K}.

\item{\bf MINOS} With $\delta_{CP}=0$ the best fit result is $2\sin^2\theta_{23}\sin^2\theta_{13} =0.041^{+0.047}_{-0.031}(0.079^{+0.071}_{-0.053})$ for normal (inverted) hierarchy and $\theta_{13}=0$ is disfavored at 89$\%$ C.L. \cite{MINOS}.

\item{\bf Double Chooz} The early result from Double Chooz reactor electron antineutrino disappearance experiment is $3.7^\circ<\theta_{13}<12^\circ$ at 90$\%$ C.L. \cite{DoubleChooz1}. The updated results are $\sin^22\theta_{13}$ = 0.109 $\pm$ 0.030(stat) $\pm$ 0.025(syst) (i.e. the central value $\theta_{13}=9.639^\circ$) and excluding the no-oscillation hypothesis at 99.8$\%$ C.L. \cite{DoubleChooz2}.

\item{\bf DayaBay} The Daya Bay collaboration presents the reactor electron antineutrino disappearance experiment result $\sin^22\theta_{13} = 0.092 \pm 0.016$ (stat) $\pm 0.005$ (syst) (i.e. the central value $\theta_{13}=8.8^\circ$) and non-zero $\theta_{13}$ with a significance of 5.2 standard deviations \cite{DayaBay1}. Recent updated result is $\sin^22\theta_{13} = 0.089 \pm 0.010$ (stat) $\pm0.005$ (syst) (i.e. the central value $\theta_{13}=8.7^\circ$) with $\theta_{13}=0$ disfavored at 7.7 $\sigma$ \cite{DayaBay2} \cite{DayaBay3}.

\item{\bf RENO} The result from RENO experiment is $\sin^22\theta_{13} = 0.113 \pm 0.013$ (stat) $ \pm 0.019$ (syst) (i.e. the central value $\theta_{13}=9.821^\circ$)\cite{RENO}.

\end{itemize}

For convenience of discussion, an updated global analysis on neutrino oscillation data \cite{GlobalFit} is re-presented in Table \ref{GlobalFit}, and we single out the mixing angles and represent them in degrees in Table \ref{AngleFit}.
\begin{table}[t]
\caption{\label{GlobalFit} The update global fit results of three neutrino oscillation, where $\Delta m^2$ is defined as $m_3^2-(m_1^2+m_2^2)/2$ and $\delta m^2=m_2^2-m_1^2$. }
\begin{ruledtabular}
\begin{tabular}{lcccc}
Parameter & Best fit & 1$\sigma$ range & 2$\sigma$ range & 3$\sigma$ range \\
\hline
$\delta m^2$/$10^{-5}$$\mathrm{eV}^2$(NH or IH) & 7.54 & 7.32--7.80 & 7.15--8.00 & 6.99--8.18 \\
\hline
$\sin^2\theta_{12}/10^{-1}$(NH or IH) & 3.07 & 2.91--3.25 & 2.75--3.42 & 2.59--3.59 \\
\hline
$\Delta m^2/10^{-3}\mathrm{eV}^2$(NH) & 2.43 & 2.33--2.49 & 2.27--2.55 & 2.19--2.62 \\
$\Delta m^2/10^{-3}\mathrm{eV}^2$(IH) & 2.42 & 2.31--2.49 & 2.26--2.53 & 2.17--2.61 \\
\hline
$\sin^2\theta_{13}/10^{-2}$(NH) & 2.41 & 2.16--2.66 & 1.93--2.90 & 1.69--3.13 \\
$\sin^2\theta_{13}/10^{-2}$(IH) & 2.44 & 2.19--2.67 & 1.94--2.91 &1.71--3.15 \\
\hline
$\sin^2\theta_{23}/10^{-1}$(NH) & 3.86 & 3.65--4.10 & 3.48--4.48 & 3.31--6.37 \\
$\sin^2\theta_{23}/10^{-1}$(IH) & 3.92 & 3.70--4.31 &3.53--4.84$\oplus$5.43--6.41 & 3.35--6.63 \\
\hline
$\delta/\pi$(NH) & 1.08 & 0.77--1.36 & -- & -- \\
$\delta/\pi$(IH) & 1.09 & 0.83--1.47 & -- & -- \\
\end{tabular}
\end{ruledtabular}
\end{table}

\begin{table}
\caption{\label{AngleFit} The mixing angles from neutrino oscillation fit results in \cite{GlobalFit} (in degree).}
\begin{ruledtabular}
\begin{tabular}{lcccc}
Parameter & Best fit & 1$\sigma$ range & 2$\sigma$ range & 3$\sigma$ range \\
\hline
$\theta_{12}$(NH or IH) & 33.6 & 32.6--34.8 & 31.6--35.8 & 30.6--36.8 \\
\hline
$\theta_{13}$(NH) & 8.93 & 8.45--9.39 & 7.99--9.80 & 7.47--10.2  \\
$\theta_{13}$(IH) & 8.99 & 8.51--9.40 & 8.01--9.82 & 7.51--10.2  \\
\hline
$\theta_{23}$(NH) & 38.4 & 37.2--39.8 & 36.2--42.0 & 35.1--53.0  \\
$\theta_{23}$(IH) & 38.8 & 37.5--41.0 & 36.5--42.0$\oplus$47.5--53.2 & 35.4--54.5  \\
\hline
$\delta$(NH) & 194.4 & 138.6--244.8 & -- & --\\
$\delta$(IH) & 196.2 & 149.4--264.6 & -- & --
\end{tabular}
\end{ruledtabular}
\end{table}

Analyzing the PMNS matrix, one notices an obvious symmetry, but not exact. If writing it in an ideal form which has an exact symmetry,  there are various textures which have different symmetric patterns. In other words,
some phenomenologically assigned forms for the mixing matrix $U_{\rm{PMNS}}$ explicitly manifest flavor symmetries while the practical form of the matrix  implies that the symmetric structures should be spontaneously or explicitly broken. Synthesizing the proposals for the symmetric textures existing in literature, there are nine in total such ansatzes  (1) Tri-Bimaximal Mixing (TBM) \cite{tribimaximal};
(2) Democratic Mixing (DM) \cite{democratic};
(3) Bimaximal Mixing (BM) \cite{bimaximal};
(4) Golden Ratio Mixing-1 (GRM1) \cite{GoldenRatioMixing1};
(5) Golden Ratio Mixing-2 (GRM2) \cite{GoldenRatioMixing2};
(6) Hexagonal Mixing (HM) \cite{HexagonalMixing};
(7) Tetra-Maximal Mixing (TMM) \cite{Tetramaximal};
(8) Toorop-Feruglio-Hagedorn Mixing-1 (TFH1) \cite{TFH1, TFH2, TFH3};
(9) Toorop-Feruglio-Hagedorn Mixing-2 (TFH2) \cite{TFH1, TFH2, TFH3}.
We list the explicit forms of these patterns in section II.

Some of the matrix forms listed above require zero-$\theta_{13}$ which is in obvious contradiction to the newly measured value. It is shown that all those forms can be modified with  perturbative rotations into the form of real PMNS matrix which is consistent with data.

In this work, we explicitly show how a perturbative rotation transforms the the matrix into a one with a sizable $\theta_{13}$  and practical $\theta_{12}$, $\theta_{23}$. Our numerical analyses are shown via several tables and figures. Then we make some discussions in the last section.

\section{The symmetric textures of the mixing matrix}

Here we list all the nine symmetric textures proposed in literature:
\begin{equation}
 U_{\rm{TBM}}=\left(
\begin{array}{ccc}
 \sqrt{\frac{2}{3}} & \frac{1}{\sqrt{3}} & 0 \\
 -\frac{1}{\sqrt{6}} & \frac{1}{\sqrt{3}} &
   \frac{1}{\sqrt{2}} \\
 \frac{1}{\sqrt{6}} & -\frac{1}{\sqrt{3}} &
   \frac{1}{\sqrt{2}}
\end{array}
\right),\label{TBM}
\end{equation}
\begin{equation}
 U_{\rm{DM}}=\left(
\begin{array}{ccc}
 \frac{1}{\sqrt{2}} & \frac{1}{\sqrt{2}} & 0 \\
 -\frac{1}{\sqrt{6}} & \frac{1}{\sqrt{6}} &
   \sqrt{\frac{2}{3}} \\
 \frac{1}{\sqrt{3}} & -\frac{1}{\sqrt{3}} &
   \frac{1}{\sqrt{3}}
\end{array}
\right),
\end{equation}
\begin{equation}
 U_{\rm{BM}}=\left(
\begin{array}{ccc}
 \frac{1}{\sqrt{2}} & \frac{1}{\sqrt{2}} & 0 \\
 -\frac{1}{2} & \frac{1}{2} & \frac{1}{\sqrt{2}} \\
 \frac{1}{2} & -\frac{1}{2} & \frac{1}{\sqrt{2}}
\end{array}
\right),
\end{equation}
\begin{equation}
 U_{\rm{GRM1}}=\left(
\begin{array}{ccc}
 \sqrt{\frac{1}{2} \left(1+\frac{1}{\sqrt{5}}\right)} &
   \sqrt{\frac{2}{5+\sqrt{5}}} & 0 \\
 -\frac{1}{\sqrt{5+\sqrt{5}}} & \frac{1}{2}\sqrt{1+\frac{1}{\sqrt{5}}}   & \frac{1}{\sqrt{2}} \\
 \frac{1}{\sqrt{5+\sqrt{5}}} & -\frac{1}{2}
   \sqrt{1+\frac{1}{\sqrt{5}}} & \frac{1}{\sqrt{2}}
\end{array}
\right),
\end{equation}
\begin{equation}
 U_{\rm{GRM2}}=\left(
\begin{array}{ccc}
 \frac{1}{4} \left(1+\sqrt{5}\right) & \frac{1}{2}
   \sqrt{\frac{1}{2} \left(5-\sqrt{5}\right)} & 0 \\
 -\frac{1}{4} \sqrt{5-\sqrt{5}} & \frac{1+\sqrt{5}}{4
   \sqrt{2}} & \frac{1}{\sqrt{2}} \\
 \frac{1}{4} \sqrt{5-\sqrt{5}} & -\frac{1+\sqrt{5}}{4
   \sqrt{2}} & \frac{1}{\sqrt{2}}
\end{array}
\right),
\end{equation}
\begin{equation}
 U_{\rm{HM}}=\left(
\begin{array}{ccc}
 \frac{\sqrt{3}}{2} & \frac{1}{2} & 0 \\
 -\frac{1}{2 \sqrt{2}} & {1\over2}\sqrt{3\over2} &
   \frac{1}{\sqrt{2}} \\
 \frac{1}{2 \sqrt{2}} & -{1\over2}\sqrt{3\over2} &
   \frac{1}{\sqrt{2}}
\end{array}
\right),
\end{equation}
\begin{equation}
 U_{\rm{TMM}}=\left(
\begin{array}{ccc}
 \frac{1}{2} \left(1+\frac{1}{\sqrt{2}}\right) &
   \frac{1}{2} & \frac{1}{2}
   \left(1-\frac{1}{\sqrt{2}}\right) \\
 -\frac{1-i \left(1-\sqrt{2}\right)}{2 \sqrt{2}} &
   \frac{1}{2} \left(1-\frac{i}{\sqrt{2}}\right) &
   \frac{1+i \left(1+\frac{1}{\sqrt{2}}\right)}{2 \sqrt{2}}
   \\
 \frac{1+i \left(1-\sqrt{2}\right)}{2 \sqrt{2}} &
   -\frac{1}{2} \left(1+\frac{i}{\sqrt{2}}\right) &
   -\frac{1-i \left(1+\frac{1}{\sqrt{2}}\right)}{2 \sqrt{2}}
\end{array}
\right),
\end{equation}
\begin{eqnarray}
U_{\rm{TFH1}}=\left(
\begin{array}{ccc}
{{3+\sqrt{3}}\over6} & {1\over{\sqrt{3}}} & {{-3+\sqrt{3}}\over6}\\
{{-3+\sqrt{3}}\over6} & {1\over{\sqrt{3}}} & {{3+\sqrt{3}}\over6}\\
{1\over{\sqrt{3}}} & -{1\over{\sqrt{3}}} & {1\over{\sqrt{3}}}
\end{array}
\right),
\end{eqnarray}
\begin{eqnarray}
U_{\rm{TFH2}}=\left(
\begin{array}{ccc}
{{3+\sqrt{3}}\over6} & {1\over{\sqrt{3}}} & {{3-\sqrt{3}}\over6}\\
-{1\over{\sqrt{3}}} & {1\over{\sqrt{3}}} & {1\over{\sqrt{3}}}\\
{{3-\sqrt{3}}\over6} & -{1\over{\sqrt{3}}} & {{3+\sqrt{3}}\over6}
\end{array}
\right).
\end{eqnarray}
It is noted that our expressions of the symmetric forms listed above may differ from those given in literature by a sign or even a phase factor in a row or column of the matrices, but obviously, an additional overall phase $e^{i\alpha}$ does not change the physics of the mixing, and moreover, our forms is more convenient to be compared with the conventional expression Eq.(\ref{PMNS2}) adopted by the PDG \cite{PDG}.

\section{The minimal modifications to these patterns}

As well known, the eigenstates of weak interaction are not that of the mass Hamiltonian, thus for physical processes one should rotate the weak basis into the mass basis.  The unitary transformation between the two bases is expressed as a 3$\times$3 matrix: the CKM matrix for quarks and PMNS matrix for leptons. The PMNS matrix manifests a not-exact regulation. It is supposed that the exact symmetric texture is originating from a symmetry at high energy scale and breaking it leads to the practical matrix which keeps an approximate symmetric pattern.

Our goal is to break the symmetric matrix by a perturbation.

Generally speaking, the perturbation can be realized by transforming the symmetric form with two different unitary matrices $U^{\dagger}_{\rm{L}}\cdot V_{\rm{PMNS}}\cdot U_{\rm{R}}$. The unitary matrices $U_{\rm{L/R}}$ are just three-dimensional rotations and can be a combination of the following matrices which are rotations about three independent axes:
\begin{equation}
 P_{x}=\left(
\begin{array}{ccc}
 c_x & e^{-i \delta_x } s_x & 0 \\
 -e^{i \delta_x } s_x & c_x & 0 \\
 0 & 0 & 1
\end{array}
\right)\,,
P_{y}=\left(
\begin{array}{ccc}
 1 & 0 & 0 \\
 0 & c_y & e^{-i \delta_y } s_y \\
 0 & -e^{i \delta_y } s_y & c_y
\end{array}
\right)\,,
P_{z}=\left(
\begin{array}{ccc}
 c_z & 0 & e^{-i \delta_z } s_z \\
 0 & 1 & 0 \\
 -e^{i \delta_z } s_z & 0 & c_z
\end{array}
\right)\,,
\end{equation}
where $\delta_x$, $\delta_y$, $\delta_z$ are arbitrary phases and $s_x\equiv \sin x$, $c_x\equiv \cos x$ and $x$, $y$, $z$ are rotation angles. Without losing generality, we only consider the minimal modifications. In this scheme we let one of $U_L$ and $U_R$ be a unit matrix, and only the another play the role of perturbation.

In this work, we only carefully analyze the case for the Tri-bimaximal mixing and an explicit illustration on the results is presented by tables and figures, whereas the procedure of perturbing  the rest eight symmetric textures is similar, so we collect corresponding tables and figures in the attached Appendices.

There are 6 possible ways to perturb the symmetric textures: $P_x\cdot U_{\rm{TBM}}$, $P_y\cdot U_{\rm{TBM}}$, $P_z\cdot U_{\rm{TBM}}$, $U_{\rm{TBM}}\cdot P_x$, $U_{\rm{TBM}}\cdot P_y$, and $U_{\rm{TBM}}\cdot P_z$. We can obtain the real $U_{\rm PMNS}$ by adjusting the parameters in $P_x, P_y, P_z$. In Table \ref{TabTBM}, we show the trigonometric functions of the mixing angles, $T_{12}$, $T_{23}$, $S_{13}$ and the Jarlskog invariant $J_{\rm{CP}}$.

\begin{table}
\begin{center}
\begin{tabular}{|c||c|c|c|c|}\hline
TBM & $T_{12}$ & $T_{23}$ & $S_{13}$ & $J_{\rm CP}$ \\
\hline\hline
$P_x\cdot U$ & $2 \sqrt{{1 + \sin(2x) \cos \delta_x}\over{5+3 \cos(2x) - 4 \sin(2x) \cos \delta_x}}$ & $\cos x$ & ${1\over{\sqrt{2}}}\sin x$ & ${1\over{12}} \sin (2x) \sin \delta_x$ \\
\hline
$P_y\cdot U$ & ${1\over\sqrt{2}}$ & $\sqrt{{1 + \sin (2y)\cos\delta_y}\over{1 - \sin(2y)\cos\delta_y}}$ & 0 & 0 \\
\hline
$P_z\cdot U$ & $2\sqrt{{1-\sin(2z)\cos\delta_z}\over{5+3\cos(2z) +4\sin(2z)\cos\delta_z}}$ & $\sec z$ & ${1\over\sqrt{2}}\sin z$ & ${1\over{12}}\sin(2z)\sin\delta_z$ \\
\hline
$U\cdot P_x$ & $\sqrt{{3 - \cos(2x) + 2\sqrt{2}\sin(2x)\cos\delta_x}\over{3 + \cos(2x) - 2\sqrt{2}\sin(2x)\cos\delta_x}}$ & 1 & 0 & 0 \\
\hline
$U\cdot P_y$ & ${1\over\sqrt{2}}\cos y$ & $\sqrt{{5 + \cos(2y) + 2\sqrt{6}\sin(2y)\cos\delta_y}\over{5 + \cos(2y) - 2\sqrt{6}\sin(2y)\cos\delta_y}}$ & ${1\over\sqrt{3}}\sin y$ & ${1\over{6\sqrt{6}}}\sin(2y)\sin\delta_y$ \\
\hline
$U\cdot P_z$ & ${1\over\sqrt{2}}\sec z$ & $\sqrt{{6 + 3\cos(2z) - 3\sqrt{3}\sin(2z)\cos\delta_z}\over{6 + 3\cos(2z) + 3\sqrt{3}\sin(2z)\cos\delta_z}}$ & $\sqrt{2\over3}\sin z$ & ${1\over{6\sqrt{3}}}\sin(2z)\sin\delta_z$ \\
\hline
\end{tabular}
\begin{quote}
\caption{The results of $T_{12}$, $T_{23}$, $S_{13}$ and $J_{\rm CP}$ as perturbing TBM.}
\label{TabTBM}
\end{quote}
\end{center}
\end{table}

\clearpage
\section{Numerical Analyses}

In this section, we analyze the numerical results obtained from the formulation derived above. In fact, the procedure for perturbing all these nine symmetric mixing patterns are analogous, so we take the Tri-Bimaximal mixing as an example and present the corresponding results of the rest ones in Appendices \ref{AppenB}.

Our strategy is following: in the equations presented in last section, we let the left side $T_{ij}$ be the experimentally measured values which are based on a global fit of the neutrino oscillations and listed  in Table \ref{AngleFit}, while the right side is the formulas we derived by perturbing the symmetric forms. Equating the two sides, we obtain several relations between the model parameters, meanwhile we take into account the experimental errors. Plotting them in a figure (Fig. \ref{CPxUtbm}, for example), we have three curves which respectively satisfy the relations for $T^{exp}_{12},\; T^{exp}_{13},\; T^{exp}_{23}$. With the experimental errors, the three curves expand into three contour bands whose boundaries correspond to the error tolerance. We will observe the diagrams and see if they have overlapping regions. If there exists a common region(s) for the model parameters where all the three equations are satisfied simultaneously, we would say, this scheme is plausible, instead, if there is no such a common region, the scheme is not successful and must be abandoned. For instance, in the case of $P_x\cdot U_{TBM}$, we have
\begin{eqnarray}
T_{12}^{exp}&=&2 \sqrt{{1 + \sin(2x) \cos \delta_x}\over{5+3 \cos(2x) - 4 \sin(2x) \cos \delta_x}}, \\
T_{23}^{exp}&=&\cos x, \\
S_{13}^{exp}&=&{1\over{\sqrt{2}}}\sin x,
\end{eqnarray}
where the superscript "exp" refers to the experimental data.
Solving these equations, we obtain three curves which correspond to relations between the model parameters $x$ and $\delta_x$ as shown in Fig. \ref{CPxUtbm}. Due to the experimental errors, the curves expand into bands. The rest schemes are similar and we will not respectively discuss the results with different perturbation ansatzes in every detail, but show them in the following sections and appendices.

For more explicitly demonstrating the fitting effects, we provide the scatter plots. In the plots we set $\theta_{12}$, $\theta_{23}$, $\theta_{13}$ and $J_{CP}$ as horizontal and vertical axes alternatively, then mark the experimental data of the corresponding quantities, and each of them spreads into a band whose width is 3 standard deviations (3-$\sigma$). There is an overlapping region where both experimental data are satisfied within 3$\sigma$s. Then we plot our theoretical predictions by letting the model (perturbation) parameters scan their whole allowed ranges (for example, for $P_x\cdot U_{\rm TBM}$, $0^{\circ}\leq x\leq 180^{\circ}$ and $0^{\circ}\leq \delta_x\leq 180^{\circ}$). If the theoretically predicted values which are calculated with a given perturbation ansatz (the red dots) fall into the overlapping region, it means that the equation about the model parameters has solutions which coincide with the data at least within 3-$\sigma$ tolerance. If there are not red dots in the region, the model fails to provide a solution, so that does not work at all. Then even though in all the four diagrams solutions for the model parameters seem to exist, we have to investigate if the solutions  provided by the four scatter plots correspond to the same model parameter region. Indeed, the answer resides in the curved band diagrams. Whereas, the scatter plots can offer some detailed information about the mixing angles and $J_{\rm CP}$ which will be measured in the future experiments.

\subsection{$P_x\cdot U_{\rm{TBM}}$}
\begin{figure}
\centering
\includegraphics[bb=0 0 550 550, height=8.5cm, width=8.5cm, angle=0]{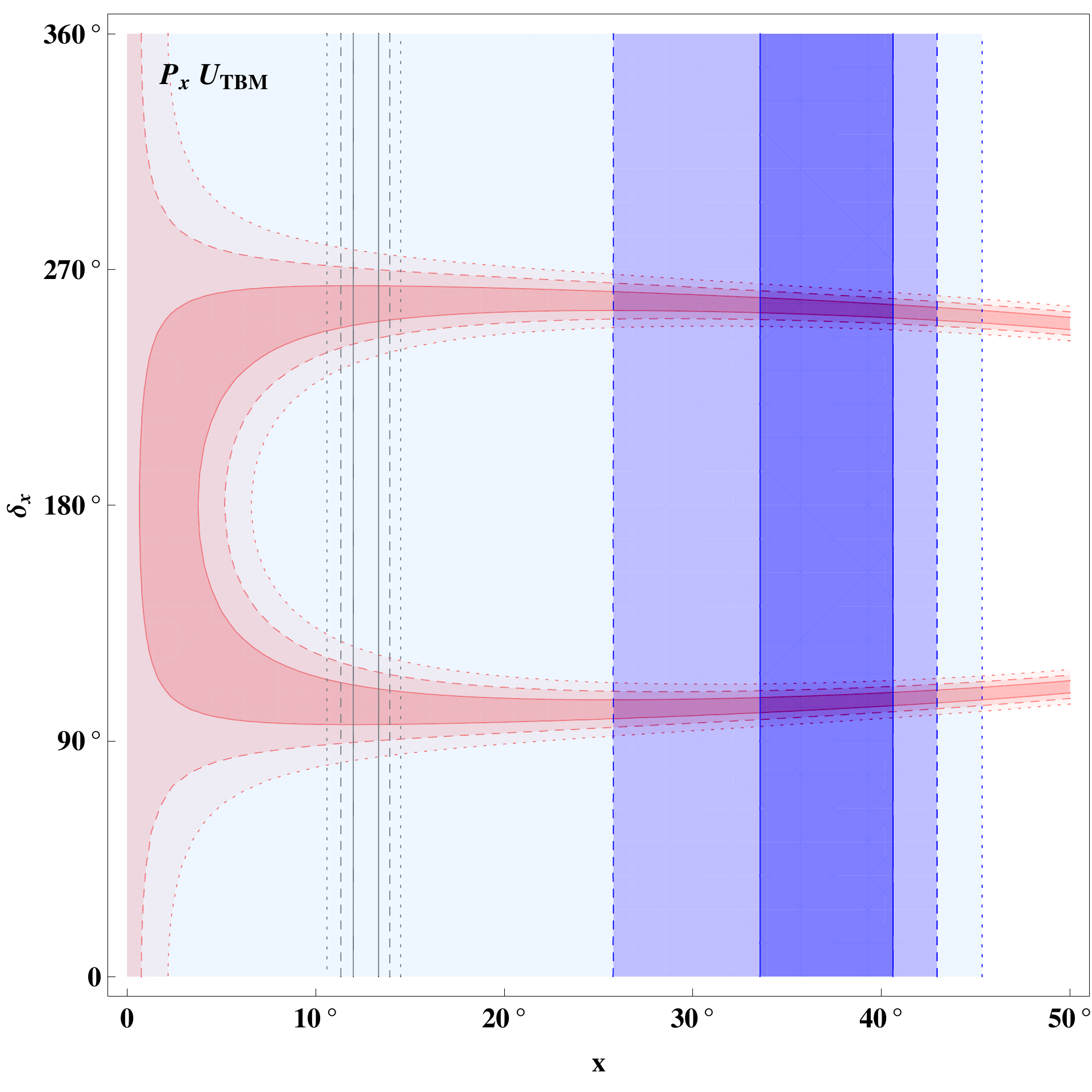}
\caption{The solutions corresponding to the measured $\theta_{12}$, $\theta_{23}$, and $\theta_{13}$ for $P_x\cdot U_{\rm{TBM}}$ in the parameter space of $x - \delta_x$. The red, light red, and pink regions are corresponding to the 1-$\sigma$, 2-$\sigma$ and 3-$\sigma$ tolerance levels of $\theta_{12}$ (data from Table \ref{AngleFit}) which are divided by red sold, dashed and dotted lines respectively. The blue, light blue and nattier blue regions are corresponding to $\theta_{23}$ are for 1-$\sigma$, 2-$\sigma$ and 3-$\sigma$ tolerance levels of $\theta_{23}$ (Table \ref{AngleFit}) which are divided by blue sold, dashed and dotted lines respectively. The 1-$\sigma$, 2-$\sigma$ and 3-$\sigma$ ranges of $\theta_{13}$ (without a special color mark in the diagram) are divided by black solid, dashed and dotted lines, respectively. Similar diagrams for other schemes in this work are labeled with these conventions. (including the color convention.)}
\label{CPxUtbm}
\end{figure}

\begin{figure}
\centering
\includegraphics[bb=0 110 550 550, height=12.5cm, width=15.5cm, angle=0]{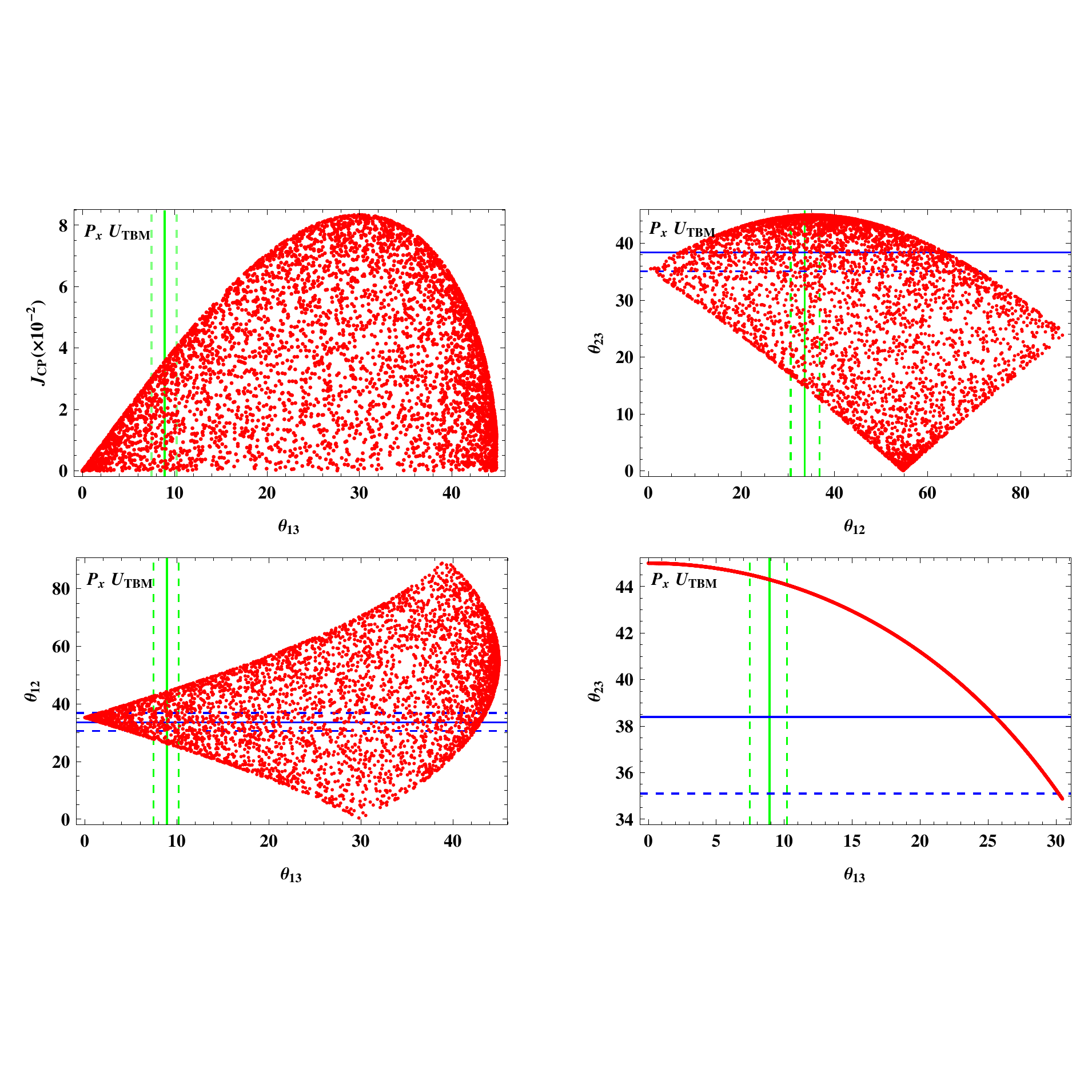}
\caption{The scatter plots for $\theta_{12}$, $\theta_{23}$, $\theta_{13}$ and $J_{\rm CP}$ with the $P_x\cdot U_{\rm{TBM}}$ anzatz. The central values and 3-$\sigma$ ranges (by fitting the data shown in Table \ref{AngleFit}) of the three mixing angles are labeled by solid lines and dashed lines, and green for horizontal axis and blue for ordinate one, respectively. (Color convention.)}
\label{FigPxUtbm}
\end{figure}

We present the curved bands of $\theta_{12}$, $\theta_{23}$, and $\theta_{13}$ in Fig. \ref{CPxUtbm} where the perturbation variables $x$ and $\delta_{x}$ serve as the two perpendicular coordinate axes. The bands are obtained by fitting the data presented in Table \ref{AngleFit}. The red, light red, and pink regions correspond to 1-$\sigma$, 2-$\sigma$ and 3-$\sigma$ ranges of $\theta_{12}$, and these three regions are divided by red sold, dashed, and dotted lines, respectively. Similarly, the blue, light blue, and nattier blue regions are for 1-$\sigma$, 2-$\sigma$, and 3-$\sigma$ ranges of $\theta_{23}$ whose boundaries are marked by blue sold, dashed, and dotted lines, respectively. The 1-$\sigma$, 2-$\sigma$ and 3-$\sigma$ ranges of $\theta_{13}$ are divided by black solid, dashed and dotted lines. In this work, all the curved band diagrams are labelled under this convention.

The scatter plots among the three mixing angles $\theta_{12}$, $\theta_{23}$, $\theta_{13}$ and Jarlskog invariant $J_{\rm CP}$ are shown in Fig. \ref{FigPxUtbm}. In the perturbation ansatz $P_x\cdot U_{\rm{TBM}}$, $J_{\rm CP}$ varies in a range $0\sim 4\times 10^{-2}$. Many points fall in the 3-$\sigma$ overlapping region of $\theta_{12}-\theta_{23}$ and $\theta_{13}-\theta_{12}$ whereas for $\theta_{13}-\theta_{23}$ the points squeeze on a line which is far away from the central value of $\theta_{23}$, as long as we require the points not to deviate from the central value of $\theta_{13}$ by more than 3$\sigma$s. It means that simultaneously fitting these two mixing angles is difficult with the $P_x\cdot U_{TBM}$ ansatz, at least not very optimistic.

By the scatter plots, it is noted that  $\theta_{23}$ does not exceed $45^\circ$ within the 3-$\sigma$ ranges of $\theta_{12}$ and $\theta_{13}$. For the $P_x\cdot U_{\rm{TBM}}$ scheme the conclusion $\theta_{23}<45^\circ$  does not change in the whole perturbation parameter space of $x$ and $\delta_x$, i.e., $x, \delta_x\in ( 0^\circ, 360^\circ )$. The constraint $\theta_{23}<45^\circ$ can also be seen from a correlation listed in Table \ref{TabTBM} as
\begin{eqnarray}
\tan^2\theta_{23}=1-2\sin^2\theta_{13}
\end{eqnarray}
which indicates that zero $\theta_{13}$ results in $\theta_{23}=45^\circ$ or vice versa and non-zero $\theta_{13}$ requires  $\theta_{23}$ to be less than $45^\circ$.

\subsection{$P_z\cdot U_{\rm{TBM}}$}

\begin{figure}
\centering
\includegraphics[bb=0 0 550 550, height=8.5cm, width=8.5cm, angle=0]{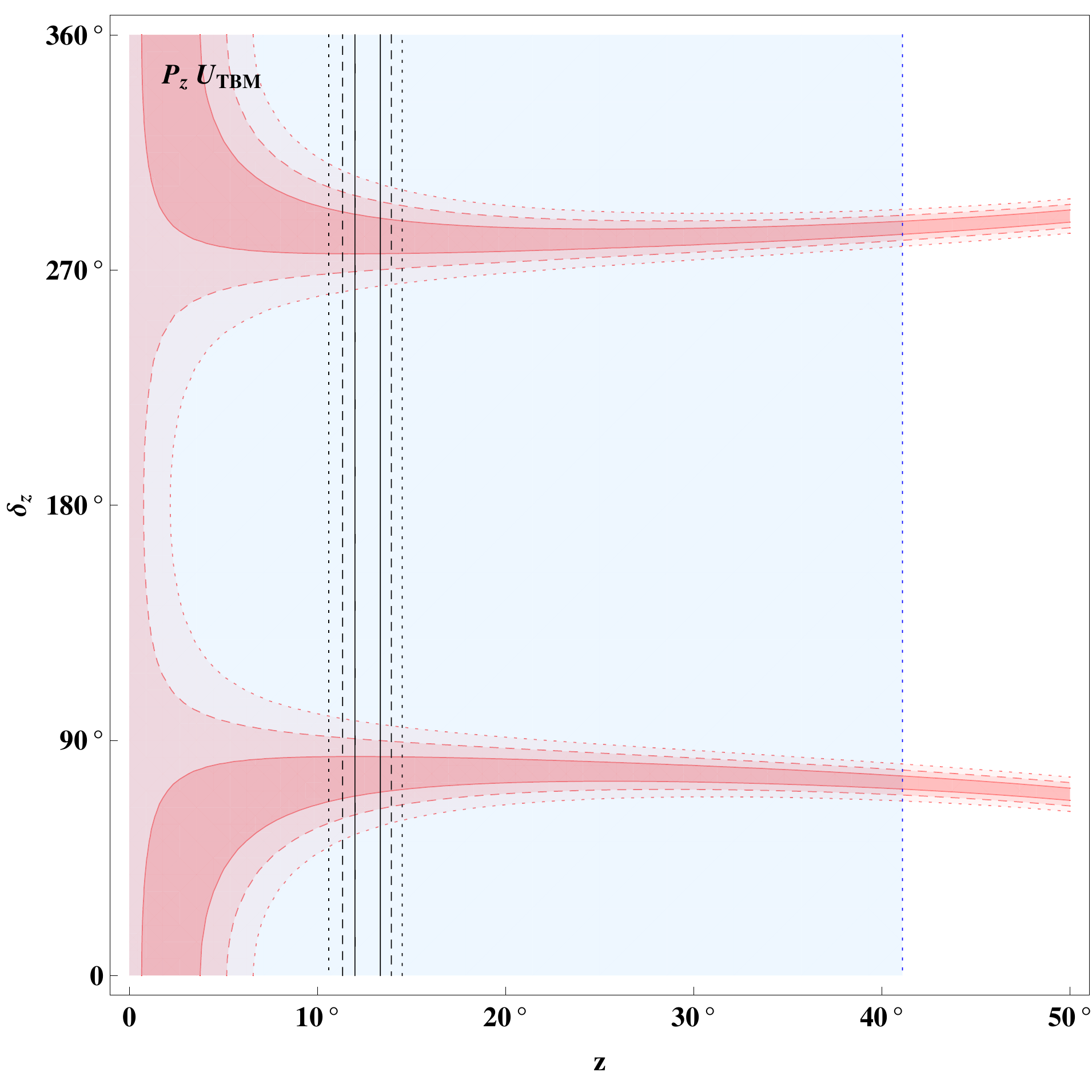}
\caption{The solutions corresponding to the measured $\theta_{12}$, $\theta_{23}$, and $\theta_{13}$ for $P_z\cdot U_{\rm{TBM}}$ in the parameter space of $z - \delta_z$. Caption is the same as displayed in Fig. \ref{CPxUtbm}.} \label{CPzUtbm}
\end{figure}

The curved bands of $\theta_{12}$, $\theta_{23}$ and $\theta_{13}$ in the whole space of the perturbation variables $z$ and $\delta_z$ are presented in Fig. \ref{CPzUtbm}. It is noted that the curved band for $\theta_{12}$ overlaps with that of $\theta_{13}$ within 1-$\sigma$ while its overlap with $\theta_{23}$ band is 2-$\sigma$s from its central value with $z\in (0^\circ, 50^\circ)$ and $\delta_z\in (0^\circ, 360^\circ)$. This leads to a conclusion that the  perturbation ansatz $P_z\cdot U_{\rm{TBM}}$ is more difficult to accommodate the experimental values of three mixing angles simultaneously compared to $P_x\cdot U_{\rm{TBM}}$.

\begin{figure}
\centering
\includegraphics[bb=0 110 550 550, height=12.5cm, width=15.5cm, angle=0]{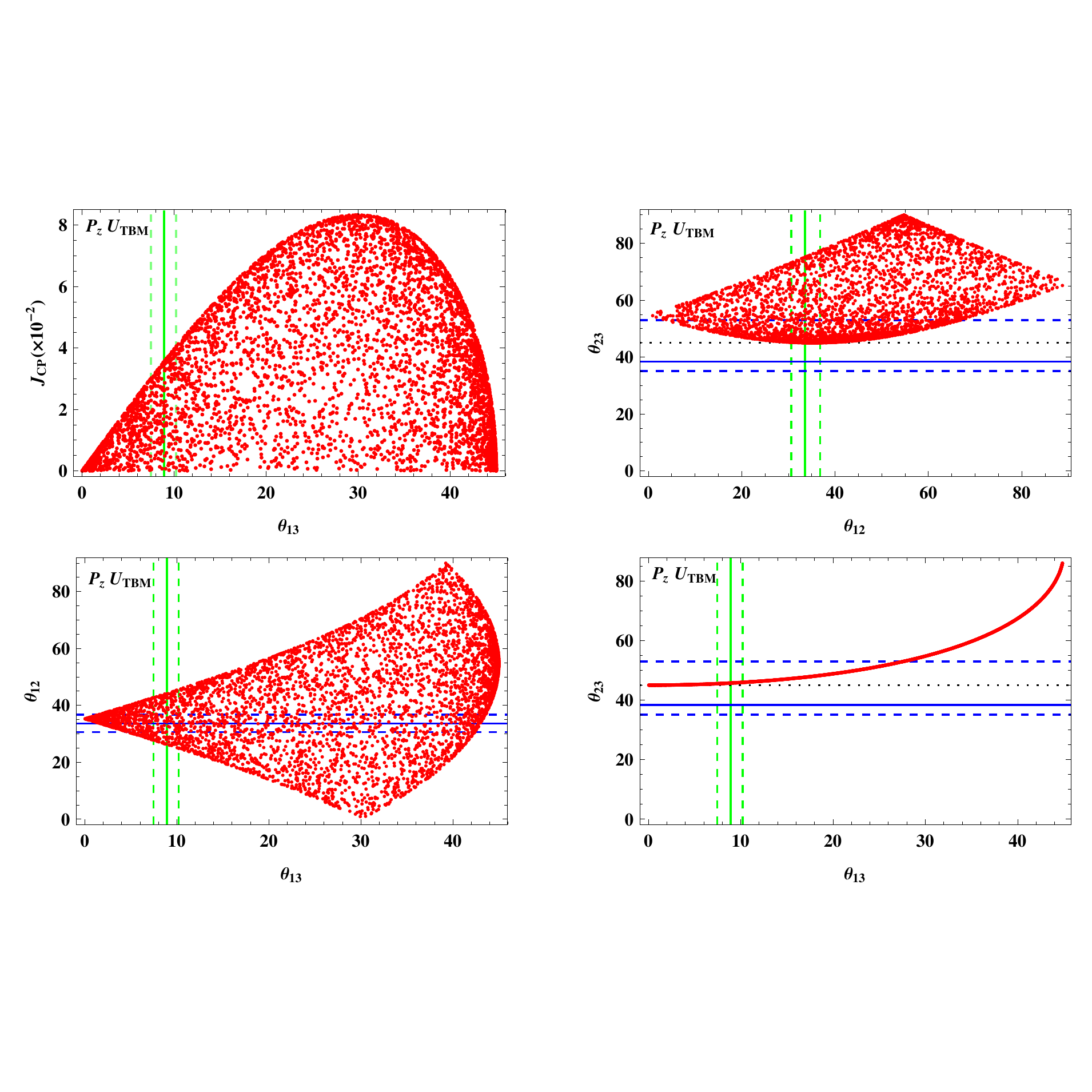}
\caption{The scatter plots for $\theta_{12}$, $\theta_{23}$, $\theta_{13}$ and $J_{\rm CP}$ with the  $P_z\cdot U_{\rm{TBM}}$ ansatz. Caption is the same as displayed in Fig. \ref{FigPxUtbm}.
} \label{FigPzUtbm}
\end{figure}

In Fig. \ref{FigPzUtbm} we present the scatter plots among the three mixing angles $\theta_{12}$, $\theta_{23}$, $\theta_{13}$ and Jarlskog invariant $J_{CP}$. The perturbation ansatz $P_z\cdot U_{\rm TBM}$ provides an upper limit $4\times 10^{-2}$ for $J_{\rm CP}$.
There are many points lie in the 3-$\sigma$ overlapping region of  $\theta_{13}-\theta_{12}$ while for $\theta_{12}-\theta_{23}$ and $\theta_{13}-\theta_{23}$, our points fall far away from the central value of $\theta_{23}$.

From $T_{23}$ and $S_{13}$ in Table \ref{TabTBM}, we have a correlation
\begin{eqnarray}
\tan^2\theta_{23}={1\over{1-2\sin^2\theta_{13}}},
\end{eqnarray}
which manifests that a zero-$\theta_{13}$ leads to $\theta_{23}=45^\circ$ or vice versa, while a non-zero $\theta_{13}$ determines  $\theta_{23}>45^\circ$.

\subsection{$U_{\rm{TBM}}\cdot P_y$}

\begin{figure}
\centering
\includegraphics[bb=0 0 550 550, height=8.5cm, width=8.5cm, angle=0]{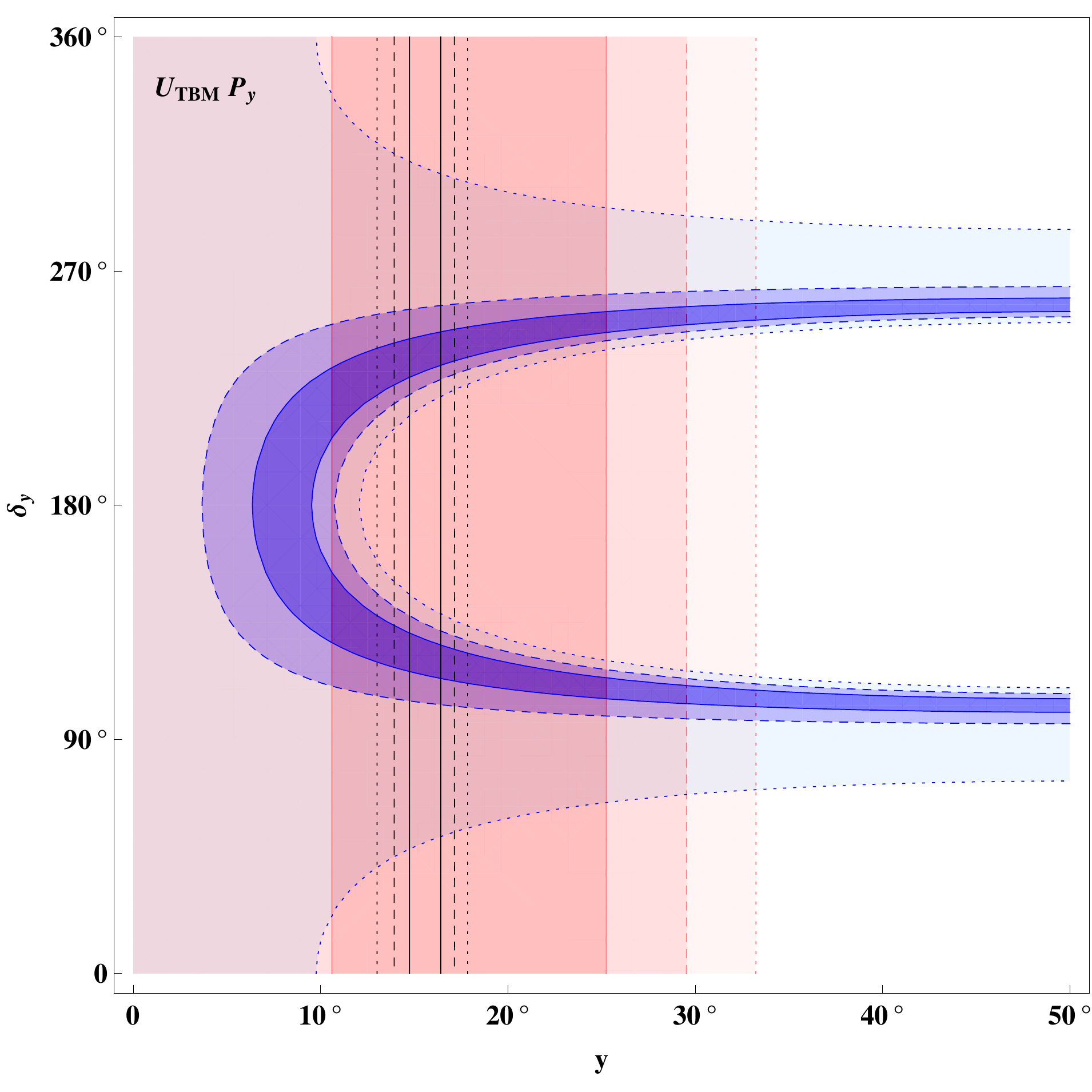}
\caption{The solutions corresponding to the measured $\theta_{12}$, $\theta_{23}$, and $\theta_{13}$ for the $U_{\rm{TBM}}\cdot P_y$ ansatz in the space of parameters $y - \delta_y$. Caption is the same as displayed in Fig. \ref{CPxUtbm}.} \label{CUtbmPy}
\end{figure}

\begin{figure}
\centering
\includegraphics[bb=0 110 550 550, height=12.5cm, width=15.5cm, angle=0]{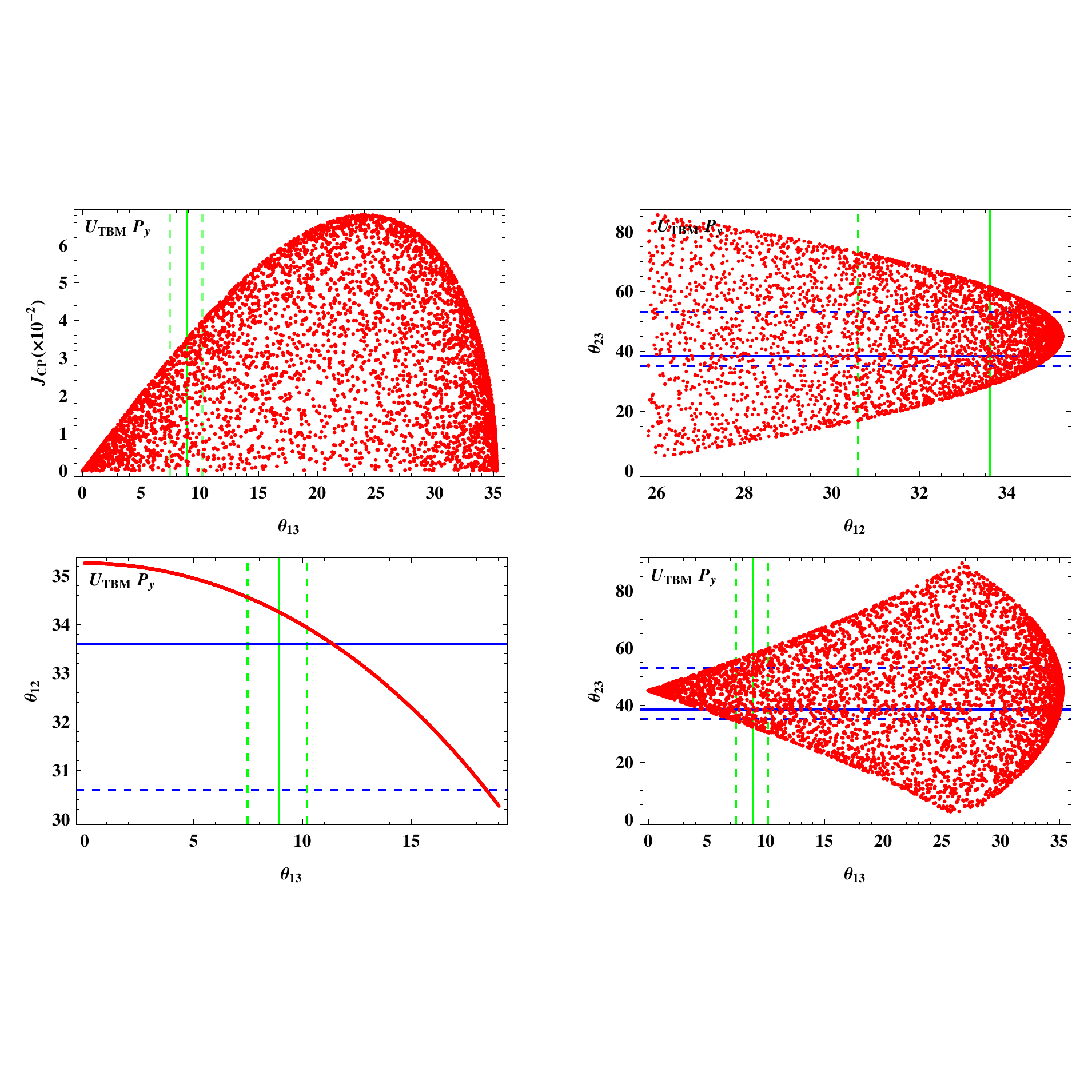}
\caption{The scatter plots for $\theta_{12}$, $\theta_{23}$, $\theta_{13}$ and $J_{\rm CP}$ for $U_{\rm{TBM}}\cdot P_y$. Caption is the same as displayed in Fig. \ref{FigPxUtbm}.} \label{FigUtbmPy}
\end{figure}

For $U_{\rm{TBM}}\cdot P_y$ the curved bands of the three mixing angles  are shown in Fig. \ref{CUtbmPy}. It is noted that $\theta_{12}$, $\theta_{23}$ and $\theta_{13}$ share an overlapping region within 1-$\sigma$. It indicates that the perturbation ansatz $U_{\rm TBM}\cdot P_y$ provides a plausible scheme to accommodate all the experimental values of three mixing angles.

The scatter plots among the three mixing angles and Jarlskog invariant in the perturbation ansatz $U_{\rm TBM}\cdot P_y$ are shown in Fig. \ref{FigUtbmPy}. The $\theta_{13}-J_{\rm CP}$ presents an upper limit of $J_{\rm CP}$ approximately $4\times 10^{-2}$. There are large amounts of points lying in the 3-$\sigma$ overlapping regions of $\theta_{12}-\theta_{23}$ and $\theta_{13}-\theta_{23}$ while for $\theta_{13}-\theta_{12}$, points squeeze on a line. Even though the line deviates from the crossing point of the central values of $\theta_{13}$ and $\theta_{12}$, this line does pass through the 1-$\sigma$ overlapping region of $\theta_{13}-\theta_{12}$.

Whether $\theta_{23}>45^\circ$ or $\theta_{23}<45^\circ$ cannot be determined in this perturbation ansatz and the solution points are observed to be symmetric about the horizontal line $\theta_{23}=45^\circ$.

\begin{figure}
\centering
\includegraphics[bb=0 184 550 360, height=6cm, width=15.5cm, angle=0]{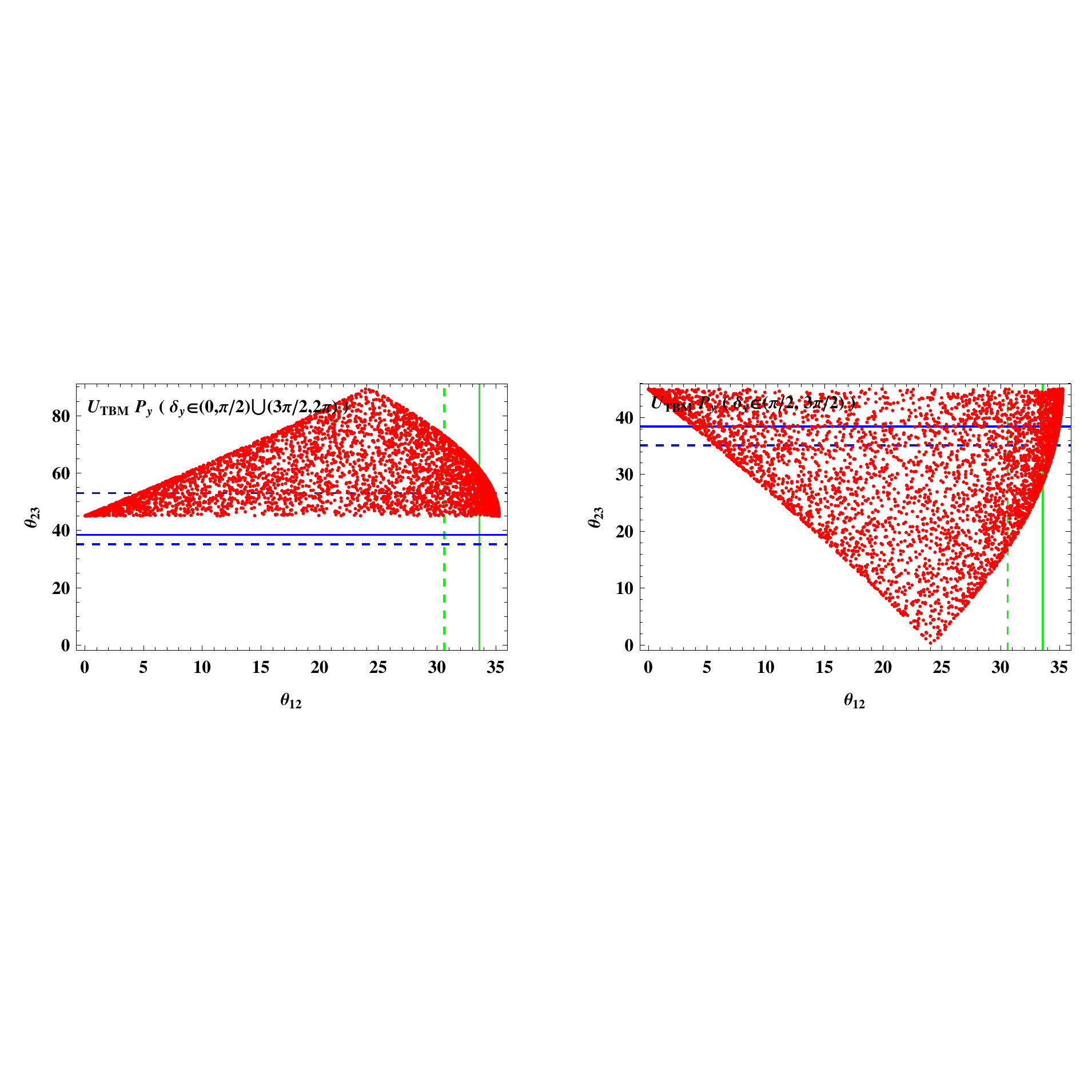}
\caption{The solutions corresponding to the measured $\theta_{12}$ and $\theta_{23}$ for $\delta_y\in(0,\pi/2)\cup(3\pi/2,2\pi)$ (left panel) and $\delta_y\in(\pi/2, 3\pi/2)$ (right panel) with $y\in(0, \pi/2)$ in the perturbation ansatz $U_{\rm{TBM}}\cdot P_z$. The central values and 3-$\sigma$ ranges (from the fit in Table \ref{AngleFit}) of these two mixing angles are labeled by solid lines and dashed lines, green for horizontal axis and blue for ordinate one, respectively.} \label{CUtbmPy45}
\end{figure}

In the scatter plot of $\theta_{12}-\theta_{23}$, our calculations indicate that as $\delta_y$ varies in the range $(0,\pi/2)\cup(3\pi/2,2\pi)$  $\theta_{23}>45^\circ$ whereas  $\delta_{y}\in(\pi/2,3\pi/2)$ we note  $\theta_{23}<45^\circ$. This relationship can be confirmed by scanning  the different parameter ranges of $\delta_y$ presented in Fig. \ref{CUtbmPy45}. With the ansatz $U_{\rm{TBM}}\cdot P_y$, $U_{e3}$  becomes
\begin{eqnarray}
(U_{\rm{TBM}}\cdot P_y)_{e3}={1\over\sqrt{3}}e^{-i\delta_y}\sin y,
\label{UtbmPye3}
\end{eqnarray}
and $U_{e3}$  from PMNS matrix in Eq. (\ref{PMNS2}) is
\begin{eqnarray}
(U_{\rm{PMNS}})_{e3}=e^{-i\delta}\sin \theta_{13}.
\label{Upmnse3}
\end{eqnarray}
In this case, it is easy to get a conclusion that as $y\in (0, \pi/2)$, $\delta_y\in (0,2\pi)$, and letting $\delta\in (0, 2\pi)$, Eq. (\ref{UtbmPye3}) and Eq. (\ref{Upmnse3}) would demand $\delta=\delta_y$. The equivalence between $\delta$ and $\delta_y$ implies that the CP phase $\delta$ determines whether $\theta_{23}>45^\circ$ or $\theta_{23}<45^\circ$ or vice versa. Table \ref{AngleFit} provides 1-$\sigma$ range of $\delta$ (the best fit value approximately is $190^\circ$) located in $(\pi/2, 3\pi/2)$, thus $\theta_{23}$ should be smaller than $45^\circ$ with the $U_{\rm TBM}\cdot P_y$ ansatz.
This implication can also be derived from the expressions of $T_{23}$ and $S_{13}$ given in Table \ref{TabTBM} as
\begin{eqnarray}
\tan^2\theta_{23}={{1 - \sin^2\theta_{13} + 2\sqrt{2}\sin\theta_{13}\sqrt{1-3\sin^2\theta_{13}}\cos\delta_y}\over{1 - \sin^2\theta_{13} - 2\sqrt{2}\sin\theta_{13}\sqrt{1-3\sin^2\theta_{13}}\cos\delta_y}}.
\end{eqnarray}
The relation indicates that $\delta_y\in(0,\pi/2)\cup(3\pi/2, 2\pi)$ requires  $\theta_{23}>45^\circ$ and   $\theta_{23}<45^\circ$ as $\delta_y\in (\pi/2, 3\pi/2)$.  Equality $\delta=\delta_y$ means that $\theta_{23}>$ or $<45^\circ$ can be determined by the CP phase $\delta$ or vice versa.

\subsection{$U_{\rm{TBM}}\cdot P_z$}

\begin{figure}
\centering
\includegraphics[bb=0 0 550 550, height=8.5cm, width=8.5cm, angle=0]{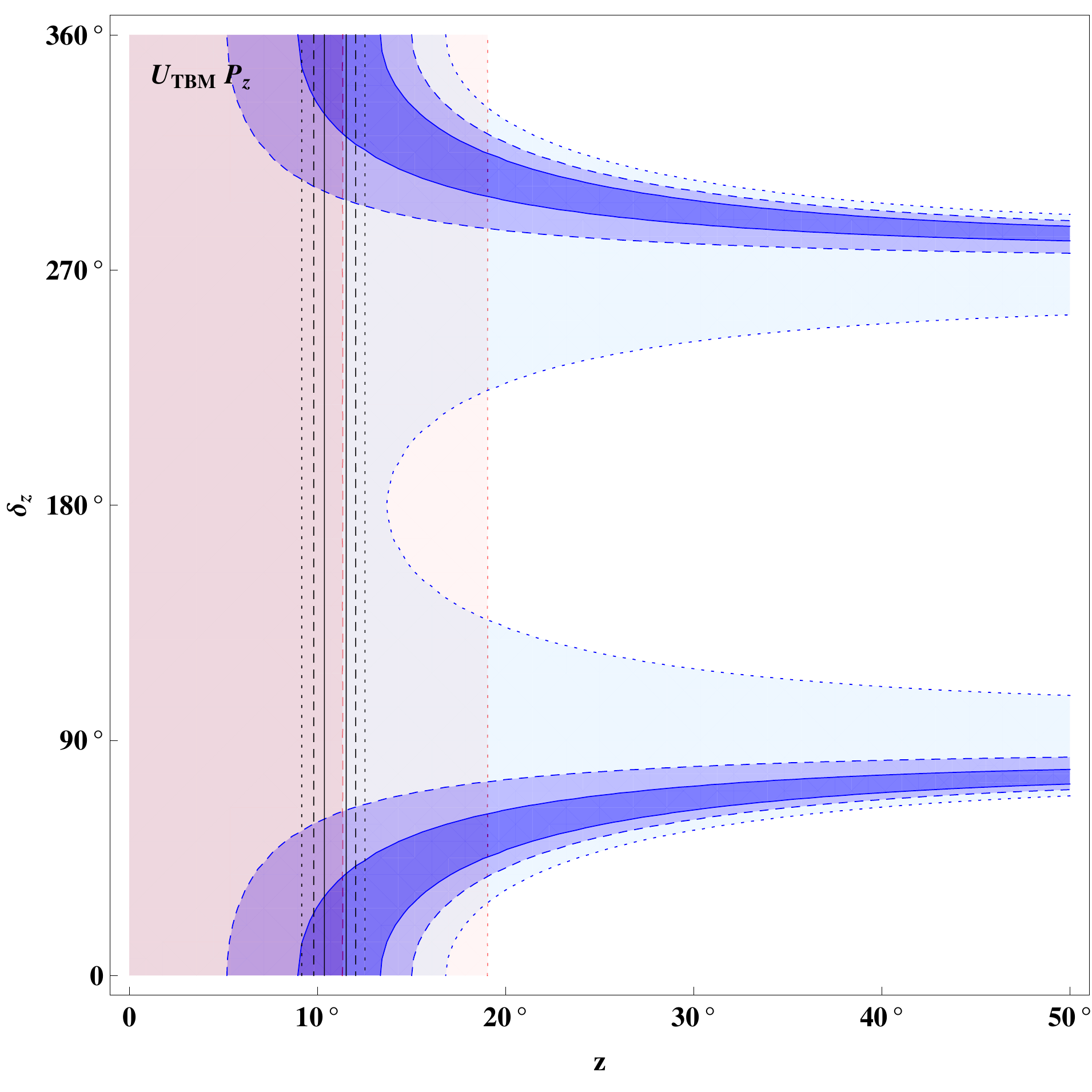}
\caption{The solutions corresponding to the measured $\theta_{12}$, $\theta_{23}$, and $\theta_{13}$ for $U_{\rm{TBM}}\cdot P_z$ in the parameter space of $z - \delta_z$. Caption is the same as displayed in Fig. \ref{CPxUtbm}.} \label{CUtbmPz}
\end{figure}

Now let us turn to the ansatz $U_{\rm{TBM}}\cdot P_z$.
The curved bands of the three mixing angles in the perturbation ansatz $U_{\rm{TBM}}\cdot P_z$ is shown in Fig. \ref{CUtbmPz}. It is obvious that the curved band for $\theta_{23}$ overlaps with that of $\theta_{13}$ within 1-$\sigma$, and $\theta_{12}$ band shares overlapping regions with $\theta_{23}$ and $\theta_{13}$ within two or three $\sigma$s.

\begin{figure}
\centering
\includegraphics[bb=0 110 550 550, height=12.5cm, width=15.5cm, angle=0]{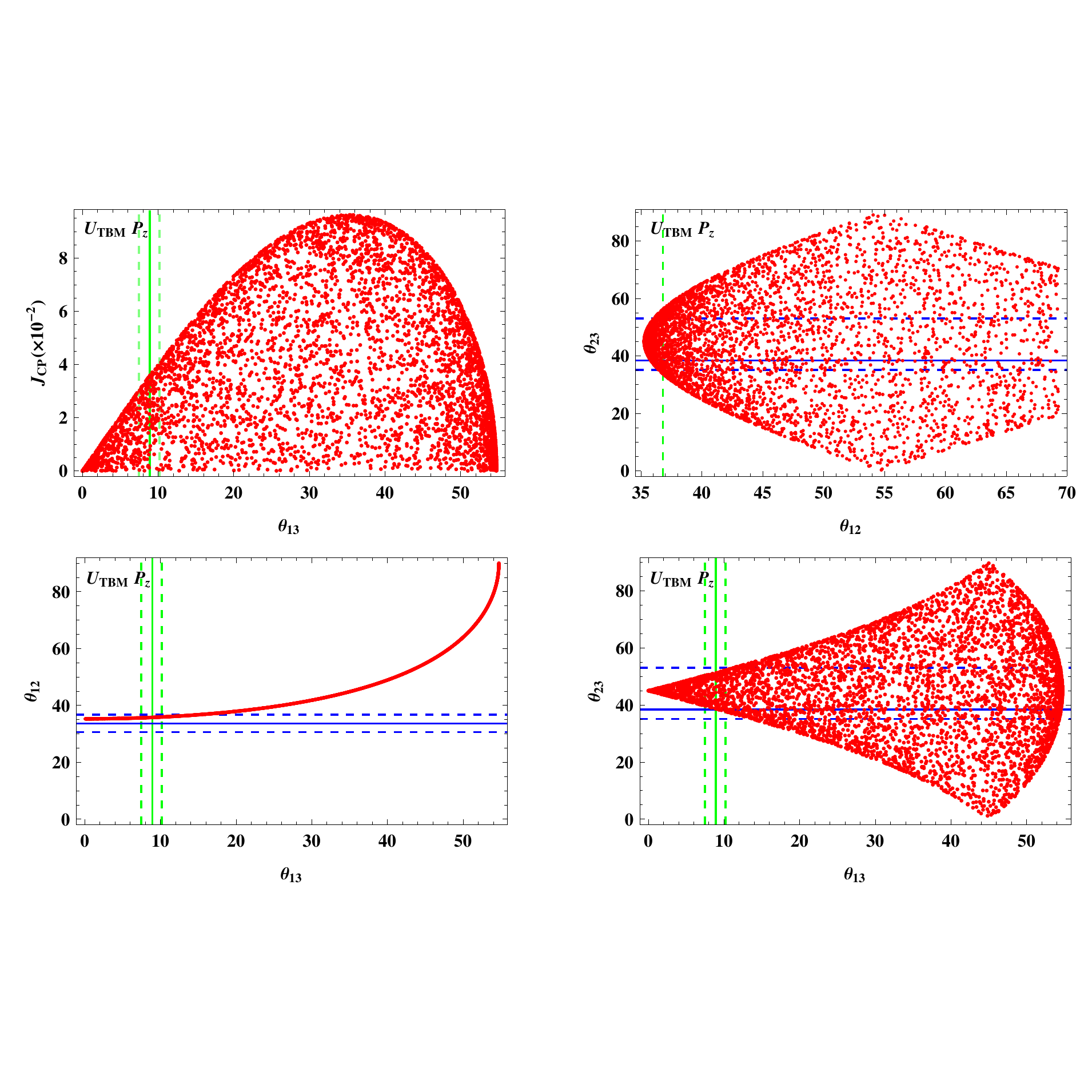}
\caption{The scatter plots for correlations among $\theta_{12}$, $\theta_{23}$, $\theta_{13}$ and $J_{\rm CP}$ for $U_{\rm{TBM}}\cdot P_z$. Caption is the same as displayed in Fig. \ref{FigPxUtbm}.} \label{FigUtbmPz}
\end{figure}

The scatter plots among these three mixing angles and Jarlskog invariant $J_{\rm CP}$ are shown in Fig. \ref{FigUtbmPz}. Under the perturbation ansatz $U_{\rm TBM}\cdot P_z$, the Jarlskog invariant possesses its upper limit approximately $4\times 10^{-2}$.
Plenty of points lie in the 3-$\sigma$ overlapping region of $\theta_{13}-\theta_{23}$ while for $\theta_{12}-\theta_{23}$ and $\theta_{13}-\theta_{12}$, the points are far away from the central value of $\theta_{12}$.

\begin{figure}
\centering
\includegraphics[bb=0 184 550 360, height=6cm, width=15.5cm, angle=0]{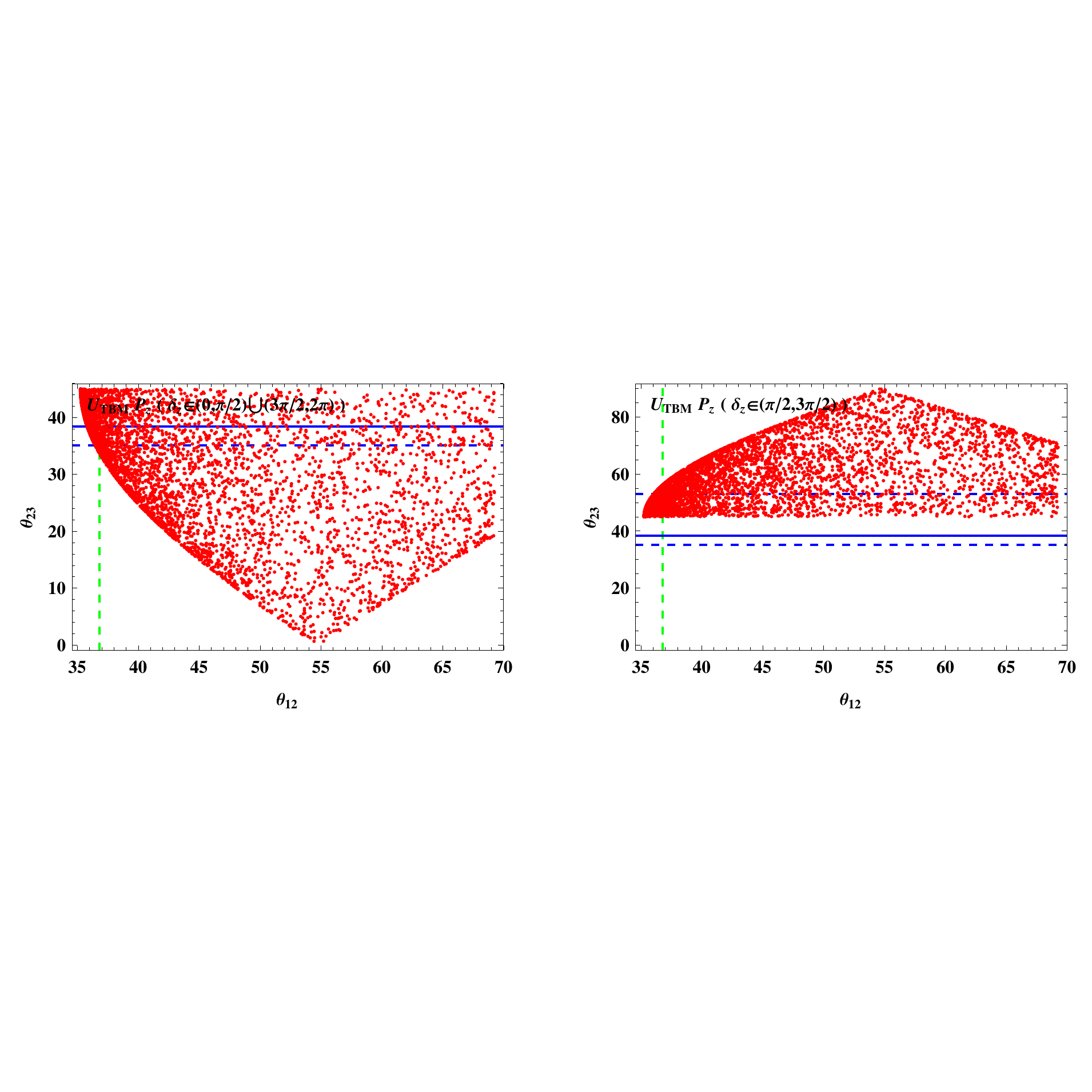}
\caption{The scatter plot for correlation between $\theta_{12}$ and $\theta_{23}$ for $\delta_z\in(0,\pi/2)\cup(3\pi/2,2\pi)$ (left panel) and $\delta_z\in(\pi/2, 3\pi/2)$ (right panel) with $y\in(0, 1.3)$ in the perturbation mode $U_{\rm{TBM}}\cdot P_z$. The central values and 3-$\sigma$ ranges (from the fit result in Table \ref{AngleFit}) of these two mixing angles are labeled by solid lines and dashed lines, green for horizontal axis and blue for ordinate one, respectively.} \label{CUtbmPz45}
\end{figure}

Similarly to the ansatz $U_{\rm{TBM}}\cdot P_y$, $\delta_z$ varies in a range $(0,\pi/2)\cup(3\pi/2,2\pi)$, $\theta_{23}<45^\circ$ while $\delta_{z}\in(\pi/2,3\pi/2)$, $\theta_{23}>45^\circ$. The whole range $0\sim 2\pi$ of $\delta_z$ is scanned and we find that if $z\in(0, 1.3)$ and $\delta_z$ resides in first and fourth quadrants, $\theta_{23}<45^{\circ}$; whereas if $\delta_z$ is in second and third quadrants
$\theta_{23}>45^{\circ}$
as shown in Fig. \ref{CUtbmPz45}. If $z>1.3$, there exist  mirror-symmetric diagrams to the corresponding ones. In the perturbation ansatz $U_{\rm{TBM}}\cdot P_z$,  $U_{e3}$  is
\begin{eqnarray}
(U_{\rm{TBM}}\cdot P_z)_{e3}={1\over\sqrt{3}}e^{-i\delta_z}\sin z,
\label{UtbmPze3}
\end{eqnarray}
and $U_{e3}$  in the PMNS matrix is shown in Eq. (\ref{Upmnse3}). With $z\in (0, 1.3)$, $\delta_y\in (0,2\pi)$, and $\delta\in (0, 2\pi)$, Eq. (\ref{UtbmPze3}) and Eq. (\ref{Upmnse3}) lead to $\delta=\delta_z$. The equivalence between $\delta$ and $\delta_z$ leads to that the CP phase $\delta$ determines whether $\theta_{23}>45^\circ$ or $\theta_{23}<45^\circ$ or vice versa. With the fit value of $\delta$ in Table \ref{AngleFit}, we observe that $\theta_{23}$ is  larger than $45^\circ$ which is different from the conclusion made by the  $U_{\rm{TBM}}\cdot P_y$ ansatz. This can also be derived from the expressions of $T_{23}$ and $S_{13}$ given in Table \ref{TabTBM},
\begin{eqnarray}
\tan^2\theta_{23}={{3 - \sin^2\theta_{13} - 3\sqrt{2}\sin\theta_{13}\sqrt{1-{3\over2}\sin^2\theta_{13}}
\cos\delta_z}\over{3 - \sin^2\theta_{13} + 3\sqrt{2}\sin\theta_{13}\sqrt{1-{3\over2}\sin^2\theta_{13}}
\cos\delta_z}},
\end{eqnarray}
from which, the ranges of $\delta_z\in(0,\pi/2)\cup(3\pi/2, 2\pi)$  and $\delta_z\in(\pi/2, 3\pi/2)$ determine $\theta_{23}<45^\circ$ or  $>45^\circ$.

\section{Discussion and Conclusion}

The recent experiments determine a non-zero $\theta_{13}$ which is in contrary to the prediction made by most of the symmetric textures for the lepton mixing matrix.  Even though the real PMNS matrix deviates from the  symmetric form, an approximate symmetry is obvious. Moreover, it is believed that the symmetric texture is resulted in from the physics at higher energy scales and is broken during its evolution to lower energy scales. To investigate what physics is at high energy scale, we would study what mechanism breaks the symmetry.  Following the schemes given in literature, we adopt the perturbation to deform the symmetric texture into the real PMNS matrix. Because the original symmetry is approximately retained, a perturbation may be a suitable choice. In this work we perturb the nine given matrix textures which possess symmetric patterns by various ansatzes. Owing to the similarity  in all the cases, we take the Tri-bimaximal mixing pattern as an example to exhibit how this perturbation method applies. We summarize the results in Table \ref{Tab9P}. Various ansatzes by which the three mixing angles receive corrections are carefully analyzed and their effects are marked by a tick $\surd$ or a cross $\times$ to note if the ansatz is favored or disfavored. The subscripts $S$, $L$, $SL$ of ${\theta_{23}}_S$, ${\theta_{23}}_L$ and ${\theta_{23}}_{SL}$ in Table IV imply that $\theta_{23}$ acquires values smaller, larger, and smaller or larger than $45^\circ$, respectively. And the mark $\otimes$ signifies the situation in which the three mixing angles acquire corrections and $\theta_{13}$ is non-zero, but the ansatz cannot make the theoretical values to be consistent with that data in Table \ref{AngleFit}. We also mark the best perturbation ansatz with a star symbol $\star$ in the table.

\begin{table}
\begin{center}
\begin{tabular}{|c||c|c|c|c|c|c|}\hline
Constant Pattern & $P_x\cdot U$ & $P_y\cdot U$ & $P_z\cdot U$ & $U\cdot P_x$ & $U\cdot P_y$ & $U\cdot P_z$ \\
\hline\hline
TBM & $\surd ({\theta_{23}}_S)$ & $\times$ & $\surd ({\theta_{23}}_L)$ & $\times$ & $\surd ({\theta_{23}}_{SL})\star$ & $\surd ({\theta_{23}}_{SL})$\\
\hline
DM & $\otimes$ & $\times$ & $\otimes$ & $\times$ & $\otimes$ & $\otimes$ \\
\hline
BM & $\surd({\theta_{23}}_S)$ & $\times$ & $\surd({\theta_{23}}_L)$ & $\times$ & $\otimes$ & $\otimes$ \\
\hline
GRM1 & $\surd({\theta_{23}}_S)$ & $\times$ & $\surd({\theta_{23}}_L)$ & $\times$ & $\surd({\theta_{23}}_{SL})$ & $\surd({\theta_{23}}_{SL})$ \\
\hline
GRM2 & $\surd({\theta_{23}}_S)$ & $\times$ & $\surd({\theta_{23}}_L)$ & $\times$ & $\surd({\theta_{23}}_{SL})$ & $\surd({\theta_{23}}_{SL})$ \\
\hline
HM & $\surd({\theta_{23}}_S)$ & $\times$ & $\surd({\theta_{23}}_L)$ & $\times$ & $\otimes$ & $\otimes$ \\
\hline
TMM & $\surd({\theta_{23}}_S)$ & $\times$ & $\surd({\theta_{23}}_{SL})$ & $\times$ & $\otimes$ & $\otimes$ \\
\hline
TFH1 & $\otimes$ & $\times$ & $\surd({\theta_{23}}_L)$ & $\times$ & $\otimes$ & $\surd({\theta_{23}}_{SL})$ \\
\hline
TFH2 & $\surd({\theta_{23}}_S)$ & $\times$ & $\otimes$ & $\times$ & $\otimes$ & $\surd({\theta_{23}}_{SL})$ \\
\hline
\end{tabular}
\begin{quote}
\caption{The list of the results for perturbing the nine symmetric mixing textures with $P_x$, $P_y$, and $P_z$. The mark $\surd$ denotes that the corrections from the perturbation scheme make all the three mixing angles to be consistent with data, the mark $\times$ means that in this ansatz, the calculated $\theta_{13}$ still remains as exact zero, so such ansatz does not work at all, and $\otimes$ for the cases that the three corrected mixing angles could not be compatible with the data presented in Table \ref{AngleFit} simultaneously. $\theta_{23}>45^\circ$, $\theta_{23}<45^\circ$, and $\theta_{23}>$ or $<45^\circ$ are marked by subscripts $S$, $L$, and $SL$, respectively. The best perturbation ansatz is labeled with $\star$.}
\label{Tab9P}
\end{quote}
\end{center}
\end{table}

Alternative perturbation schemes have also been proposed.
Let us still take the Tri-bimaximal mixing as an example to illustrate this new scheme: $U_{\rm{TBM}} = R_{23}(45^\circ) R_{13}(0^\circ) R_{12}(\arctan(1/\sqrt{2})) = R_{23}(45^\circ) R_{12}(\arctan(1/\sqrt{2}))$, inserting perturbation matrices between $R_{23}$ and $R_{12}$, namely, $R_{23}\cdot P \cdot R_{12}$ where $P$ is a suitable perturbation $3\times 3$ matrix. However, such a perturbation ansatz cannot provide feasible mixing angles to be consistent with the experimental data. Therefore such schemes are not phenomenologically favorable.

It is observed that the relationship between $\theta_{23}$ and $\delta_{i}$ (or $\delta$) ($i=x, y, z$) could fix $\theta_{23}>45^\circ$ or $\theta_{23}<45^\circ$, thus more precise measurements on $\theta_{23}$ constrain the range of the CP phase.

The equality $\delta=\delta_i$ $(i=x, y, z)$ is derived as $i$ is  constrained in the first quadrant $i\in (0, \pi/2)$, but when it is in the second quadrant $i\in (\pi/2, \pi)$, the result would not remain the same. From our analytical equations of $J_{\rm CP}$, $J_{\rm CP}$ is proportional to  $\sin(2i)\sin\delta_i$ (for the TMM case, this relationship does not exist) and from standard form of the PMNS matrix (\ref{PMNS2}) and definition of the Jarlskog invariant (\ref{JcpD}), we determine $J_{\rm CP}\sim \sin \delta$. We  present a relation between $J_{\rm CP}(\delta)$ and the perturbation parameters in Table \ref{TabJdelta}.

\begin{table}
\begin{center}
\begin{tabular}{|c||c|c|c|c|c|}\hline
$J_{\rm CP}$ $(\delta)$ & $\delta_i\in (0,\pi/2)$ & $\delta_i\in (\pi/2, \pi)$ & $\delta_i\in (\pi, 3\pi/2)$ & $\delta_i\in (3\pi/2, 2\pi)$ \\
\hline\hline
$i\in (0,\pi/2)$ & $+$ $(\delta<\pi)$ & $+$ $(\delta<\pi)$ & $-$ $(\delta>\pi)$ & $-$ $(\delta>\pi)$ \\
\hline
$i\in (\pi/2, \pi)$ & $-$ $(\delta>\pi)$ & $-$ $(\delta<\pi)$ & $+$ $(\delta<\pi)$ & $+$ $(\delta<\pi)$ \\
\hline
\end{tabular}
\begin{quote}
\caption{The relationship between $J_{\rm CP}$ $(\delta)$ and perturbation parameters $i$, $\delta_i$ ($i=x, y, z$), where a plus $+$ for a positive $J_{\rm CP}$ and minus $-$ for a negative one. }
\label{TabJdelta}
\end{quote}
\end{center}
\end{table}

Our numerical analysis indicates that the Tri-bimaximal model (TBM) is a more favorable texture that may accommodate a sizable $\theta_{13}$ after a perturbative correction. With the perturbation,  $\theta_{23}$ and $\theta_{13}$ deviate from $45^\circ$ and $0^\circ$ as required by the data, and it means that the $\mu-\tau$ symmetry \cite{mutau} originally embedded in the neutrino mass matrix is broken by the perturbation. Especially the $U_{\rm TBM}\cdot P_y$ provides the most plausible perturbation ansatz for the theoretical mixing angles to be consistent with the experimental values in the 1-$\sigma$ level. This indicates that the most viable correction to TBM is produced by the rotation in the $2-3$ plane, i.e. to break the $\mu-\tau$ symmetry by a perturbation. This provides us a clue for the model building in the future.

\begin{acknowledgments}
This work is supported by the National Natural Science Foundation of China under the contract No.11075079, 11135009.
\end{acknowledgments}

\appendix
\section{The analytical results of the rest eight constant mixing patterns}
\label{AppenA}

In this section, we list the results of $T_{12}$, $T_{23}$, $S_{13}$ and $J_{\rm CP}$ after perturbations to DM (Table \ref{TabDM}), BM (Table \ref{TabBM}), GRM1 (Table \ref{TabGRM1a}, \ref{TabGRM1b}, \ref{TabGRM1c}), GRM2 (Table \ref{TabGRM2a}, \ref{TabGRM2b}, \ref{TabGRM2c}), HM (Table \ref{TabHM}), TMM (Table \ref{TabTMMa}, \ref{TabTMMb}, \ref{TabTMMc}, \ref{TabTMMd}), TFH1 (Table \ref{TabTFH1a}, \ref{TabTFH1b}), TFH2 (Table \ref{TabTFH2a}, \ref{TabTFH2b}).

\begin{table}
\begin{center}
\begin{tabular}{|c||c|c|c|c|}\hline
DM & $T_{12}$ & $T_{23}$ & $S_{13}$ & $J_{\rm CP}$ \\
\hline\hline
$P_x\cdot U$ & $\sqrt{{6 + 3\cos(2x) + 3\sqrt{3}\sin(2x)\cos\delta_x}\over{6 + 3\cos(2x) - 3\sqrt{3}\sin(2x)\cos\delta_x}}$ & $\sqrt{2}\cos x$ & $\sqrt{2\over3}\sin x$ & ${1\over{6\sqrt{3}}}\sin(2x)\sin\delta_x$ \\
\hline
$P_y\cdot U$ & 1 & $\sqrt{{3 + \cos(2y) + 2\sqrt{2}\sin(2y)\cos\delta_y}\over{3 - \cos(2y) - 2\sqrt{2}\sin(2y)\cos\delta_y}}$ & 0 & 0 \\
\hline
$P_z\cdot U$ & $\sqrt{{5 + \cos(2z) - 2\sqrt{6}\sin(2z)\cos\delta_z}\over{5 + \cos(2z) + 2\sqrt{6}\sin(2z)\cos\delta_z}}$ & $\sqrt{2}\sec z$ & ${1\over\sqrt{3}}\sin z$ & ${1\over{6\sqrt{6}}}\sin(2z)\sin\delta_z$ \\
\hline
$U\cdot P_x$ & $\sqrt{{1 + \sin(2x)\cos\delta_x}\over{1 - \sin(2x)\cos\delta_x}}$ & $\sqrt{2}$ & 0 & 0 \\
\hline
$U\cdot P_y$ & $\cos y$ & ${1\over2}\sqrt{{5 + 3\cos(2y) + 4\sin(2y)\cos\delta_y}\over{1-\sin(2y)\cos\delta_y}}$ & ${1\over\sqrt{2}}\sin y$ & ${1\over{12}}\sin(2y)\cos\delta_y$ \\
\hline
$U\cdot P_z$ & $\sec z$ & ${1\over2}\sqrt{{5 + 3\cos(2z) - 4\sin(2z)\cos\delta_z}\over{1 + \sin(2z)\cos\delta_z}}$ & ${1\over\sqrt{2}}\sin z$ & ${1\over{12}}\sin(2z)\sin\delta_z$ \\
\hline
\end{tabular}
\begin{quote}
\caption{Results of $T_{12}$, $T_{23}$, $S_{13}$ and $J_{\rm CP}$ after perturbation to DM.}
\label{TabDM}
\end{quote}
\end{center}
\end{table}

\begin{table}
\begin{center}
\begin{tabular}{|c||c|c|c|c|}\hline
BM & $T_{12}$ & $T_{23}$ & $S_{13}$ & $J_{\rm CP}$ \\
\hline\hline
$P_x\cdot U$ & $\sqrt{{3 + \cos(2x) + 2\sqrt{2}\sin(2x)\cos\delta_x}\over{3 + \cos(2x) - 2\sqrt{2}\sin(2x)\cos\delta_x}}$ & $\cos x$ & ${1\over\sqrt{2}}\sin x$ & ${1\over{8\sqrt{2}}}\sin(2x)\sin\delta_x$ \\
\hline
$P_y\cdot U$ & 1 & $\sqrt{{1 + \sin(2y)\cos\delta_y}\over{1 - \sin(2y)\cos\delta_y}}$ & 0 & 0 \\
\hline
$P_z\cdot U$ & $\sqrt{{3 + \cos(2z) - 2\sqrt{2}\sin(2z)\cos\delta_z}\over{3 + \cos(2z) + 2\sqrt{2}\sin(2z)\cos\delta_z}}$ & $\sec z$ & ${1\over\sqrt{2}}\sin z$ & ${1\over{8\sqrt{2}}}\sin (2z) \sin \delta_z$ \\
\hline
$U\cdot P_x$ & $\sqrt{{1 + \sin(2x)\cos\delta_x}\over{1 - \sin(2x)\cos\delta_x}}$ & 1 & 0 & 0 \\
\hline
$U\cdot P_y$ & $\cos y$ & $\sqrt{{3 + \cos(2y) + 2\sqrt{2}\sin(2y)\cos\delta_y}\over{3 + \cos(2y) - 2\sqrt{2}\sin(2y)\cos\delta_y}}$ & ${1\over\sqrt{2}}\sin y$ & ${1\over{8\sqrt{2}}}\sin(2y)\sin\delta_y$ \\
\hline
$U\cdot P_z$ & $\sec z$ & $\sqrt{{3 + \cos(2z) - 2\sqrt{2}\sin(2z)\cos\delta_z}\over{3 + \cos(2z) + 2\sqrt{2}\sin(2z)\cos\delta_z}}$ & ${1\over\sqrt{2}}\sin z$ & ${1\over{8\sqrt{2}}}\sin(2z)\sin\delta_z$ \\
\hline
\end{tabular}
\begin{quote}
\caption{Results of $T_{12}$, $T_{23}$, $S_{13}$ and $J_{\rm CP}$ after perturbation to BM.}
\label{TabBM}
\end{quote}
\end{center}
\end{table}

\begin{table}
\begin{center}
\begin{tabular}{|c||c|}\hline
GRM1 & $T_{12}$  \\
\hline\hline
$P_x\cdot U$ & $\sqrt{{7 + \sqrt{5} + (1-\sqrt{5})\cos(2x) + 2\sqrt{2}(1+\sqrt{5})\sin(2x)\cos\delta_x}\over{2\left[4 + \sqrt{5} + (2+\sqrt{5})\cos(2x) - \sqrt{2}(1+\sqrt{5})\sin(2x)\cos\delta_x \right]}}$ \\
\hline
$P_y\cdot U$ & ${2\over{1+\sqrt{5}}}$  \\
\hline
$P_z\cdot U$ & $\sqrt{{7 + \sqrt{5} + (1-\sqrt{5})\cos(2z) - 2\sqrt{2}(1+\sqrt{5})\sin(2z)\cos\delta_z}\over{2[4 + \sqrt{5} + (2+\sqrt{5})\cos(2z)+\sqrt{2}(1+\sqrt{5})\sin(2z)\cos\delta_z}]}$ \\
\hline
$U\cdot P_x$ & $\sqrt{{5 + \sqrt{5} - (1+\sqrt{5})\cos(2x) + 2(1+\sqrt{5})\sin(2x)\cos\delta_x}\over{5 + \sqrt{5} + (1+\sqrt{5})\cos(2x) - 2(1+\sqrt{5})\sin(2x)\cos\delta_x}}$  \\
\hline
$U\cdot P_y$ & ${2\over{1+\sqrt{5}}}\cos y$ \\
\hline
$U\cdot P_z$ & ${2\over{1+\sqrt{5}}}\sec z$ \\
\hline
\end{tabular}
\begin{quote}
\caption{Results of $T_{12}$ after perturbation to GRM1.}
\label{TabGRM1a}
\end{quote}
\end{center}
\end{table}

\begin{table}
\begin{center}
\begin{tabular}{|c||c|}\hline
GRM1 & $T_{23}$ \\
\hline\hline
$P_x\cdot U$ & $\cos x$  \\
\hline
$P_y\cdot U$ & $\sqrt{{1 + \sin(2y)\cos\delta_y}\over{1 - \sin(2y)\cos\delta_y}}$  \\
\hline
$P_z\cdot U$ & $\sec z$  \\
\hline
$U\cdot P_x$ & 1 \\
\hline
$U\cdot P_y$ & $\sqrt{{15 + \sqrt{5} + (5-\sqrt{5})\cos(2y) + 2\sqrt{10(5+\sqrt{5})}\sin(2y)\cos\delta_y}\over{15 + \sqrt{5} + (5-\sqrt{5})\cos(2y) - 2\sqrt{10(5+\sqrt{5})}\sin(2y)\cos\delta_y}}$  \\
\hline
$U\cdot P_z$ & $\sqrt{{(5 + 7\sqrt{5})+(5 + 3\sqrt{5})\cos(2z) -
 2\sqrt{10(5 + \sqrt{5})}\sin(2z)\cos\delta_z}\over{(5 + 7\sqrt{5}) + (5 + 3\sqrt{5})\cos(2z) +
 2\sqrt{10(5 + \sqrt{5})}\sin(2z)\cos\delta_z}}$ \\
\hline
\end{tabular}
\begin{quote}
\caption{Results of $T_{23}$ after perturbation to GRM1.}
\label{TabGRM1b}
\end{quote}
\end{center}
\end{table}

\begin{table}
\begin{center}
\begin{tabular}{|c||c|c|}\hline
GRM1 & $S_{13}$ & $J_{\rm CP}$ \\
\hline\hline
$P_x\cdot U$ & ${1\over\sqrt{2}}\sin x$ & ${1\over{4\sqrt{10}}}\sin(2x)\sin\delta_x$ \\
\hline
$P_y\cdot U$ & 0 & 0 \\
\hline
$P_z\cdot U$ & ${1\over\sqrt{2}}\sin z$ & ${1\over4\sqrt{10}}\sin(2z)\sin\delta_z$ \\
\hline
$U\cdot P_x$ & 0 & 0 \\
\hline
$U\cdot P_y$ & $\sqrt{2\over{5+\sqrt{5}}}\sin y$ & ${1\over{2\sqrt{10(5+\sqrt{5})}}}\sin(2y)\sin\delta_y$ \\
\hline
$U\cdot P_z$ & $\sqrt{{1\over2}(1+{1\over\sqrt{5}})}\sin z$ & ${1\over4}\sqrt{{1\over{10}}(1+{1\over\sqrt{5}})} \sin(2z)\sin\delta_z$ \\
\hline
\end{tabular}
\begin{quote}
\caption{Results of $S_{13}$ and $J_{\rm CP}$ after perturbation to GRM1.}
\label{TabGRM1c}
\end{quote}
\end{center}
\end{table}

\begin{table}
\begin{center}
\begin{tabular}{|c||c|}\hline
GRM1 & $T_{12}$  \\
\hline\hline
$P_x\cdot U$ & $\sqrt{{13 - \sqrt{5} + (7-3\sqrt{5})\cos(2x) + 2(1+\sqrt{5})\sqrt{5-\sqrt{5}}\sin(2x)\cos\delta_x}\over{11 + \sqrt{5} + (1+3\sqrt{5})\cos(2x) - 2(1+\sqrt{5})\sqrt{5-\sqrt{5}}\sin(2x)\cos\delta_x}}$ \\
\hline
$P_y\cdot U$ & ${{\sqrt{10-2\sqrt{5}}}\over{1+\sqrt{5}}}$  \\
\hline
$P_z\cdot U$ & $\sqrt{{13 - \sqrt{5} + (7-3\sqrt{5})\cos(2z) - 2(1+\sqrt{5})\sqrt{5-\sqrt{5}}\sin(2z)\cos\delta_z}\over{11 + \sqrt{5} + (1+3\sqrt{5})\cos(2z) + 2(1+\sqrt{5})\sqrt{5-\sqrt{5}}\sin(2z)\cos\delta_z}}$ \\
\hline
$U\cdot P_x$ & $\sqrt{{8 - 2(\sqrt{5}-1)\cos(2x) + (1+\sqrt{5})\sqrt{10-2\sqrt{5}}\sin(2x)\cos\delta_x}\over{8 + 2(\sqrt{5}-1)\cos(2x) - (1+\sqrt{5})\sqrt{10-2\sqrt{5}}\sin(2x)\cos\delta_x}}$  \\
\hline
$U\cdot P_y$ & ${{\sqrt{10-2\sqrt{5}}}\over{1+\sqrt{5}}}\cos y$ \\
\hline
$U\cdot P_z$ & ${{\sqrt{10-2\sqrt{5}}}\over{1+\sqrt{5}}}\sec z$ \\
\hline
\end{tabular}
\begin{quote}
\caption{Results of $T_{12}$ after perturbation to GRM2.}
\label{TabGRM2a}
\end{quote}
\end{center}
\end{table}

\begin{table}
\begin{center}
\begin{tabular}{|c||c|}\hline
GRM1 & $T_{23}$ \\
\hline\hline
$P_x\cdot U$ & $\cos x$  \\
\hline
$P_y\cdot U$ & $\sqrt{{1+\sin(2y)\cos\delta_y}\over{1-\sin(2y)\cos\delta_y}}$  \\
\hline
$P_z\cdot U$ & $\sec z$  \\
\hline
$U\cdot P_x$ & 1 \\
\hline
$U\cdot P_y$ & $\sqrt{{11 + \sqrt{5} + (5-\sqrt{5})\cos(2y) + 4(1+\sqrt{5})\sin(2y)\cos\delta_y}\over{11 + \sqrt{5} + (5-\sqrt{5})\cos(2y) - 4(1+\sqrt{5})\sin(2y)\cos\delta_y}}$  \\
\hline
$U\cdot P_z$ & $\sqrt{{13 - \sqrt{5} + (3+\sqrt{5})\cos(2z) - 4\sqrt{10-2\sqrt{5}}\sin(2z)\cos\delta_z}\over{13 - \sqrt{5} + (3+\sqrt{5})\cos(2z) + 4\sqrt{10-2\sqrt{5}}\sin(2z)\cos\delta_z}}$ \\
\hline
\end{tabular}
\begin{quote}
\caption{Results of $T_{23}$ after perturbation to GRM2.}
\label{TabGRM2b}
\end{quote}
\end{center}
\end{table}

\begin{table}
\begin{center}
\begin{tabular}{|c||c|c|}\hline
GRM1 & $S_{13}$ & $J_{\rm CP}$ \\
\hline\hline
$P_x\cdot U$ & ${1\over\sqrt{2}}\sin x$ & ${1\over{64}} \sqrt{5-\sqrt{5}} (1+\sqrt{5}) \sin(2x) \sin\delta_x$ \\
\hline
$P_y\cdot U$ & 0 & 0 \\
\hline
$P_z\cdot U$ & ${1\over\sqrt{2}}\sin z$ & ${1\over{64}}\sqrt{5-\sqrt{5}}(1+\sqrt{5})
\sin(2z)\sin\delta_z$ \\
\hline
$U\cdot P_x$ & 0 & 0 \\
\hline
$U\cdot P_y$ & ${1\over2}\sqrt{{1\over2}(5-\sqrt{5})}\sin y$ & ${\sqrt{5}\over{32}}\sin(2y)\sin\delta_y$ \\
\hline
$U\cdot P_z$ & ${{1+\sqrt{5}}\over4}\sin z$ & ${1\over{64}}(3+\sqrt{5})\sqrt{{1\over2}(5-\sqrt{5})} \sin(2z)\sin\delta_z$ \\
\hline
\end{tabular}
\begin{quote}
\caption{Results of $S_{13}$ and $J_{\rm CP}$ after perturbation to GRM2.}
\label{TabGRM2c}
\end{quote}
\end{center}
\end{table}

\begin{table}
\begin{center}
\begin{tabular}{|c||c|c|c|c|}\hline
BM & $T_{12}$ & $T_{23}$ & $S_{13}$ & $J_{\rm CP}$ \\
\hline\hline
$P_x\cdot U$ & $\sqrt{{5 - \cos(2x) + 2\sqrt{6}\sin(2x)\cos\delta_x}\over{7 + 5\cos(2x) - 2\sqrt{6}\sin(2x)\cos\delta_x}}$ & $\cos x$ & ${1\over\sqrt{2}}\sin x$ & ${1\over{16}}\sqrt{3\over2}\sin(2x)\sin\delta_x$ \\
\hline
$P_y\cdot U$ & ${1\over\sqrt{3}}$ & $\sqrt{{1 + \sin(2y)\cos\delta_y}\over{1 - \sin(2y)\cos\delta_y}}$ & 0 & 0 \\
\hline
$P_z\cdot U$ & $\sqrt{{5 - \cos(2z) - 2\sqrt{6}\sin(2z)\cos\delta_z}\over{7 + 5\cos(2z) + 2\sqrt{6}\sin(2z)\cos\delta_z}}$ & $\sec z$ & ${1\over\sqrt{2}}\sin z$ & ${1\over{16}}\sqrt{3\over2}\sin(2z)\sin\delta_z$ \\
\hline
$U\cdot P_x$ & ${1\over\sqrt{3}}\tan x$ & 0 & 0 & 0 \\
\hline
$U\cdot P_y$ & ${1\over\sqrt{3}}\cos y$ & $\sqrt{{7 + \cos(2y) + 4\sqrt{3}\sin(2y)\cos\delta_y}\over{7 + \cos(2y) - 4\sqrt{3}\sin(2y)\cos\delta_y}}$ & ${1\over2}\sin y$ & ${\sqrt{3}\over{32}}\sin(2y)\sin\delta_y$ \\
\hline
$U\cdot P_z$ & ${1\over\sqrt{3}}\sec z$ & $\sqrt{{5 + 3\cos(2z) - 4\sin(2z)\cos\delta_z}\over{5 + 3\cos(2z) + 4\sin(2z)\cos\delta_z}}$ & ${\sqrt{3}\over2}\sin z$ & ${3\over{32}}\sin(2z)\sin\delta_z$ \\
\hline
\end{tabular}
\begin{quote}
\caption{Results of $T_{12}$, $T_{23}$, $S_{13}$ and $J_{\rm CP}$ after perturbation to HM.}
\label{TabHM}
\end{quote}
\end{center}
\end{table}

\begin{table}
\begin{center}
\begin{tabular}{|c||c|}\hline
BM & $T_{12}$ \\
\hline\hline
$P_x\cdot U$ & $\sqrt{\frac{2 \sin (2 x) \left(2 \cos \delta _x-\sqrt{2} \sin \delta _x\right)-\cos (2 x)+5}{\sqrt{2} \sin (2 x) \sin \delta _x-2 \left(1+\sqrt{2}\right) \sin (2 x) \cos \delta _x+\frac{1}{2} \left(1+6 \sqrt{2}\right) \cos (2 x)+\frac{1}{2} \left(11+2 \sqrt{2}\right)}}$ \\
\hline
$P_y\cdot U$ & $2-\sqrt{2}$ \\
\hline
$P_z\cdot U$ & $\sqrt{\frac{-2 \sqrt{2} \sin (2 z) \sin \delta _z-4 \sin (2 z) \cos \delta _z-\cos (2 z)+5}{\sin (2 z)
   \left[\sqrt{2} \sin \delta _z+2 \left(1+\sqrt{2}\right) \cos \delta _z\right]+\frac{1}{2} \left(1+6
   \sqrt{2}\right) \cos (2 z)+\frac{1}{2} \left(11+2 \sqrt{2}\right)}}$ \\
\hline
$U\cdot P_x$ & $\sqrt{\frac{2 \left(2+\sqrt{2}\right) \sin (2 x) \cos \delta _x\-\left(1+2 \sqrt{2}\right) \cos (2 x)+2 \sqrt{2}+5}{-2
   \left(2+\sqrt{2}\right) \sin (2 x) \cos \delta _x+\left(1+2 \sqrt{2}\right) \cos (2 x)+2 \sqrt{2}+5}}$ \\
\hline
$U\cdot P_y$ & $\frac{\sqrt{2 \left(\sqrt{2}-2\right) \cos \delta_y  \sin (2 y)+\left(2 \sqrt{2}-1\right) \cos (2 y)-2 \sqrt{2}+5}}{\sqrt{6+4 \sqrt{2}}}$ \\
\hline
$U\cdot P_z$ & $2 \sqrt{\frac{1}{-2 \sin (2 z) \cos \delta _z+4 \sqrt{2} \cos (2 z)+6}}$ \\
\hline
\end{tabular}
\begin{quote}
\caption{Results of $T_{12}$ after perturbation to TMM.}
\label{TabTMMa}
\end{quote}
\end{center}
\end{table}

\begin{table}
\begin{center}
\begin{tabular}{|c||c|}\hline
BM &  $T_{23}$  \\
\hline\hline
$P_x\cdot U$ & $\frac{\sqrt{-2 \sqrt{2} \sin (2 x) \sin \delta _x-4 \left(\sqrt{2}-1\right) \sin (2 x) \cos \delta _x+\left(6\sqrt{2}-1\right) \cos (2 x)-2 \sqrt{2}+11}}{\sqrt{10+4 \sqrt{2}}}$ \\
\hline
$P_y\cdot U$ & $\sqrt{{2 \left(2+\sqrt{2}\right) \sin (2 y) \sin \delta _y+\left(1+2 \sqrt{2}\right) \sin (2 y) \cos \delta _y+2
   \sqrt{2}+5}\over{-2 \left(2+\sqrt{2}\right) \sin (2 y) \sin \delta _y-\left(1+2 \sqrt{2}\right) \sin (2 y) \cos \delta
   _y+2 \sqrt{2}+5}}$ \\
\hline
$P_z\cdot U$ & $\sqrt{10+4 \sqrt{2}} \sqrt{\frac{1}{2 \sin (2 z) \left[2 \left(\sqrt{2}-1\right) \cos \delta _z-\sqrt{2} \sin \delta
   _z\right]+\left(6 \sqrt{2}-1\right) \cos (2 z)-2 \sqrt{2}+11}}$ \\
\hline
$U\cdot P_x$ & 1 \\
\hline
$U\cdot P_y$ & $\sqrt{\frac{2 \sin (2 y) \left[\left(\sqrt{2}-2\right) \cos \delta _y-\left(4+2 \sqrt{2}\right) \sin \delta
   _y\right]+\left(2 \sqrt{2}-1\right) \cos (2 y)+2 \sqrt{2}+11}{2 \sin (2 y) \left[\left(4+2 \sqrt{2}\right) \sin \delta
   _y+\left(\sqrt{2}-2\right) \cos \delta _y\right]+\left(2 \sqrt{2}-1\right) \cos (2 y)+2 \sqrt{2}+11}}$ \\
\hline
$U\cdot P_z$ & $\sqrt{\frac{-\sin (2 z) \left(\cos \delta _z-4 \sin \delta _z\right)+2 \sqrt{2} \cos (2 z)+5}{-\sin (2 z) \left(4
   \sin \delta _z+\cos \delta _z\right)+2 \sqrt{2} \cos (2 z)+5}}$ \\
\hline
\end{tabular}
\begin{quote}
\caption{Results of $T_{23}$ after perturbation to TMM.}
\label{TabTMMb}
\end{quote}
\end{center}
\end{table}

\begin{table}
\begin{center}
\begin{tabular}{|c||c|}\hline
BM & $S_{13}$  \\
\hline\hline
$P_x\cdot U$ &  $\frac{1}{4} \sqrt{\sqrt{2} \sin (2 x) \sin \delta _x+2 \left(\sqrt{2}-1\right) \sin (2 x) \cos \delta
   _x+\frac{1}{2} \left(1-6 \sqrt{2}\right) \cos (2 x)+\frac{1}{2} \left(11-2 \sqrt{2}\right)}$  \\
\hline
$P_y\cdot U$ & $\frac{1}{4} \left(2-\sqrt{2}\right)$ \\
\hline
$P_z\cdot U$ & $\frac{\sqrt{-2 \sin (2 z) \left[2 \left(\sqrt{2}-1\right) \cos \delta _z-\sqrt{2} \sin \delta _z\right]+\left(1-6
   \sqrt{2}\right) \cos (2 z)-2 \sqrt{2}+11}}{4 \sqrt{2}}$ \\
\hline
$U\cdot P_x$ & $\frac{1}{4} \left(2-\sqrt{2}\right)$ \\
\hline
$U\cdot P_y$ & ${1\over{2 \sqrt{2}}}\sqrt{\left(2-\sqrt{2}\right) \sin (2 y) \cos \delta _y+\frac{1}{2} \left(1-2 \sqrt{2}\right) \cos (2 y)+\frac{1}{2}
   \left(5-2 \sqrt{2}\right)}$ \\
\hline
$U\cdot P_z$ & ${1\over{2 \sqrt{2}}}\sqrt{\cos \delta_z  \sin (2 z)-2 \sqrt{2} \cos (2 z)+3}$ \\
\hline
\end{tabular}
\begin{quote}
\caption{Results of $S_{13}$ after perturbation to TMM.}
\label{TabTMMc}
\end{quote}
\end{center}
\end{table}

\begin{table}
\begin{center}
\begin{tabular}{|c||c|}\hline
BM & $J_{\rm CP}$ \\
\hline\hline
$P_x\cdot U$ & $\frac{1}{64} \left[\frac{1}{2} \sin (2 x) \left(5 \sqrt{2} \sin \delta _x+6 \cos \delta _x\right)+2 \cos (2x)\right]$ \\
\hline
$P_y\cdot U$ & $\frac{1}{32} \cos (2 y)$ \\
\hline
$P_z\cdot U$ & $\frac{1}{64} \left[\frac{1}{2} \sin (2 z) \left(5 \sqrt{2} \sin \delta _z-6 \cos \delta _z\right)+2 \cos (2
   z)\right]$ \\
\hline
$U\cdot P_x$ & $\frac{1}{128} \left[\left(3 \sqrt{2}-2\right) \sin (2 x) \cos \delta _x+4 \cos (2 x)\right]$ \\
\hline
$U\cdot P_y$ & $\frac{1}{64} \left[\left(1+\frac{3}{\sqrt{2}}\right) \sin (2 y) \cos \delta _y+2 \cos (2 y)\right]$ \\
\hline
$U\cdot P_z$ & $\frac{1}{32} \left[2 \sqrt{2} \cos \delta_z  \sin (2 z)+\cos (2 z)\right]$ \\
\hline
\end{tabular}
\begin{quote}
\caption{Results of $J_{\rm CP}$ after perturbation to TMM.}
\label{TabTMMd}
\end{quote}
\end{center}
\end{table}

\begin{table}
\begin{center}
\begin{tabular}{|c||c|c|}\hline
BM & $T_{12}$ & $T_{23}$ \\
\hline\hline
$P_x\cdot U$ & $\sqrt{\frac{2\left[\sin (2 x) \cos \delta _x+1\right]}{-\sin (2 x) \cos \delta _x+\sqrt{3} \cos (2 x)+2}}$ & $\sqrt{\frac{1}{2} \sin (2 x) \cos \delta _x+\frac{1}{2} \sqrt{3} \cos (2 x)+1}$  \\
\hline
$P_y\cdot U$ & $\sqrt{3}-1$ & $\sqrt{\frac{2 \left(1+\sqrt{3}\right) \sin (2 y) \cos \delta _y+\sqrt{3} \cos (2 y)+\sqrt{3}+4}{-2 \left(1+\sqrt{3}\right)
   \sin (2 y) \cos \delta _y-\sqrt{3} \cos (2 y)+\sqrt{3}+4}}$ \\
\hline
$P_z\cdot U$ & $2 \sqrt{\frac{1-\sin (2 z) \cos \delta _z}{2 \left(1+\sqrt{3}\right) \sin (2 z) \cos \delta _z+\sqrt{3} \cos (2
   z)+\sqrt{3}+4}}$ & $\sqrt{2 \left(2+\sqrt{3}\right)} \sqrt{\frac{1}{2 \left(\sqrt{3}-1\right) \cos \delta_z  \sin (2 z)+\sqrt{3} \cos (2
   z)-\sqrt{3}+4}}$ \\
\hline
$U\cdot P_x$ & $\sqrt{\frac{2 \left(1+\sqrt{3}\right) \sin (2 x) \cos \delta _x-\sqrt{3} \cos (2 x)+\sqrt{3}+4}{-2
   \left(1+\sqrt{3}\right) \sin (2 x) \cos \delta _x+\sqrt{3} \cos (2 x)+\sqrt{3}+4}}$ & $\frac{1}{2} \left(1+\sqrt{3}\right)$  \\
\hline
$U\cdot P_y$ & $\frac{\sqrt{2 \left(\sqrt{3}-1\right) \sin (2 y) \cos \delta _y+\sqrt{3} \cos (2 y)-\sqrt{3}+4}}{\sqrt{2
   \left(2+\sqrt{3}\right)}}$ & $\frac{1}{2} \sqrt{\frac{2 \left(1+\sqrt{3}\right) \sin (2 y) \cos \delta _y+\sqrt{3} \cos (2 y)+\sqrt{3}+4}{1-\sin
   (2 y) \cos \delta _y}}$ \\
\hline
$U\cdot P_z$ & $\sqrt{\frac{2}{\sin (2 z) \cos \delta _z+\sqrt{3} \cos (2 z)+2}}$ & $\sqrt{\frac{-\sin (2 z) \cos \delta _z+\sqrt{3} \cos (2 z)+2}{\sqrt{2}\left[\sin (2 z) \cos \delta
   _z+1\right]}}$ \\
\hline
\end{tabular}
\begin{quote}
\caption{Results of $T_{12}$ and $T_{23}$ after perturbation to TFH1.}
\label{TabTFH1a}
\end{quote}
\end{center}
\end{table}

\begin{table}
\begin{center}
\begin{tabular}{|c||c|c|}\hline
BM & $S_{13}$ & $J_{\rm CP}$ \\
\hline\hline
$P_x\cdot U$ & $\sqrt{-\frac{1}{6} \sin (2 x) \cos \delta _x-\frac{\cos (2 x)}{2 \sqrt{3}}+\frac{1}{3}}$ & $\frac{\sin (2 x) \sin \delta _x}{6 \sqrt{3}}$ \\
\hline
$P_y\cdot U$ & $\frac{1}{6} \left(\sqrt{3}-3\right)$ & $\frac{\sin (2 y) \sin \delta _y}{12 \sqrt{3}}$ \\
\hline
$P_z\cdot U$ & $\frac{\sqrt{-2 \left(\sqrt{3}-1\right) \cos \delta_z  \sin (2 z)-\sqrt{3} \cos (2 z)-\sqrt{3}+4}}{2 \sqrt{3}}$ & $\frac{\sin (2 z) \sin \delta _z}{12 \sqrt{3}}$ \\
\hline
$U\cdot P_x$ & $\frac{1}{6} \left(\sqrt{3}-3\right)$ & $\frac{\sin (2 x) \sin \delta _x}{12 \sqrt{3}}$ \\
\hline
$U\cdot P_y$ & $\frac{\sqrt{-2 \left(\sqrt{3}-1\right) \sin (2 y) \cos \delta _y-\sqrt{3} \cos (2 y)-\sqrt{3}+4}}{2 \sqrt{3}}$ & $\frac{\sin (2 y) \sin \delta _y}{12 \sqrt{3}}$ \\
\hline
$U\cdot P_z$ & $\frac{\sqrt{-\sin (2 z) \cos \delta _z-\sqrt{3} \cos (2 z)+2}}{\sqrt{6}}$ & $\frac{\sin (2 z) \sin \delta _z}{6 \sqrt{3}}$ \\
\hline
\end{tabular}
\begin{quote}
\caption{Results of $S_{13}$ and $J_{\rm CP}$ after perturbation to TFH1.}
\label{TabTFH1b}
\end{quote}
\end{center}
\end{table}

\begin{table}
\begin{center}
\begin{tabular}{|c||c|c|}\hline
BM & $T_{12}$ & $T_{23}$  \\
\hline\hline
$P_x\cdot U$ & $2 \sqrt{\frac{\sin (2 x) \cos \delta _x+1}{-2 \left(1+\sqrt{3}\right) \sin (2 x) \cos \delta
   _x+\sqrt{3} \cos (2 x)+\sqrt{3}+4}}$ & $\frac{\sqrt{-2 \left(\sqrt{3}-1\right) \sin (2 x) \cos \delta _x+\sqrt{3} \cos (2 x)-\sqrt{3}+4}}{\sqrt{2
   \left(2+\sqrt{3}\right)}}$   \\
\hline
$P_y\cdot U$ & $\sqrt{3}-1$ & $\sqrt{\frac{2 \left(1+\sqrt{3}\right) \sin (2 y) \cos \delta _y-\sqrt{3} \cos (2 y)+\sqrt{3}+4}{-2
   \left(1+\sqrt{3}\right) \sin (2 y) \cos \delta _y+\sqrt{3} \cos (2 y)+\sqrt{3}+4}}$ \\
\hline
$P_z\cdot U$ & $\sqrt{\frac{2\left[1-\sin (2 z) \cos \delta _z\right]}{\sin (2 z) \cos \delta _z+\sqrt{3} \cos (2 z)+2}}$ & $\sqrt{\frac{2}{-\sin (2 z) \cos \delta _z+\sqrt{3} \cos (2 z)+2}}$ \\
\hline
$U\cdot P_x$ & $\sqrt{\frac{2 \left(1+\sqrt{3}\right) \sin (2 x) \cos \delta _x-\sqrt{3} \cos (2 x)+\sqrt{3}+4}{-2
   \left(1+\sqrt{3}\right) \sin (2 x) \cos \delta _x+\sqrt{3} \cos (2 x)+\sqrt{3}+4}}$ & $\sqrt{3}-1$ \\
\hline
$U\cdot P_y$ & $\frac{\sqrt{-2 \left(\sqrt{3}-1\right) \sin (2 y) \cos \delta _y+\sqrt{3} \cos (2 y)-\sqrt{3}+4}}{\sqrt{2
   \left(2+\sqrt{3}\right)}}$ & $2 \sqrt{\frac{\sin (2 y) \cos \delta _y+1}{-2 \left(1+\sqrt{3}\right) \sin (2 y) \cos \delta
   _y+\sqrt{3} \cos (2 y)+\sqrt{3}+4}}$ \\
\hline
$U\cdot P_z$ & $\sqrt{\frac{2}{-\sin (2 z) \cos \delta _z+\sqrt{3} \cos (2 z)+2}}$ & $\sqrt{2} \sqrt{\frac{1-\sin (2 z) \cos \delta _z}{\sin (2 z) \cos \delta _z+\sqrt{3} \cos (2 z)+2}}$ \\
\hline
\end{tabular}
\begin{quote}
\caption{Results of $T_{12}$ and $T_{23}$ after perturbation to TFH2.}
\label{TabTFH2a}
\end{quote}
\end{center}
\end{table}

\begin{table}
\begin{center}
\begin{tabular}{|c||c|c|}\hline
BM & $S_{13}$ & $J_{\rm CP}$  \\
\hline\hline
$P_x\cdot U$ & $\frac{\sqrt{2 \left(\sqrt{3}-1\right) \sin (2 x) \cos \delta _x-\sqrt{3} \cos (2 x)-\sqrt{3}+4}}{2 \sqrt{3}}$ & $\frac{\sin (2 x) \sin \delta _x}{12 \sqrt{3}}$   \\
\hline
$P_y\cdot U$ & $\frac{1}{6} \left(3-\sqrt{3}\right)$ & $\frac{\sin (2 y) \sin \delta _y}{12 \sqrt{3}}$ \\
\hline
$P_z\cdot U$ & $\frac{\sqrt{\sin (2 z) \cos \delta _z-\sqrt{3} \cos (2 z)+2}}{\sqrt{6}}$ &$\frac{\sin (2 z) \sin \delta _z}{6 \sqrt{3}}$ \\
\hline
$U\cdot P_x$ & $\frac{1}{6} \left(3-\sqrt{3}\right)$ & $\frac{\sin (2 x) \sin \delta _x}{12 \sqrt{3}}$ \\
\hline
$U\cdot P_y$ & $\frac{\sqrt{2 \left(\sqrt{3}-1\right) \cos \delta_y \sin (2 y)-\sqrt{3} \cos (2 y)-\sqrt{3}+4}}{2 \sqrt{3}}$ & $\frac{\sin (2 y) \sin \delta _y}{12 \sqrt{3}}$ \\
\hline
$U\cdot P_z$ & $\frac{\sqrt{\sin (2 z) \cos \delta _z-\sqrt{3} \cos (2 z)+2}}{\sqrt{6}}$ & $\frac{\sin (2 z) \sin \delta _z}{6 \sqrt{3}}$ \\
\hline
\end{tabular}
\begin{quote}
\caption{Results of $S_{13}$ and $J_{\rm CP}$ after perturbation to TFH2.}
\label{TabTFH2b}
\end{quote}
\end{center}
\end{table}

\newpage
\section{The numerical analyses of the rest eight constant mixing patterns}
\label{AppenB}

\subsection{Democratic Mixing}
\subsubsection{$P_x\cdot U_{\rm{DM}}$}
\begin{figure}[t]
\centering
\includegraphics[bb=0 0 550 550, height=8.5cm, width=8.5cm, angle=0]{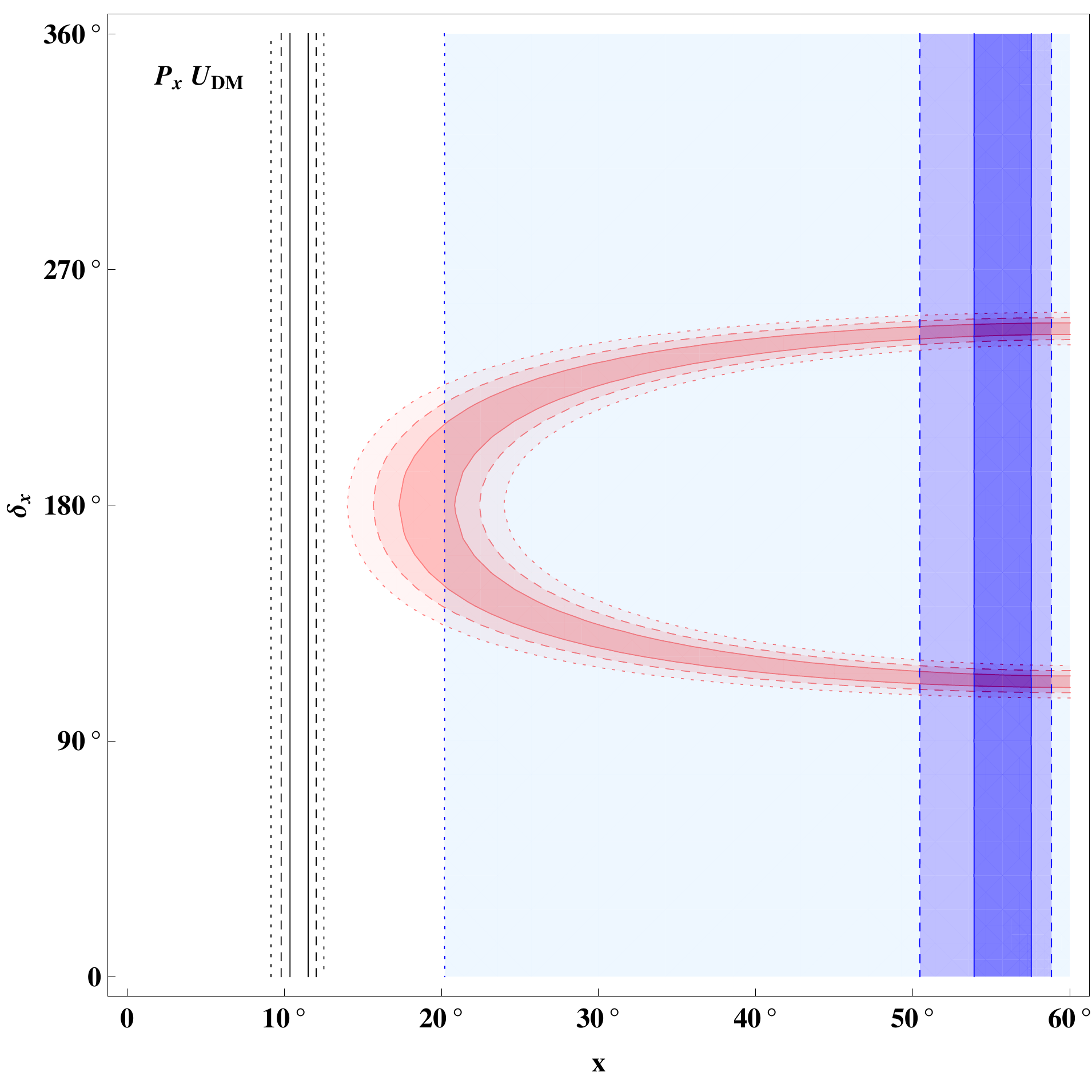}
\caption{The solutions corresponding to the measured $\theta_{12}$, $\theta_{23}$, and $\theta_{13}$ for $P_x\cdot U_{\rm{DM}}$ in the parameter space of $x - \delta_x$. Caption is the same as displayed in Fig. \ref{CPxUtbm}.} \label{CPxUdm}
\end{figure}

\clearpage
\subsubsection{$P_z\cdot U_{\rm{DM}}$}
\begin{figure}[t]
\centering
\includegraphics[bb=0 0 550 550, height=8.5cm, width=8.5cm, angle=0]{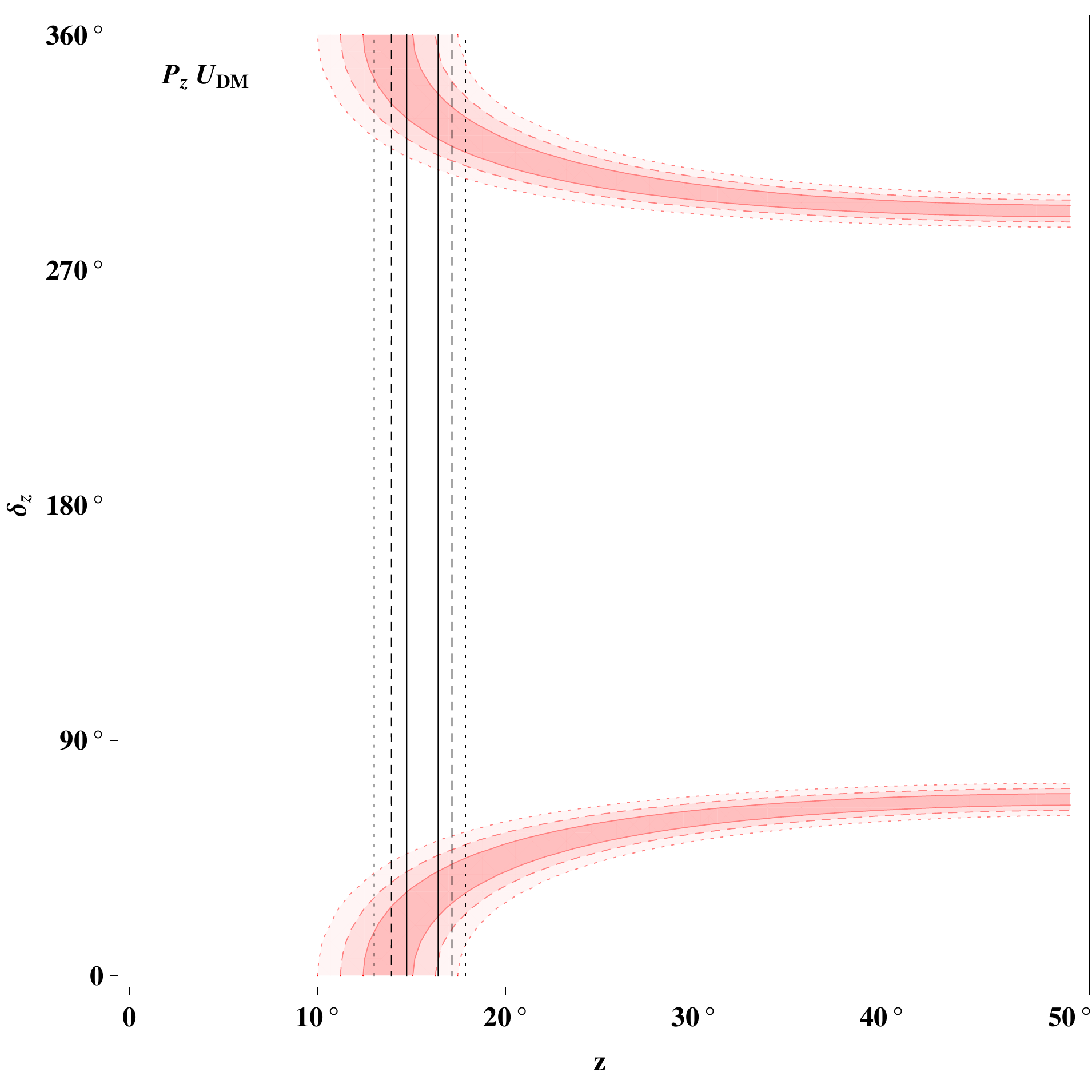}
\caption{The solutions corresponding to the measured $\theta_{12}$, $\theta_{23}$, and $\theta_{13}$ for $P_z\cdot U_{\rm{DM}}$ in the parameter space of $z - \delta_z$. Caption is the same as displayed in Fig. \ref{CPxUtbm}. } \label{CPzUdm}
\end{figure}

\clearpage
\subsubsection{$U_{\rm{DM}}\cdot P_y$}
\begin{figure}[t]
\centering
\includegraphics[bb=0 0 550 550, height=8.5cm, width=8.5cm, angle=0]{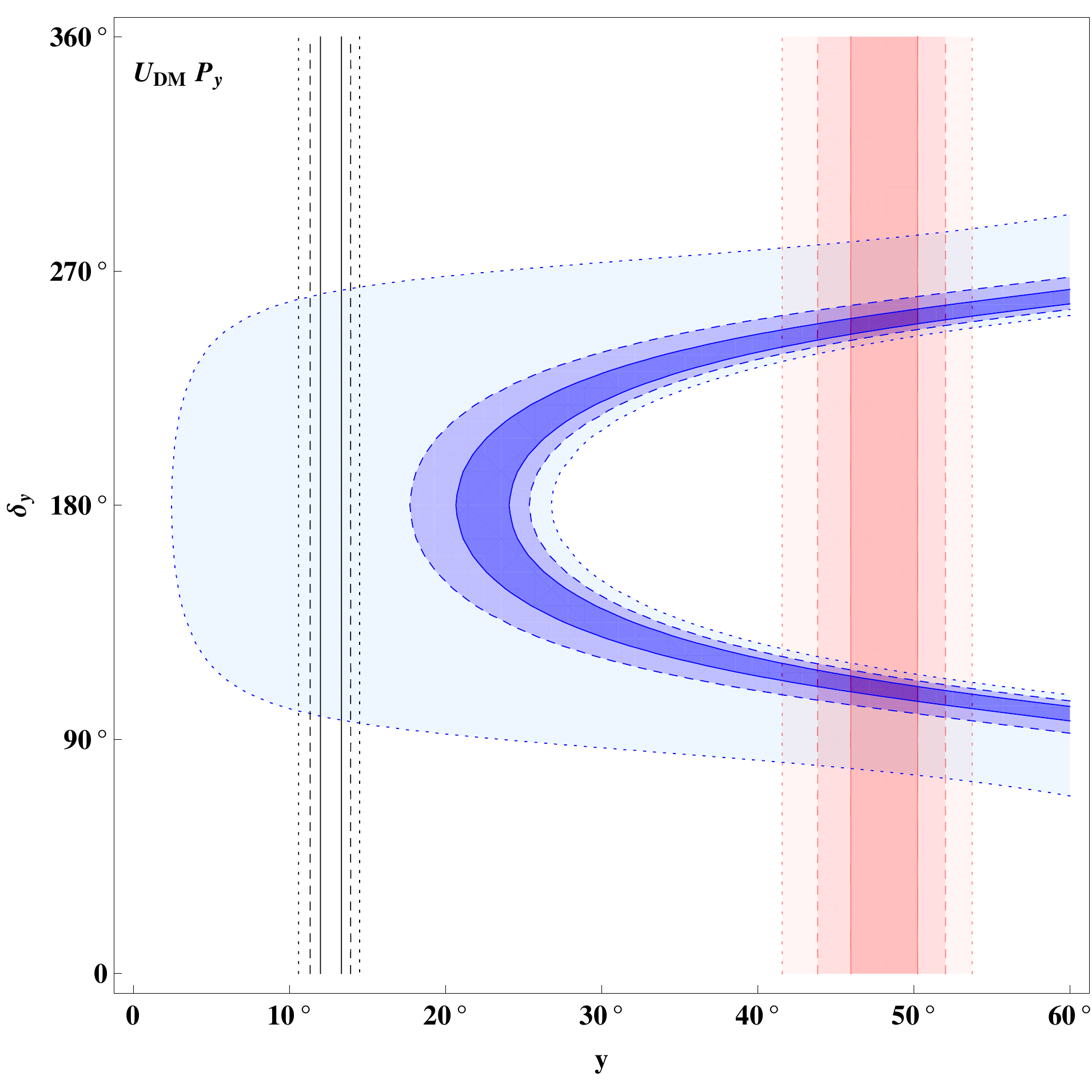}
\caption{The solutions corresponding to the measured $\theta_{12}$, $\theta_{23}$, and $\theta_{13}$ for $U_{\rm{DM}}\cdot P_y$ in the parameter space of $y - \delta_y$. Caption is the same as displayed in Fig. \ref{CPxUtbm}. } \label{CUdmPy}
\end{figure}

\clearpage
\subsubsection{$U_{\rm{DM}}\cdot P_z$}
\begin{figure}[t]
\centering
\includegraphics[bb=0 0 550 550, height=8.5cm, width=8.5cm, angle=0]{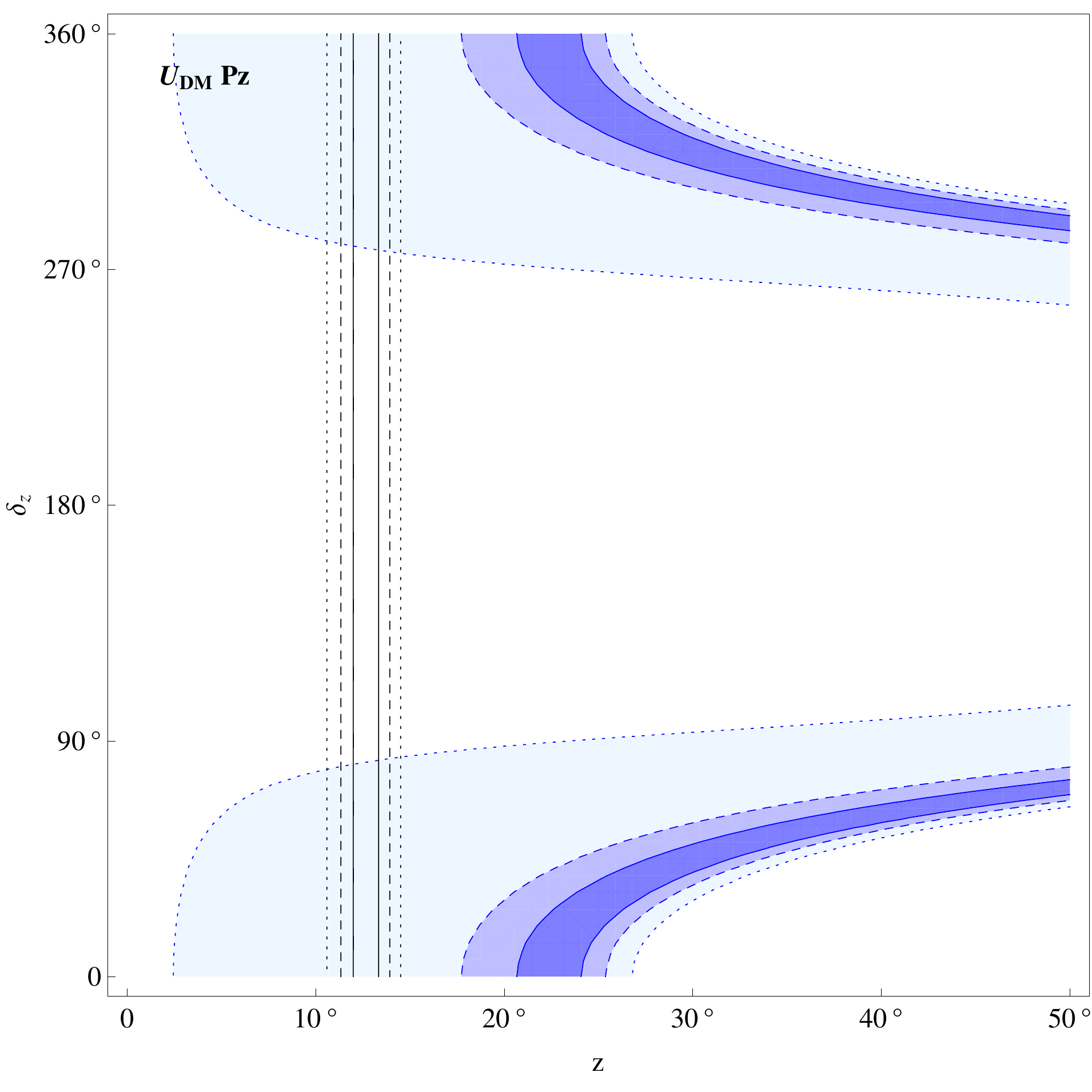}
\caption{The solutions corresponding to the measured $\theta_{12}$, $\theta_{23}$, and $\theta_{13}$ for $U_{\rm{DM}}\cdot P_z$ in the parameter space of $z - \delta_z$. Caption is the same as displayed in Fig. \ref{CPxUtbm}. } \label{CUdmPz}
\end{figure}

\clearpage
\subsection{Bimaximal Mixing}
\subsubsection{$P_x\cdot U_{\rm{BM}}$}
\begin{figure}[t]
\centering
\includegraphics[bb=0 0 550 550, height=8.5cm, width=8.5cm, angle=0]{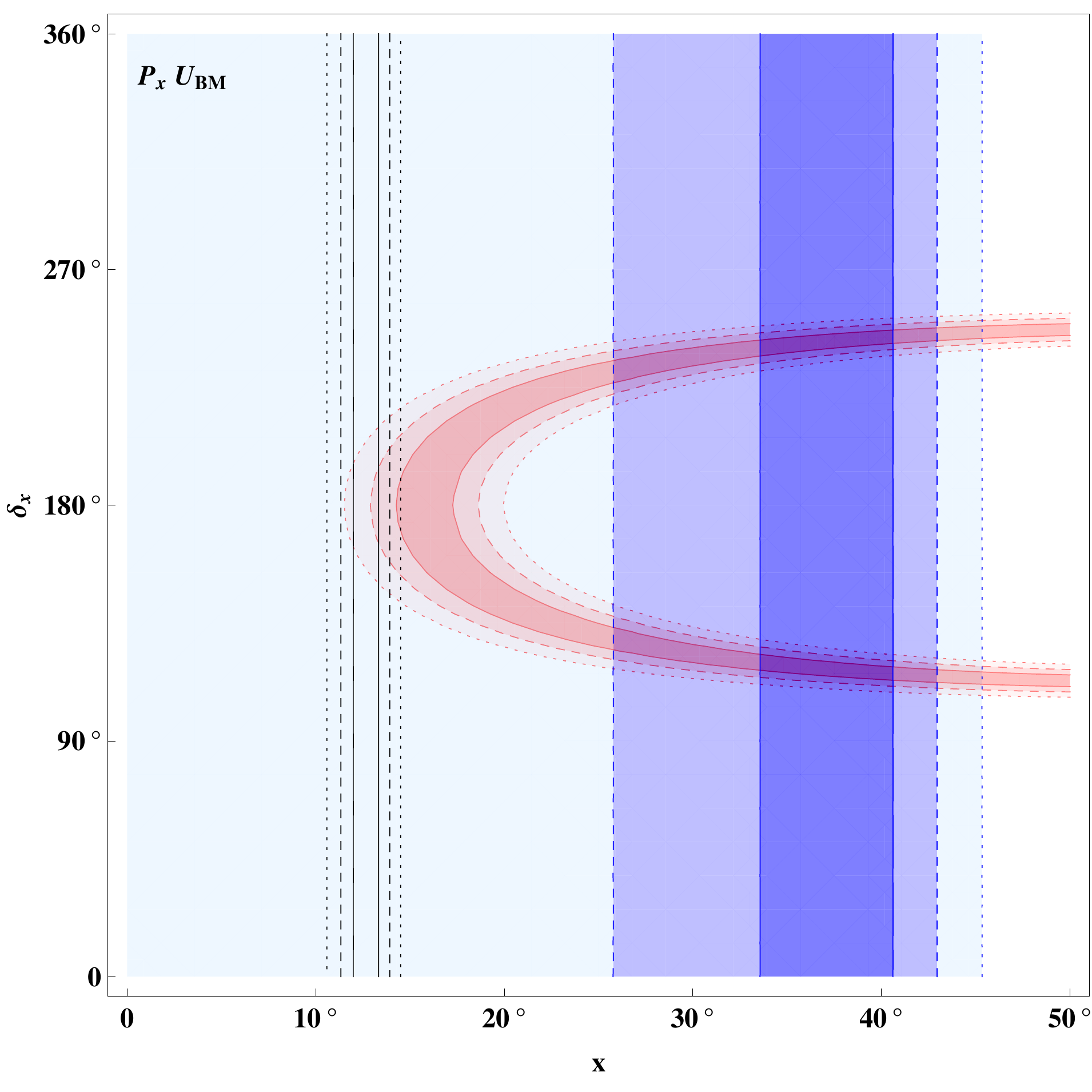}
\caption{The solutions corresponding to the measured $\theta_{12}$, $\theta_{23}$, and $\theta_{13}$ for $P_x\cdot U_{\rm{BM}}$ in the parameter space of $x - \delta_x$. Caption is the same as displayed in Fig. \ref{CPxUtbm}. } \label{CPxUbm}
\end{figure}

\clearpage
\begin{figure}[t]
\centering
\includegraphics[bb=0 110 550 550, height=12.5cm, width=15.5cm, angle=0]{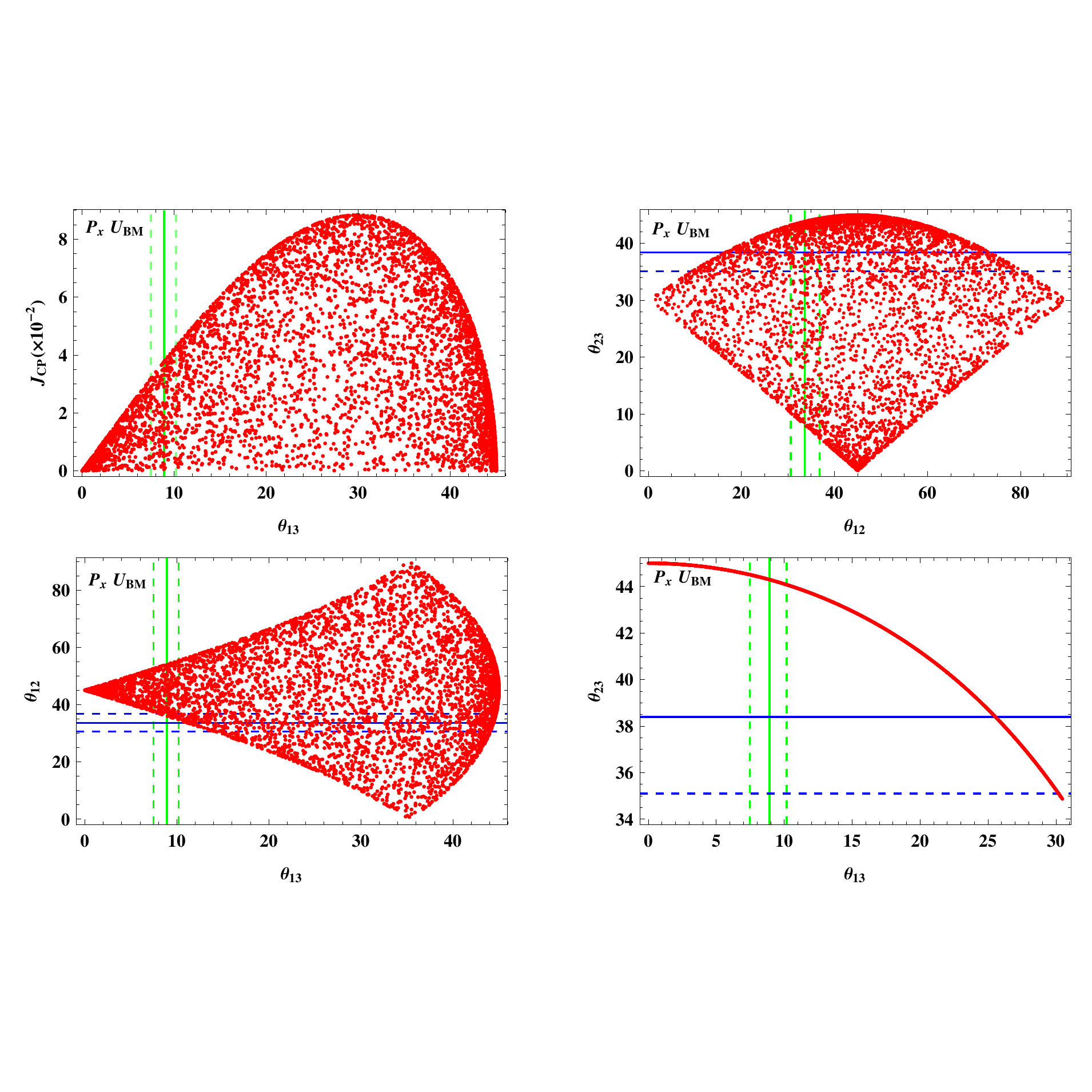}
\caption{The scatter plots for $\theta_{12}$, $\theta_{23}$, $\theta_{13}$ and $J_{\rm CP}$ with the $P_x\cdot U_{\rm{BM}}$ ansatz. Caption is the same as displayed in Fig. \ref{FigPxUtbm}.} \label{FigPxUbm}
\end{figure}

\clearpage
\subsubsection{$P_z\cdot U_{\rm{BM}}$}
\begin{figure}[t]
\centering
\includegraphics[bb=0 0 550 550, height=8.5cm, width=8.5cm, angle=0]{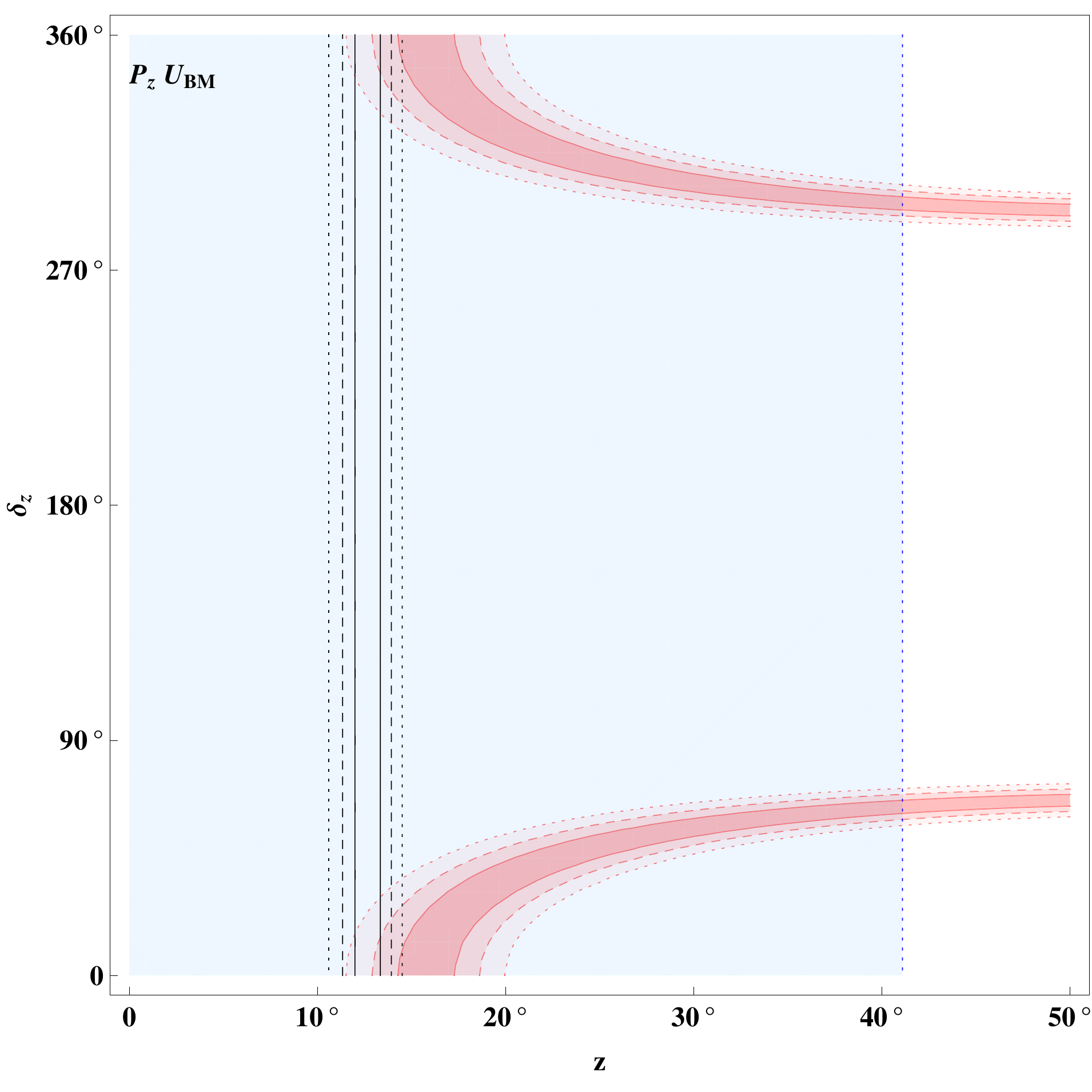}
\caption{The solutions corresponding to the measured $\theta_{12}$, $\theta_{23}$, and $\theta_{13}$ for $P_z\cdot U_{\rm{BM}}$ in the parameter space of $z - \delta_z$. Caption is the same as displayed in Fig. \ref{CPxUtbm}.} \label{CPzUbm}
\end{figure}

\clearpage
\begin{figure}[t]
\centering
\includegraphics[bb=0 110 550 550, height=12.5cm, width=15.5cm, angle=0]{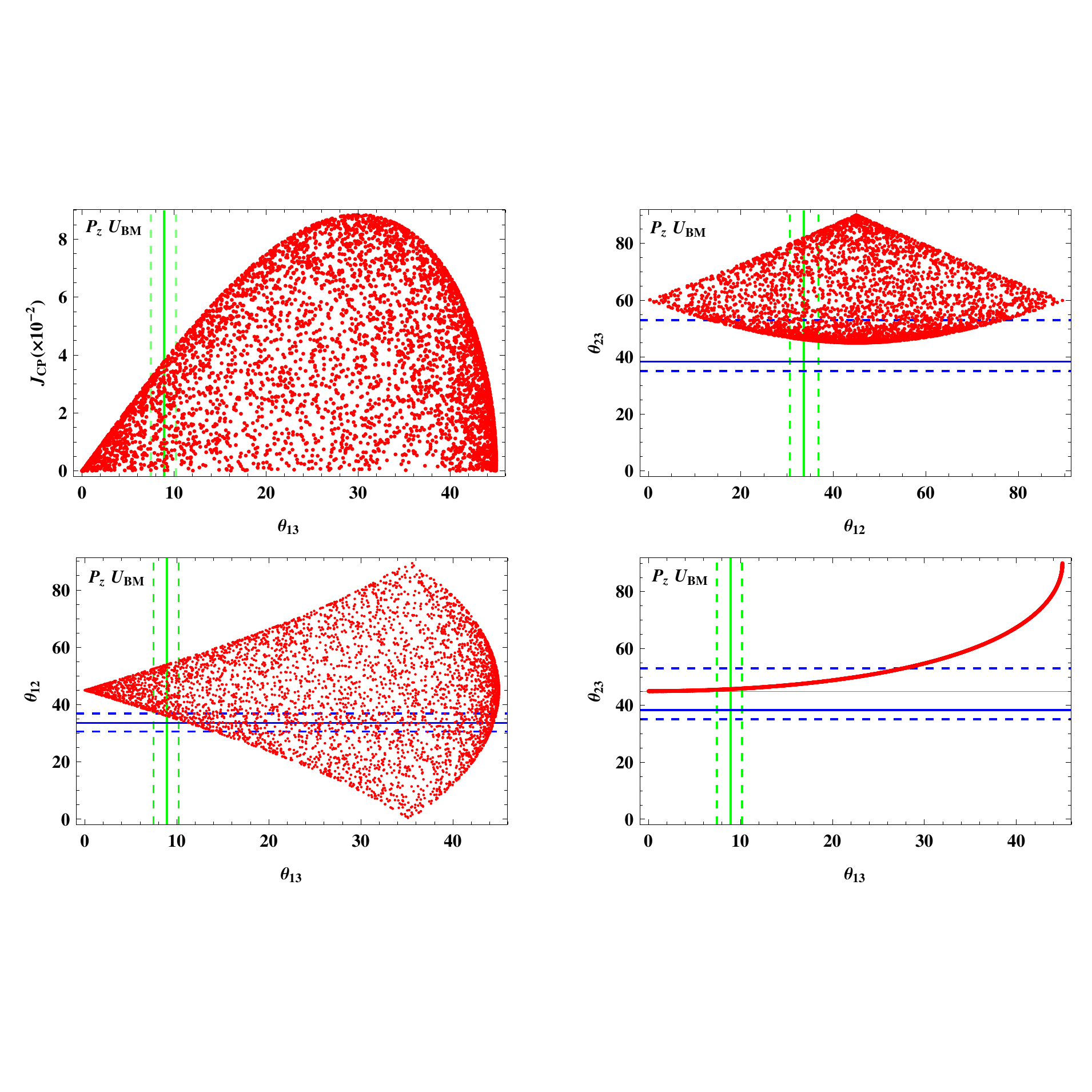}
\caption{The scatter plots for $\theta_{12}$, $\theta_{23}$, $\theta_{13}$ and $J_{\rm CP}$ with the $P_z\cdot U_{\rm{BM}}$ ansatz. Caption is the same as displayed in Fig. \ref{FigPxUtbm}.} \label{FigPzUbm}
\end{figure}

\clearpage
\subsubsection{$U_{\rm{BM}}\cdot P_y$}
\begin{figure}[t]
\centering
\includegraphics[bb=0 0 550 550, height=8.5cm, width=8.5cm, angle=0]{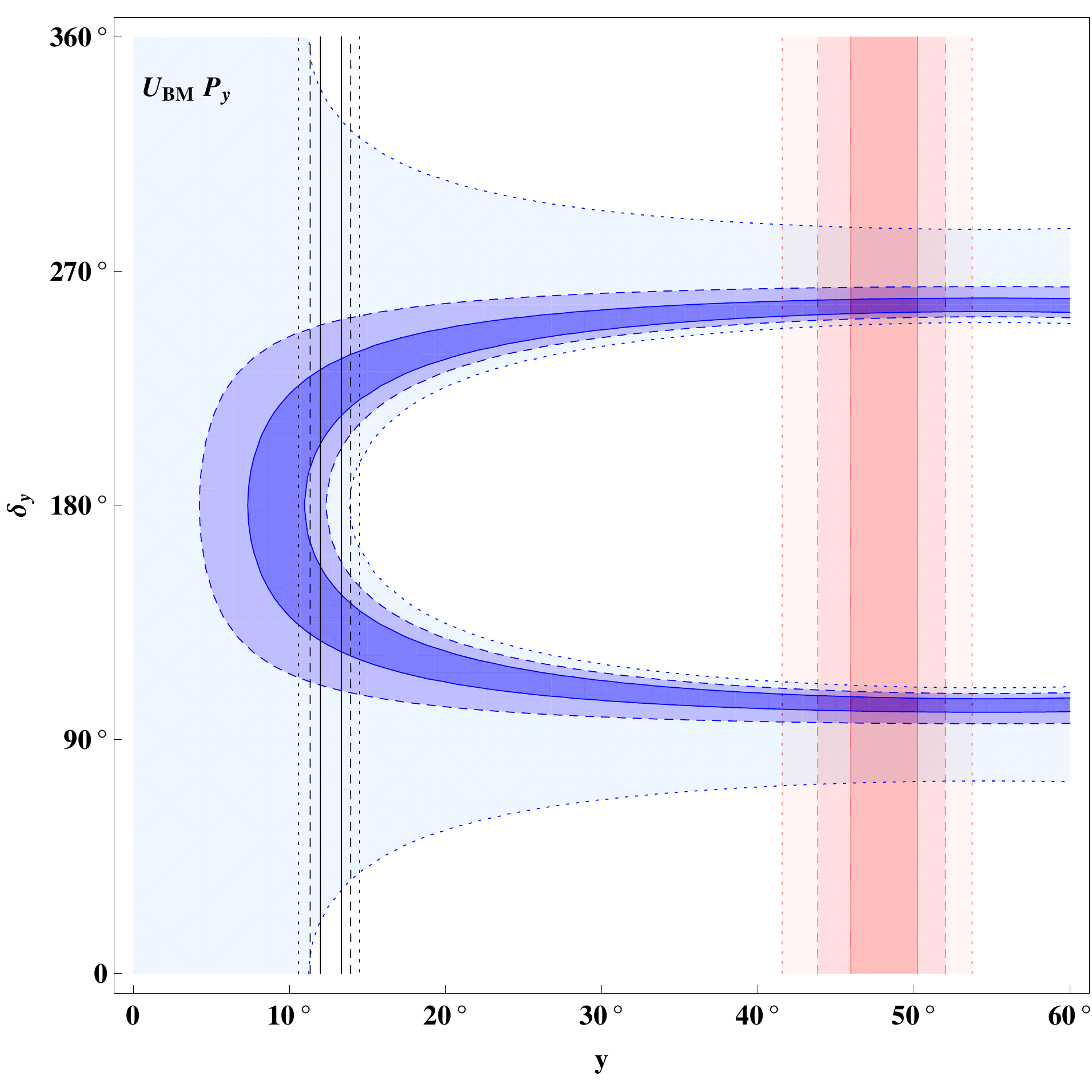}
\caption{The solutions corresponding to the measured $\theta_{12}$, $\theta_{23}$, and $\theta_{13}$ for $U_{\rm{BM}}\cdot P_y$ in the parameter space of $y - \delta_y$. Caption is the same as displayed in Fig. \ref{CPxUtbm}.} \label{CUbmPy}
\end{figure}

\clearpage
\subsubsection{$U_{\rm{BM}}\cdot P_z$}
\begin{figure}[t]
\centering
\includegraphics[bb=0 0 550 550, height=8.5cm, width=8.5cm, angle=0]{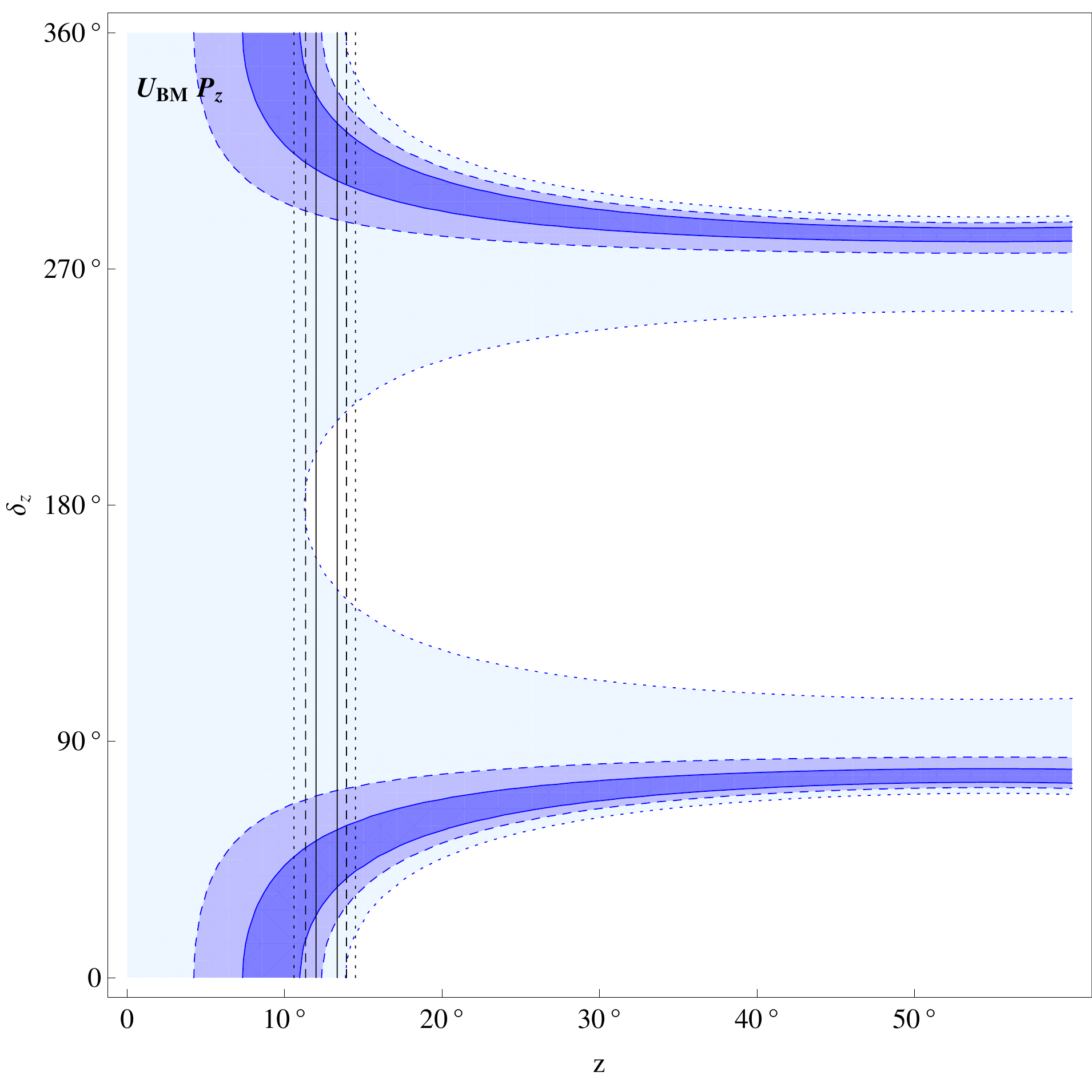}
\caption{The solutions corresponding to the measured $\theta_{12}$, $\theta_{23}$, and $\theta_{13}$ for $U_{\rm{BM}}\cdot P_z$ in the parameter space of $z - \delta_z$. Caption is the same as displayed in Fig. \ref{CPxUtbm}.} \label{CUbmPz}
\end{figure}

\clearpage
\subsection{Golden Ratio Mixing 1}
\subsubsection{$P_x\cdot U_{\rm{GRM1}}$}
\begin{figure}[t]
\centering
\includegraphics[bb=0 0 550 550, height=8.5cm, width=8.5cm, angle=0]{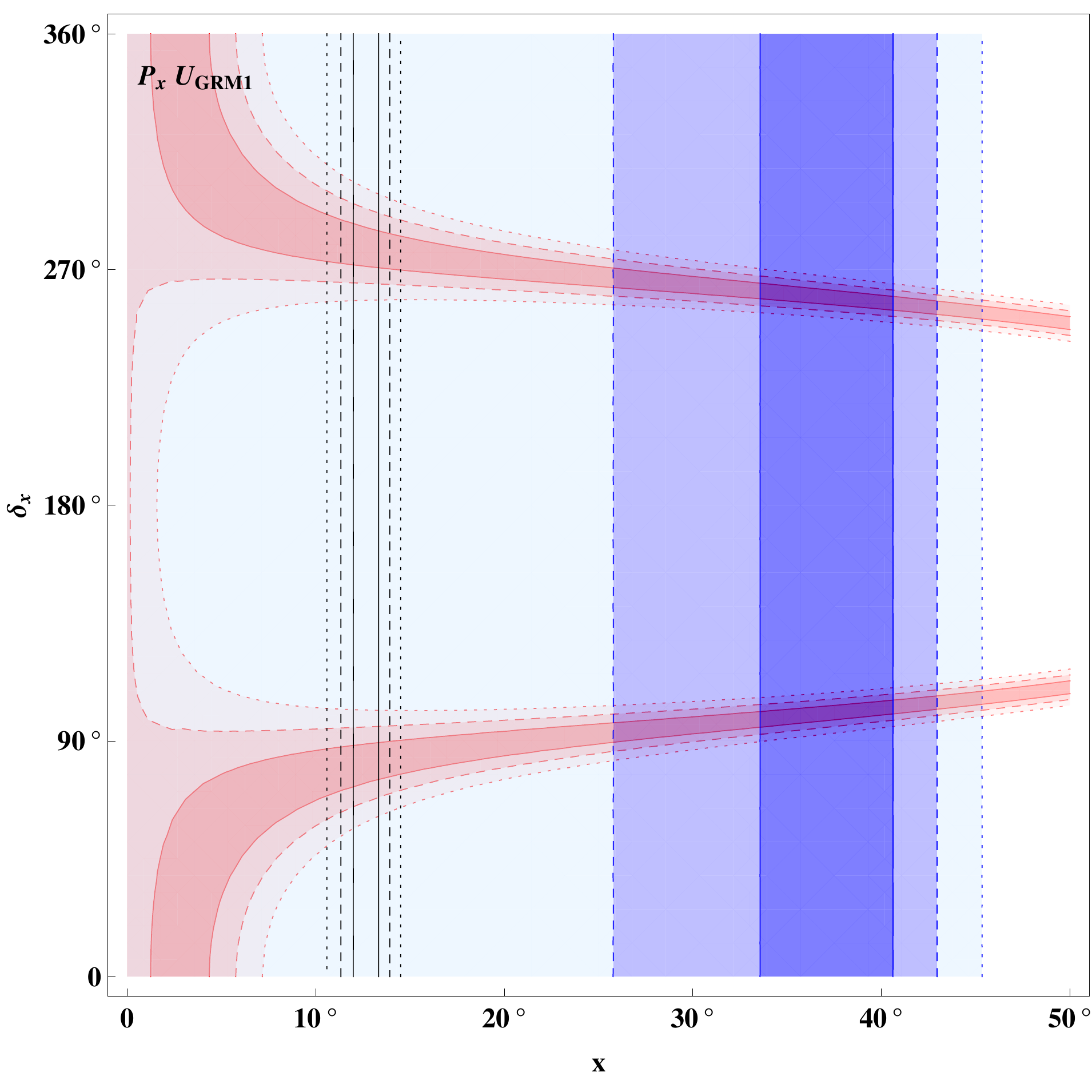}
\caption{The solutions corresponding to the measured $\theta_{12}$, $\theta_{23}$, and $\theta_{13}$ for $P_x\cdot U_{\rm{GRM1}}$ in the parameter space of $x - \delta_x$. Caption is the same as displayed in Fig. \ref{CPxUtbm}.} \label{CPxUgr1m}
\end{figure}

\clearpage
\begin{figure}[t]
\centering
\includegraphics[bb=0 110 550 550, height=12.5cm, width=15.5cm, angle=0]{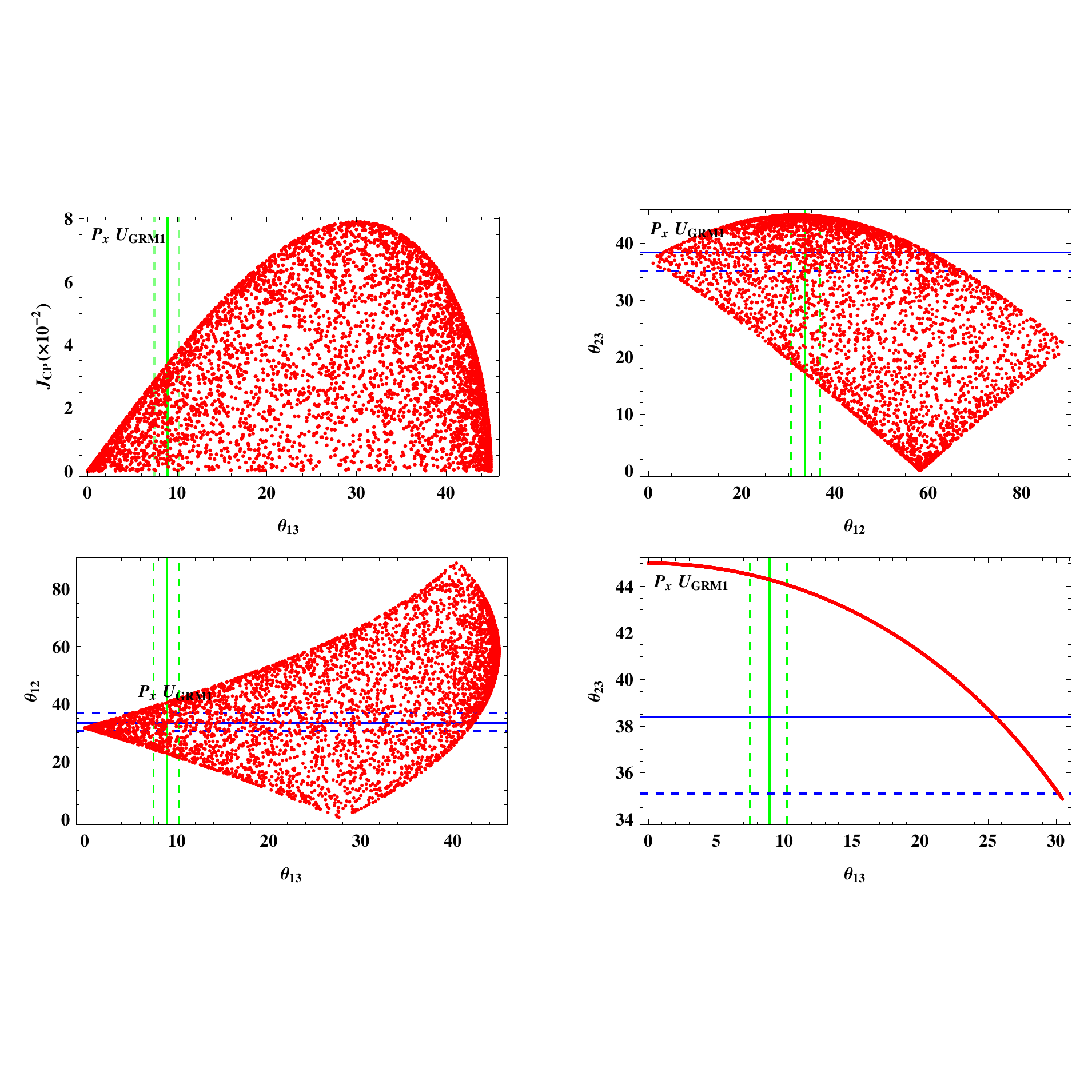}
\caption{The scatter plots for $\theta_{12}$, $\theta_{23}$, $\theta_{13}$ and $J_{\rm CP}$ with the $P_x\cdot U_{\rm{GRM1}}$ ansatz. Caption is the same as displayed in Fig. \ref{FigPxUtbm}.} \label{FigPxUgr1m}
\end{figure}

\clearpage
\subsubsection{$P_z\cdot U_{\rm{GRM1}}$}
\begin{figure}[t]
\centering
\includegraphics[bb=0 0 550 550, height=8.5cm, width=8.5cm, angle=0]{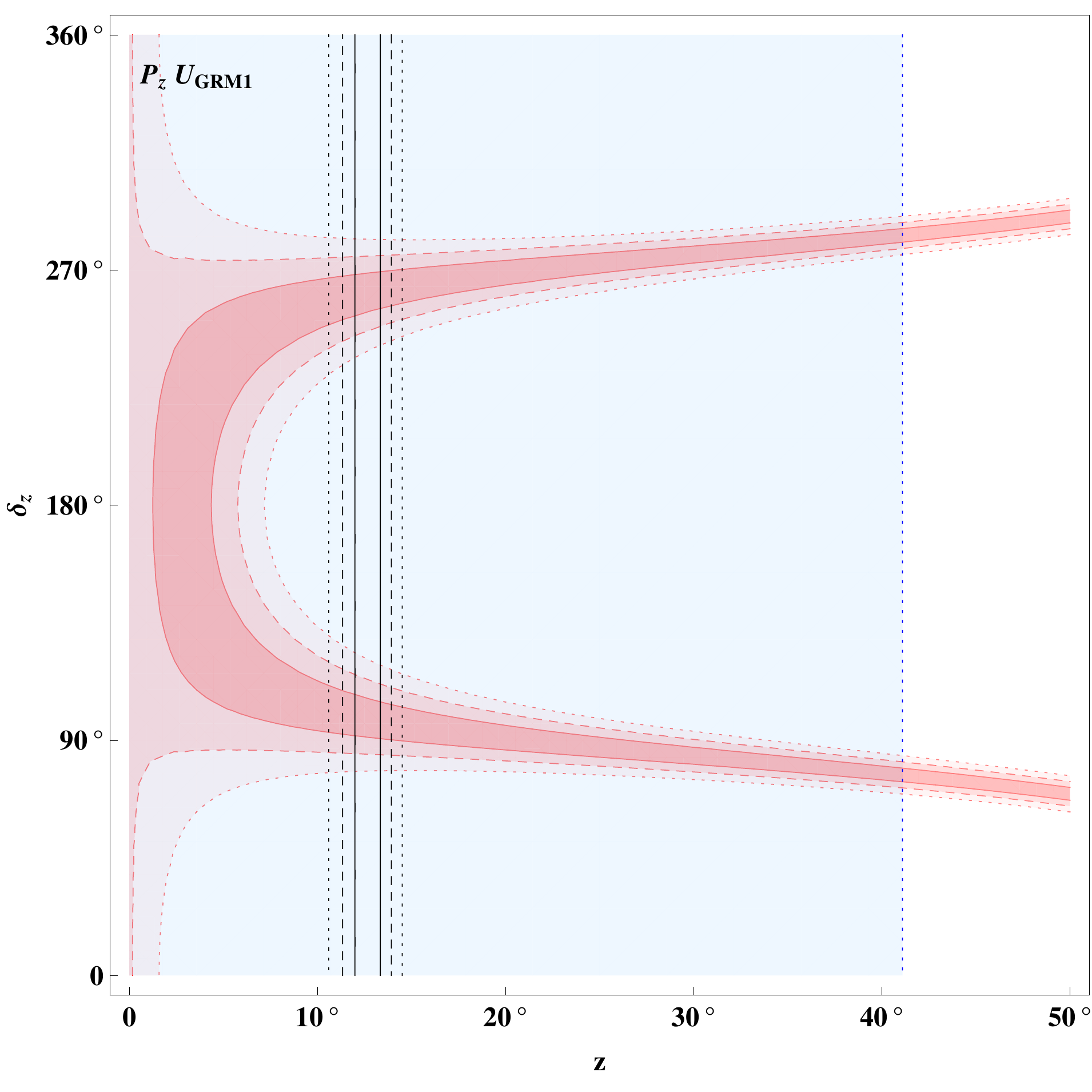}
\caption{The solutions corresponding to the measured $\theta_{12}$, $\theta_{23}$, and $\theta_{13}$ for $P_z\cdot U_{\rm{GRM1}}$ in the parameter space of $z - \delta_z$. Caption is the same as displayed in Fig. \ref{CPxUtbm}.} \label{CPzUgr1m}
\end{figure}

\clearpage
\begin{figure}[t]
\centering
\includegraphics[bb=0 110 550 550, height=12.5cm, width=15.5cm, angle=0]{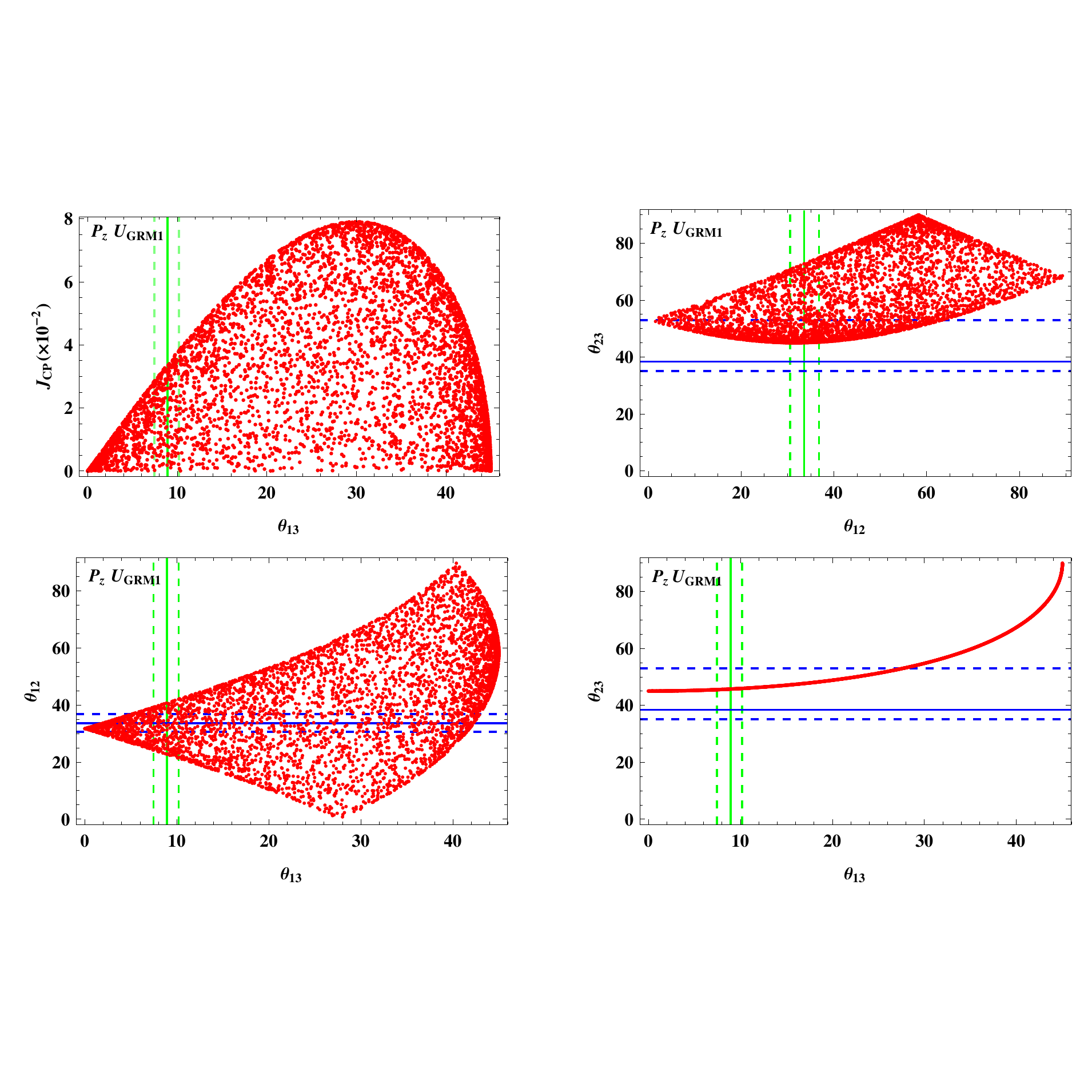}
\caption{The scatter plots for $\theta_{12}$, $\theta_{23}$, $\theta_{13}$ and $J_{\rm CP}$ with the $P_z\cdot U_{\rm{GRM1}}$. Caption is the same as displayed in Fig. \ref{FigPxUtbm}.} \label{FigPzUgr1m}
\end{figure}

\clearpage
\subsubsection{$U_{\rm{GRM1}}\cdot P_y$}
\begin{figure}[t]
\centering
\includegraphics[bb=0 0 550 550, height=8.5cm, width=8.5cm, angle=0]{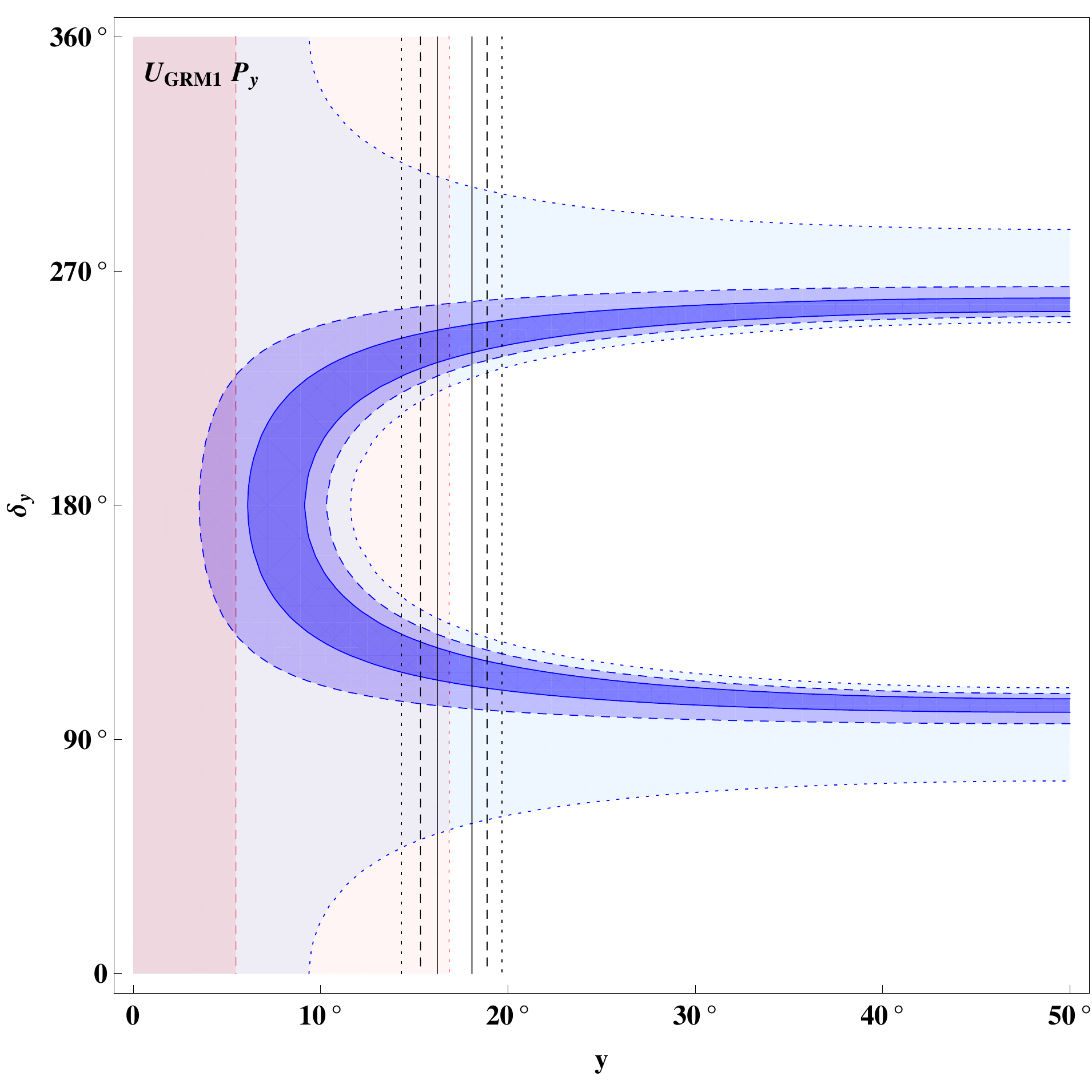}
\caption{The solutions corresponding to the measured $\theta_{12}$, $\theta_{23}$, and $\theta_{13}$ for $U_{\rm{GRM1}}\cdot P_y$ in the parameter space of $y - \delta_y$. Caption is the same  as displayed in Fig. \ref{CPxUtbm}.} \label{CUgr1mPy}
\end{figure}

\clearpage
\begin{figure}[t]
\centering
\includegraphics[bb=0 110 550 550, height=12.5cm, width=15.5cm, angle=0]{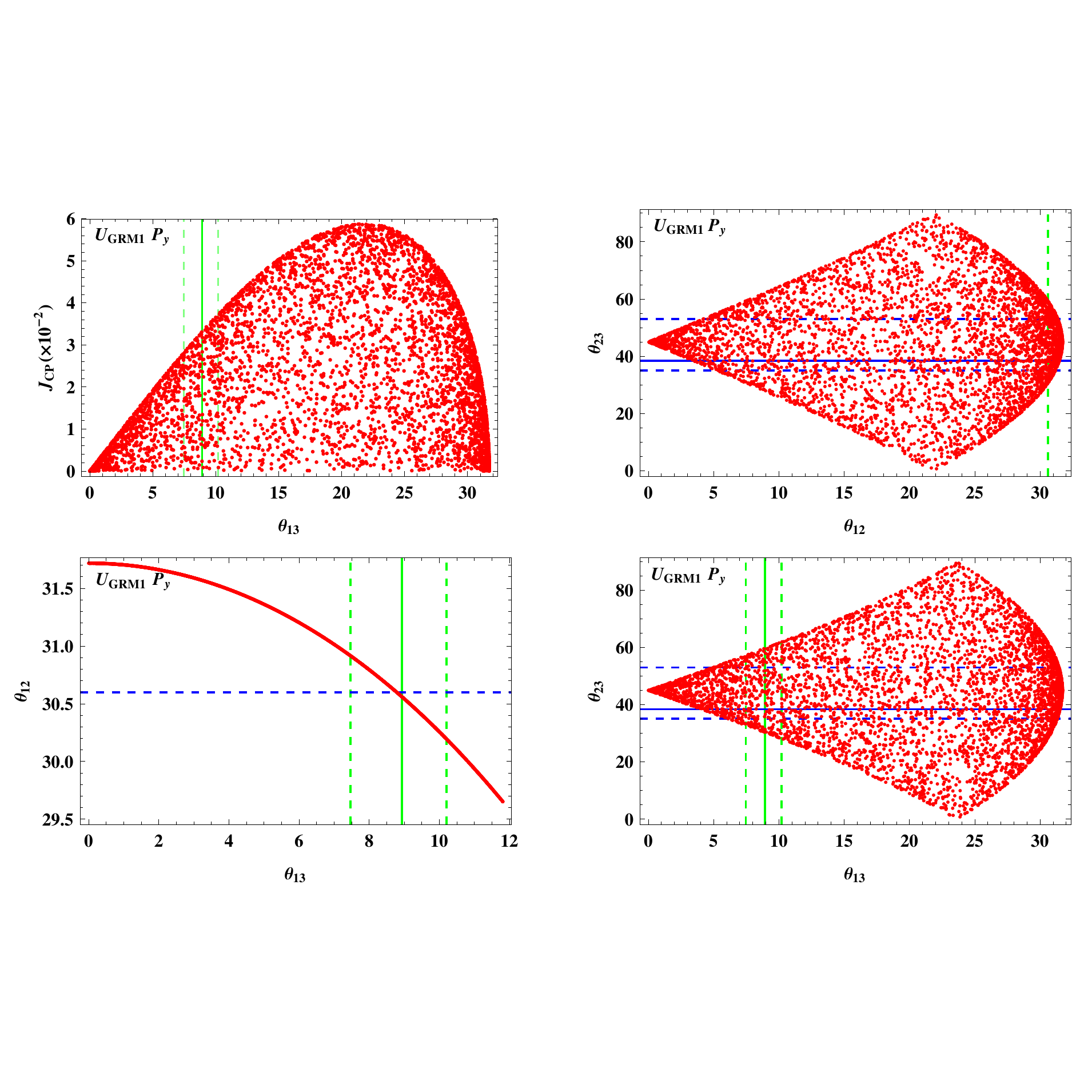}
\caption{The scatter plots for $\theta_{12}$, $\theta_{23}$, $\theta_{13}$ and $J_{\rm CP}$ with the $U_{\rm{GRM1}}\cdot P_y$ ansatz. Caption is the same as displayed in Fig. \ref{FigPxUtbm}.} \label{FigUgr1mPy}
\end{figure}

\clearpage
\subsubsection{$U_{\rm{GRM1}}\cdot P_z$}
\begin{figure}[t]
\centering
\includegraphics[bb=0 0 550 550, height=8.5cm, width=8.5cm, angle=0]{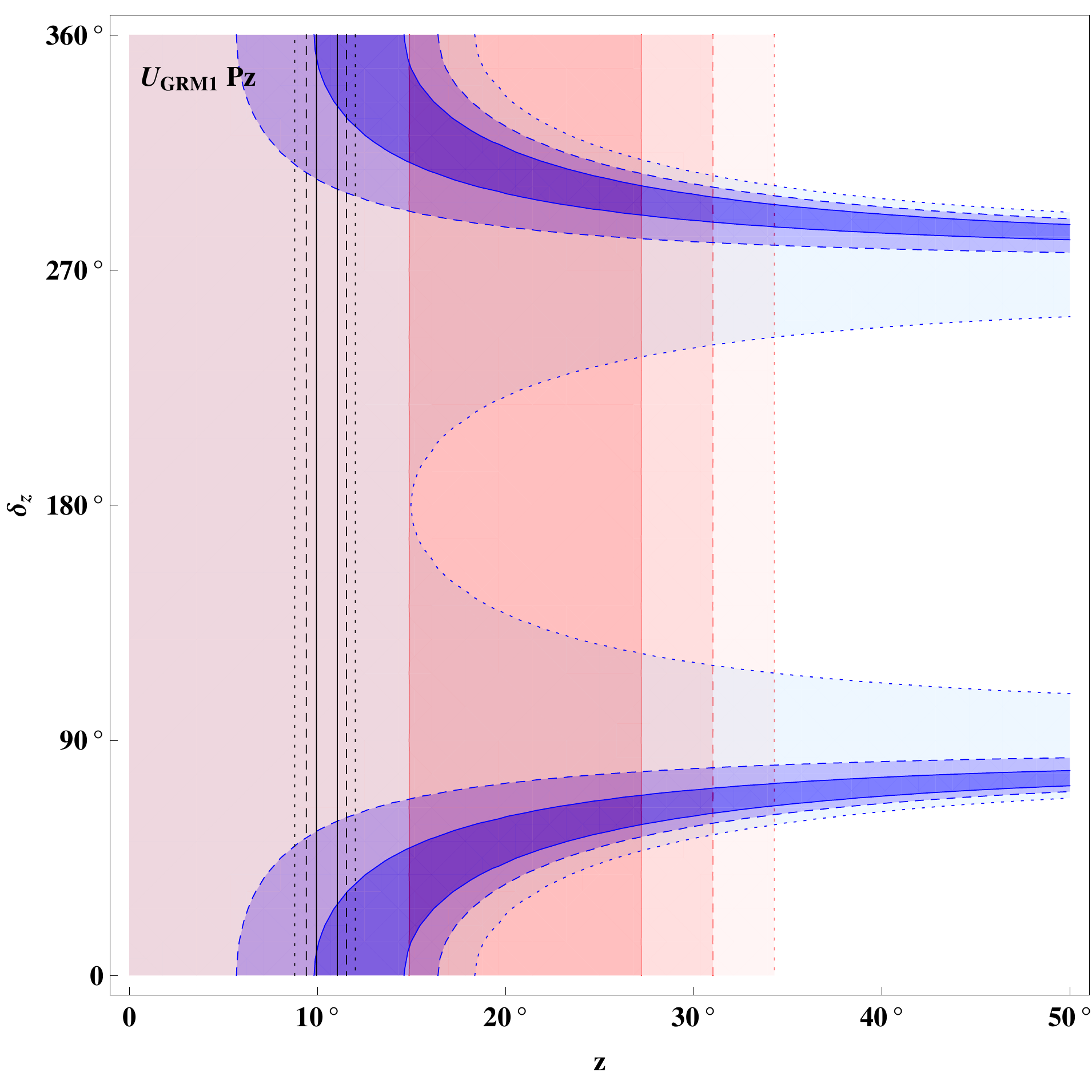}
\caption{The solutions corresponding to the measured $\theta_{12}$, $\theta_{23}$, and $\theta_{13}$ for $U_{\rm{GRM1}}\cdot P_z$ in the parameter space of $z - \delta_z$. Caption is the same as displayed in Fig. \ref{CPxUtbm}.} \label{CUgr1mPz}
\end{figure}

\clearpage
\begin{figure}[t]
\centering
\includegraphics[bb=0 110 550 550, height=12.5cm, width=15.5cm, angle=0]{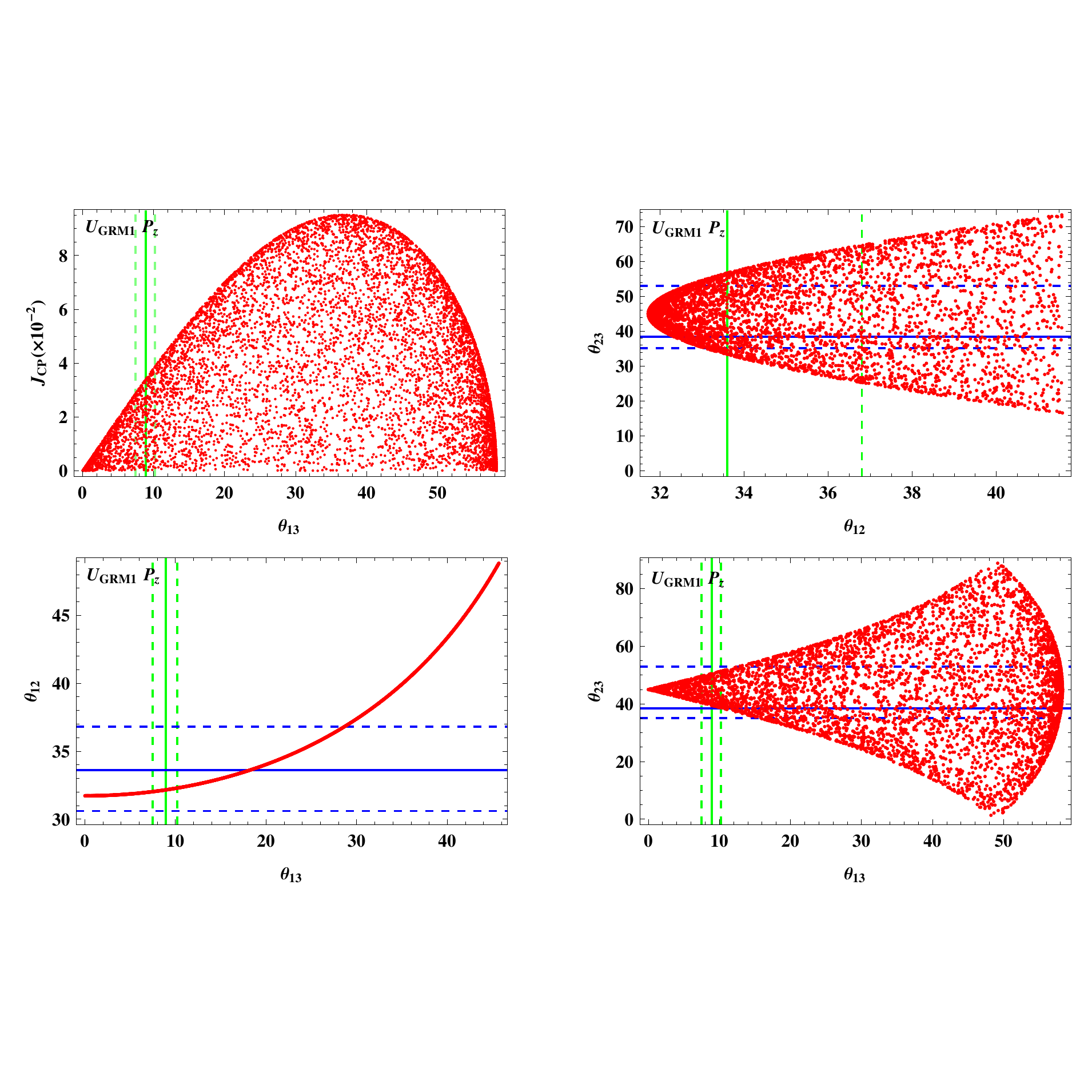}
\caption{The scatter plots for $\theta_{12}$, $\theta_{23}$, $\theta_{13}$ and $J_{\rm CP}$ with the $U_{\rm{GRM1}}\cdot P_z$ ansatz. Caption is the same as displayed in Fig. \ref{FigPxUtbm}.} \label{FigUgr1mPz}
\end{figure}

\clearpage
\subsection{Golden Ratio Mixing 2}
\subsubsection{$P_x\cdot U_{\rm{GRM2}}$}
\begin{figure}[t]
\centering
\includegraphics[bb=0 0 550 550, height=8.5cm, width=8.5cm, angle=0]{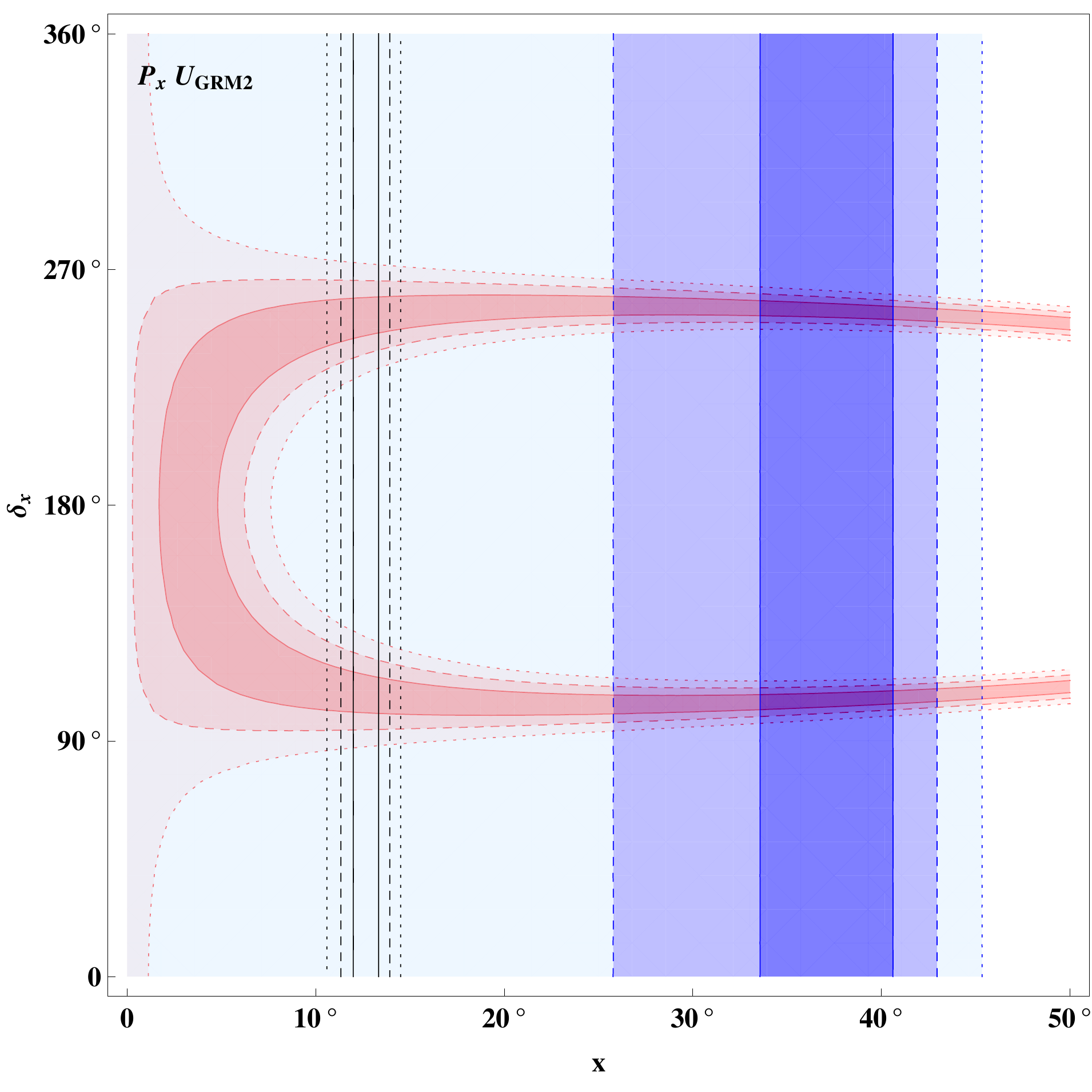}
\caption{The solutions corresponding to the measured $\theta_{12}$, $\theta_{23}$, and $\theta_{13}$ for $P_x\cdot U_{\rm{GRM2}}$ in the parameter space of $x - \delta_x$. Caption is the same as displayed in Fig. \ref{CPxUtbm}.} \label{CPxUgr2m}
\end{figure}

\clearpage
\begin{figure}[t]
\centering
\includegraphics[bb=0 110 550 550, height=12.5cm, width=15.5cm, angle=0]{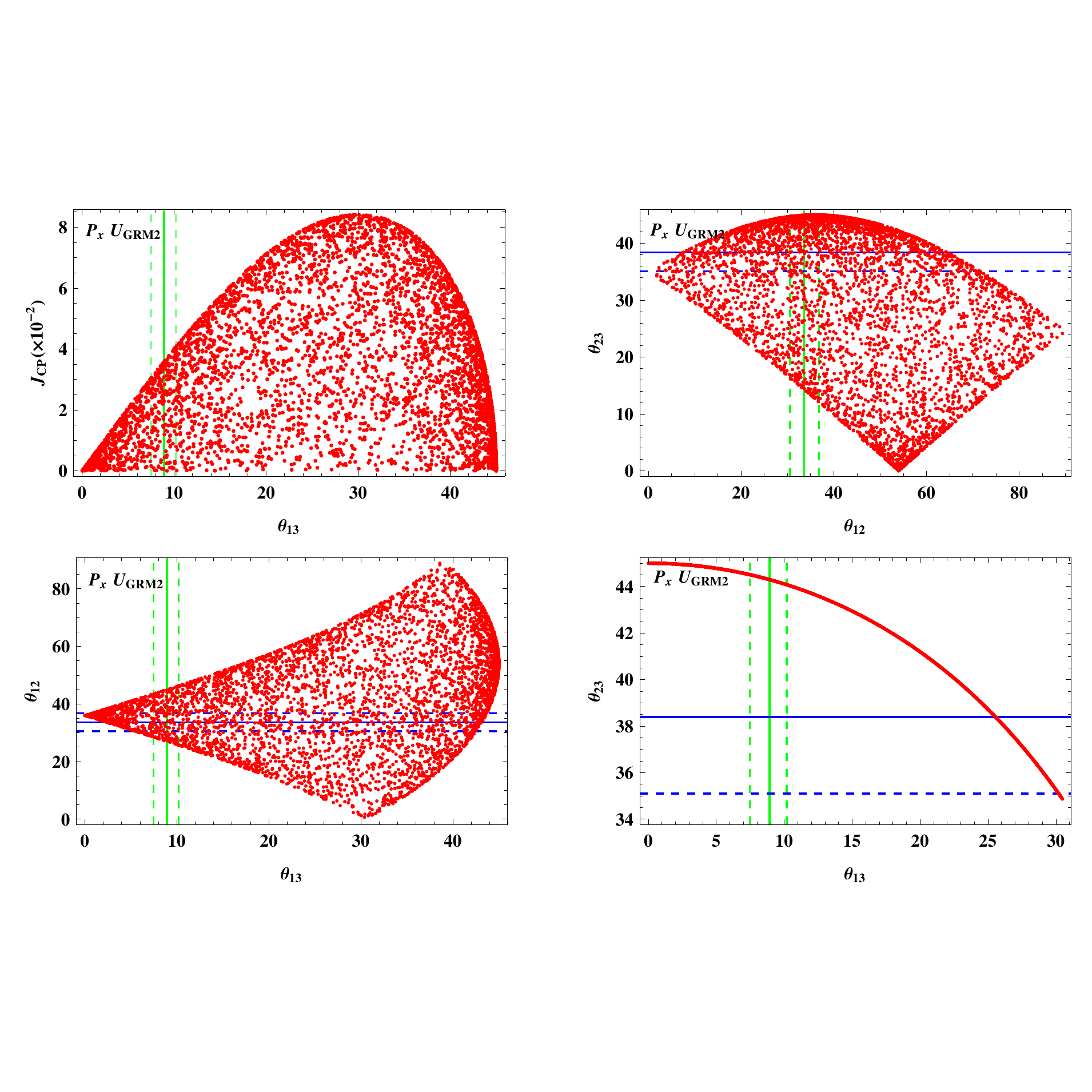}
\caption{The scatter plots for $\theta_{12}$, $\theta_{23}$, $\theta_{13}$ and $J_{\rm CP}$ with the $P_x\cdot U_{\rm{GRM2}}$ ansatz. Caption is the same as displayed in Fig. \ref{FigPxUtbm}.} \label{FigPxUgr2m}
\end{figure}

\clearpage
\subsubsection{$P_z\cdot U_{\rm{GRM2}}$}
\begin{figure}[t]
\centering
\includegraphics[bb=0 0 550 550, height=8.5cm, width=8.5cm, angle=0]{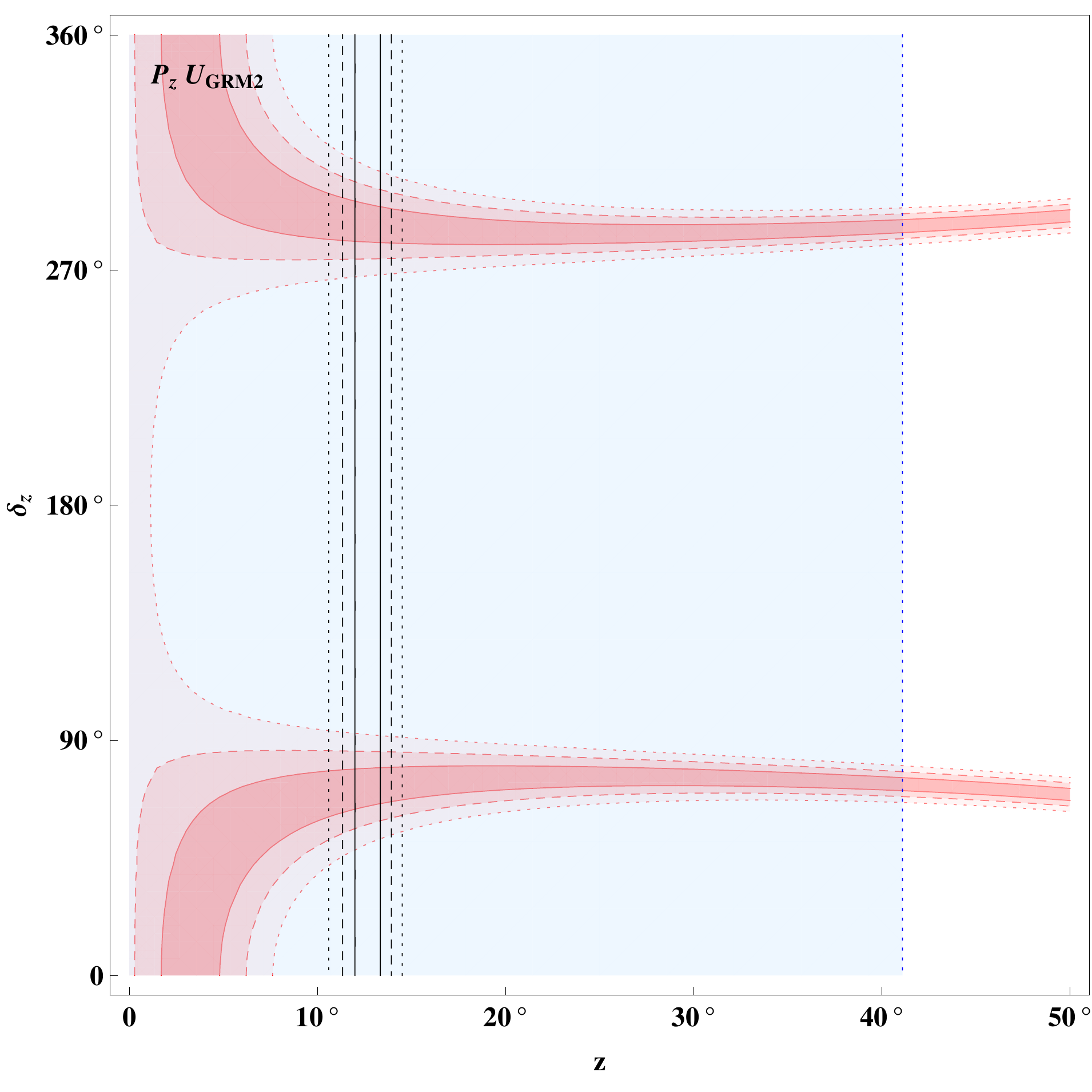}
\caption{The solutions corresponding to the measured $\theta_{12}$, $\theta_{23}$, and $\theta_{13}$ for $P_z\cdot U_{\rm{GRM2}}$ in the parameter space of $z - \delta_z$. Caption is the same as displayed in Fig. \ref{CPxUtbm}.} \label{CPzUgr2m}
\end{figure}

\clearpage
\begin{figure}[t]
\centering
\includegraphics[bb=0 110 550 550, height=12.5cm, width=15.5cm, angle=0]{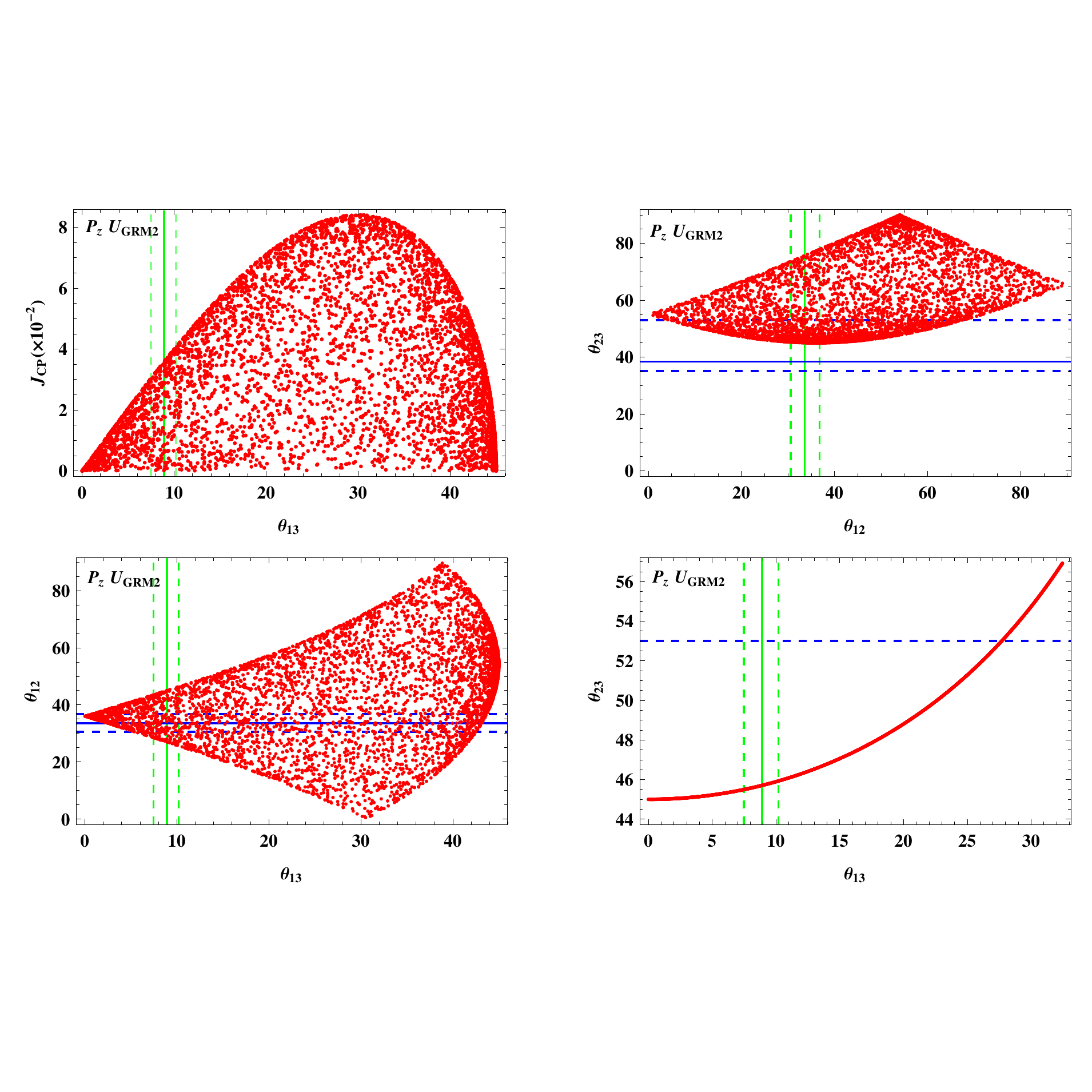}
\caption{The scatter plots for $\theta_{12}$, $\theta_{23}$, $\theta_{13}$ and $J_{\rm CP}$ with the $P_z\cdot U_{\rm{GRM2}}$ ansatz. Caption is the same as displayed in Fig. \ref{FigPxUtbm}.} \label{FigPzUgr2m}
\end{figure}

\clearpage
\subsubsection{$U_{\rm{GRM2}}\cdot P_y$}
\begin{figure}[t]
\centering
\includegraphics[bb=0 0 550 550, height=8.5cm, width=8.5cm, angle=0]{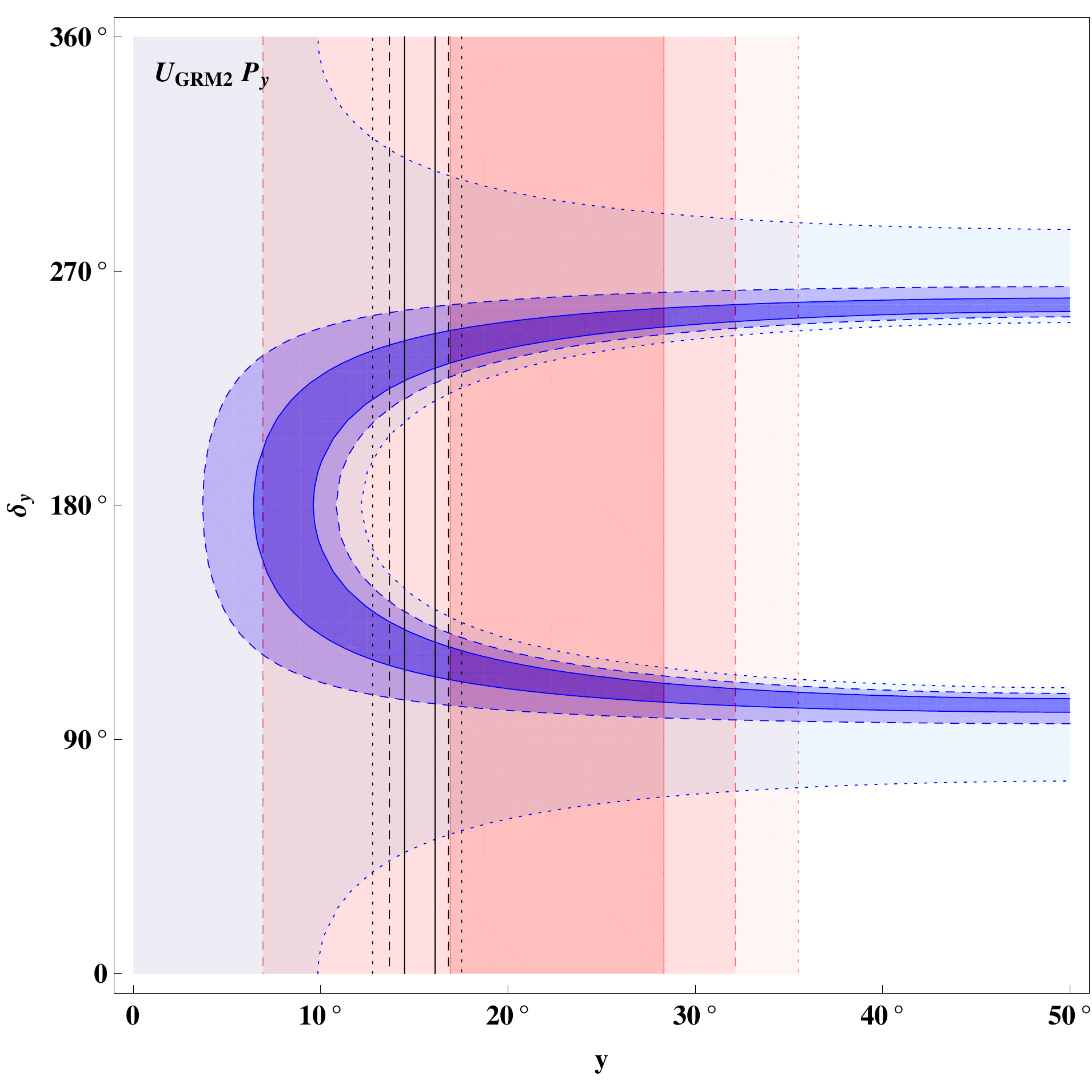}
\caption{The solutions corresponding to the measured $\theta_{12}$, $\theta_{23}$, and $\theta_{13}$ for $U_{\rm{GRM2}}\cdot P_y$ in the parameter space of $y - \delta_y$. Caption is the same as displayed in Fig. \ref{CPxUtbm}.} \label{CUgr2mPy}
\end{figure}

\clearpage
\begin{figure}[t]
\centering
\includegraphics[bb=0 110 550 550, height=12.5cm, width=15.5cm, angle=0]{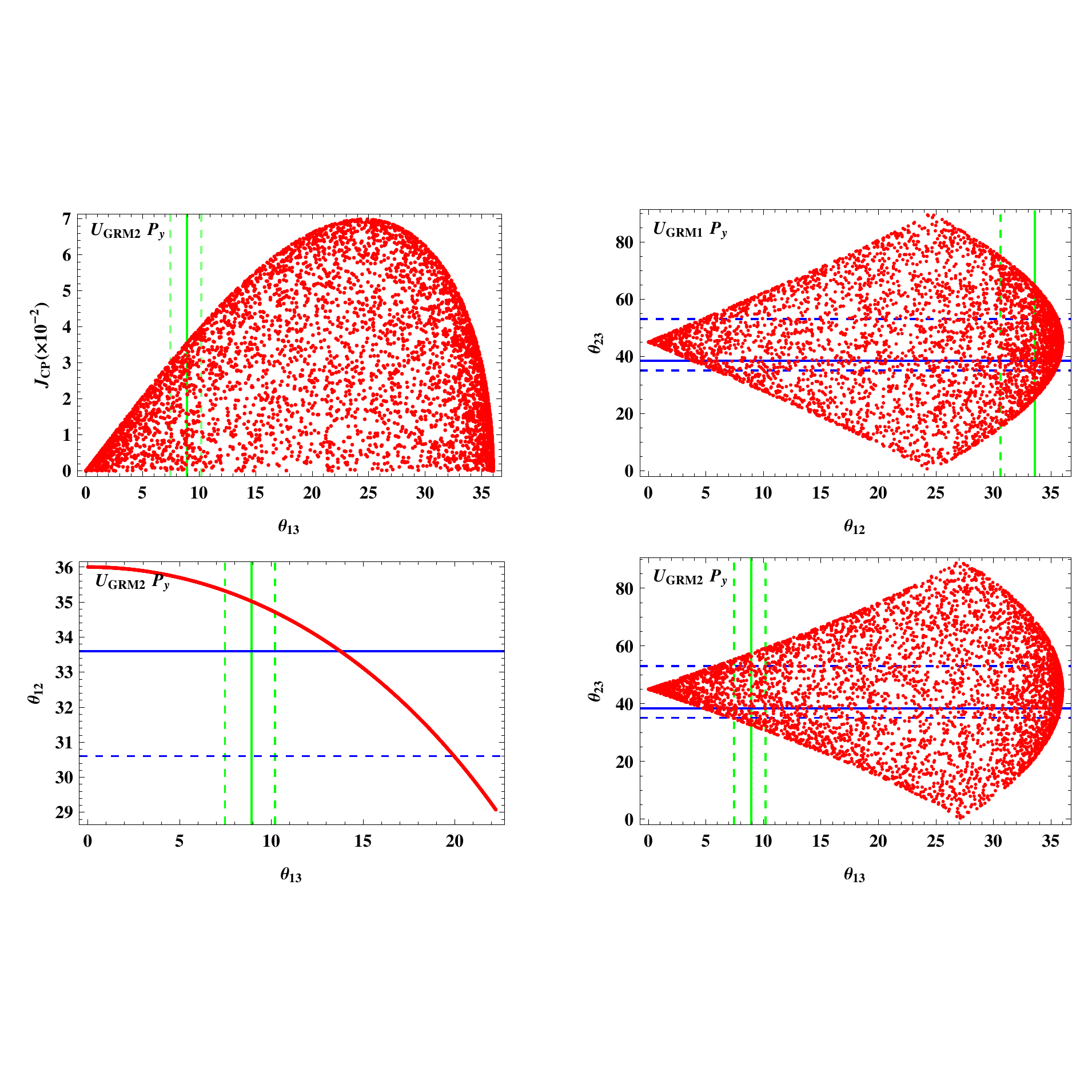}
\caption{The scatter plots for $\theta_{12}$, $\theta_{23}$, $\theta_{13}$ and $J_{\rm CP}$ with the $U_{\rm{GRM2}}\cdot P_y$ ansatz. Caption is the same as displayed in Fig. \ref{FigPxUtbm}.} \label{FigUgr2mPy}
\end{figure}

\clearpage
\subsubsection{$U_{\rm{GRM2}}\cdot P_z$}
\begin{figure}[t]
\centering
\includegraphics[bb=0 0 550 550, height=8.5cm, width=8.5cm, angle=0]{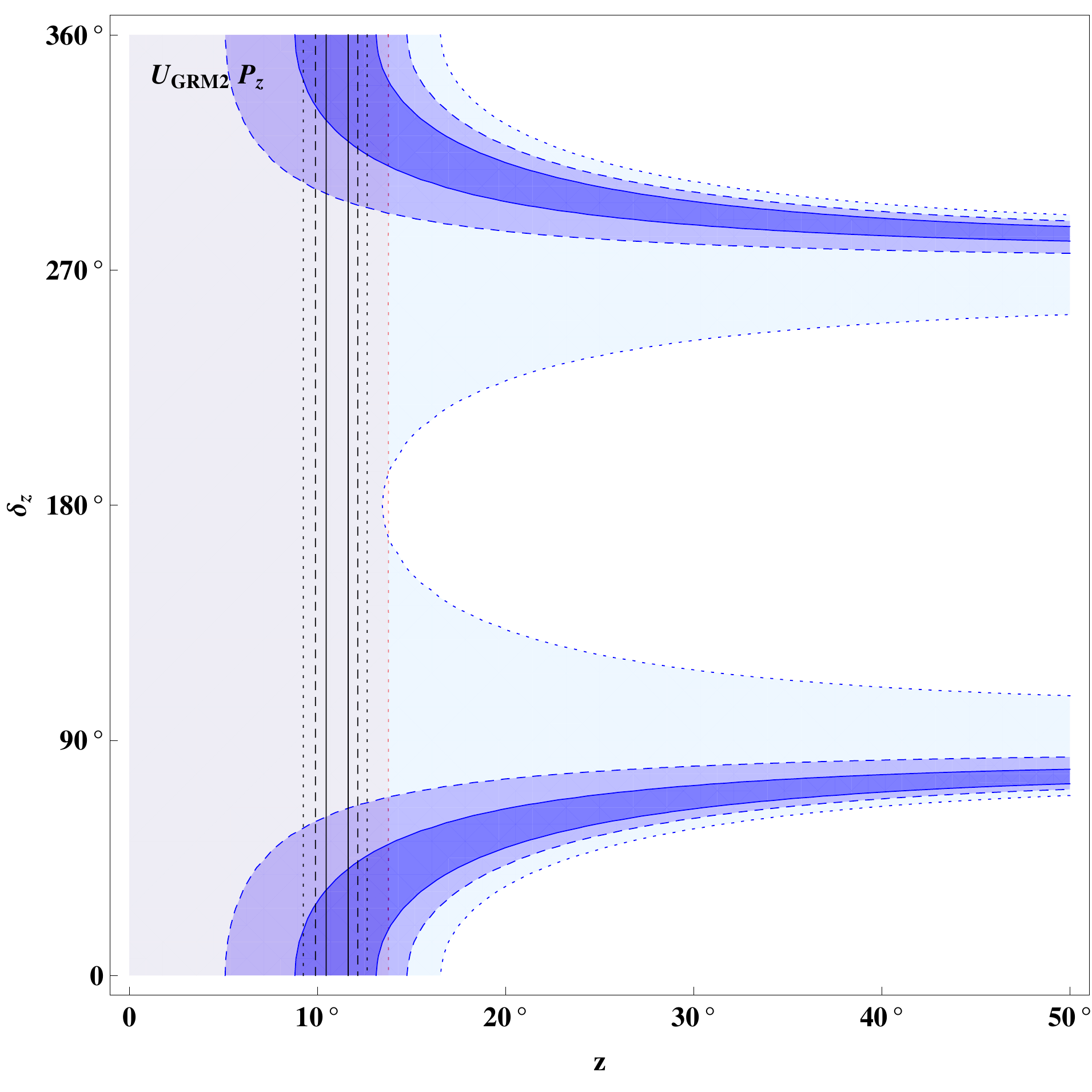}
\caption{The solutions corresponding to the measured $\theta_{12}$, $\theta_{23}$, and $\theta_{13}$ for $U_{\rm{GRM2}}\cdot P_z$ in the parameter space of $z - \delta_z$. Caption is the same as displayed in Fig. \ref{CPxUtbm}.} \label{CUgr2mPz}
\end{figure}

\clearpage
\begin{figure}[t]
\centering
\includegraphics[bb=0 110 550 550, height=12.5cm, width=15.5cm, angle=0]{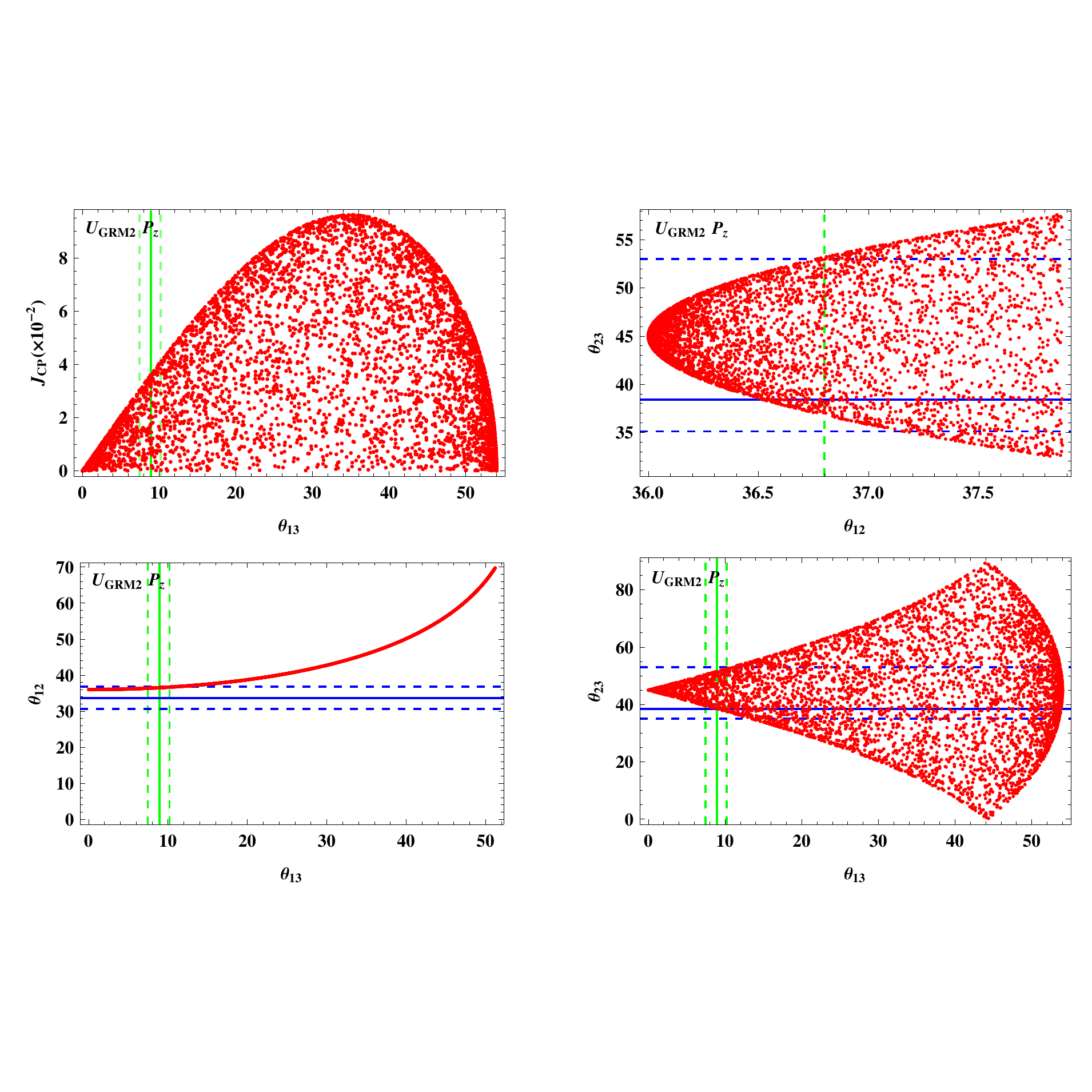}
\caption{The scatter plots for $\theta_{12}$, $\theta_{23}$, $\theta_{13}$ and $J_{\rm CP}$ with the $U_{\rm{GRM2}}\cdot P_z$ ansatz. Caption is the same  as displayed in Fig. \ref{FigPxUtbm}.} \label{FigUgr2mPz}
\end{figure}

\clearpage
\subsection{Hexagonal Mixing}
\subsubsection{$P_x\cdot U_{\rm{HM}}$}
\begin{figure}[t]
\centering
\includegraphics[bb=0 0 550 550, height=8.5cm, width=8.5cm, angle=0]{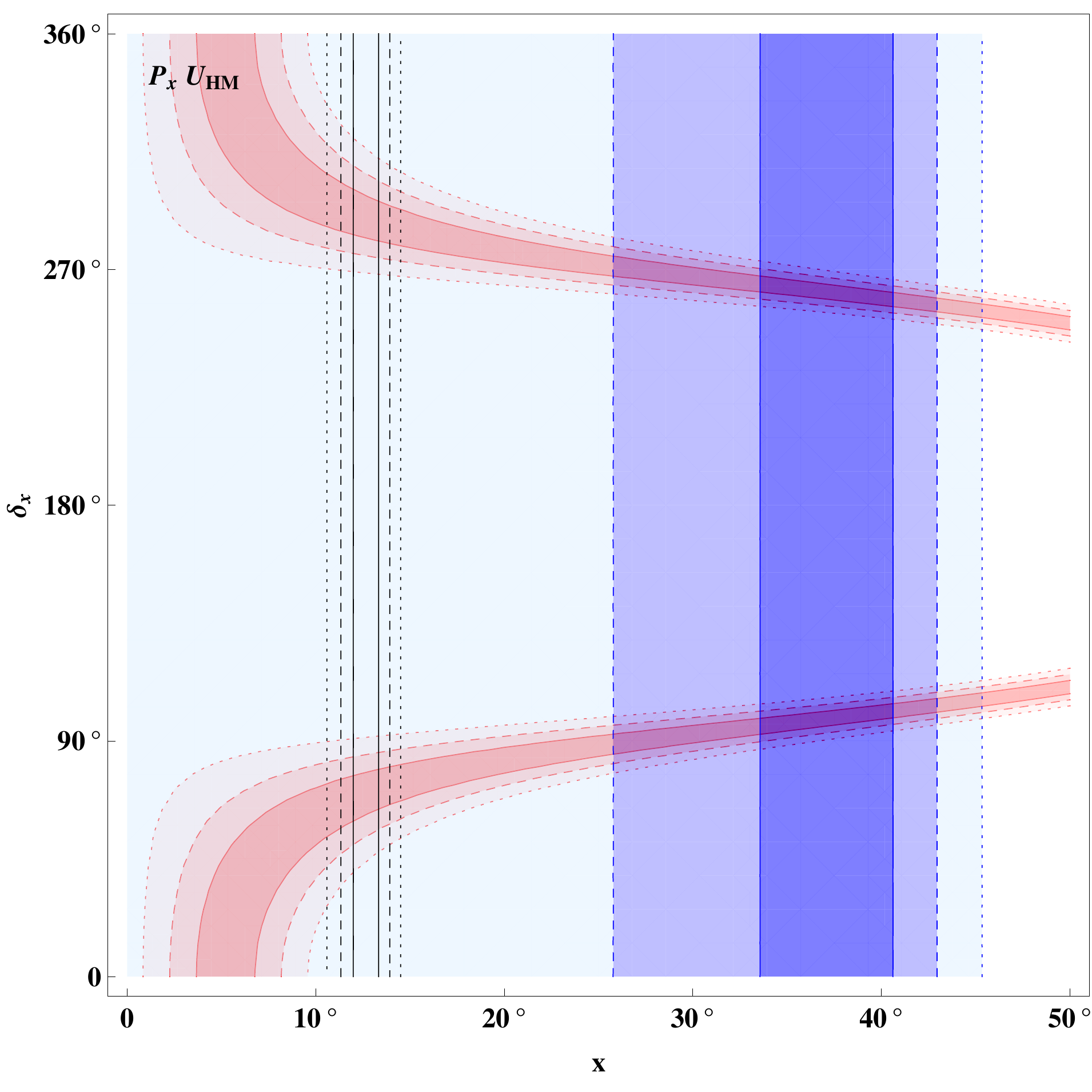}
\caption{The solutions corresponding to the measured $\theta_{12}$, $\theta_{23}$, and $\theta_{13}$ for $P_x\cdot U_{\rm{HM}}$ in the parameter space of $x - \delta_x$. Caption is the same as displayed in Fig. \ref{CPxUtbm}.} \label{CPxUhm}
\end{figure}

\clearpage
\begin{figure}[t]
\centering
\includegraphics[bb=0 110 550 550, height=12.5cm, width=15.5cm, angle=0]{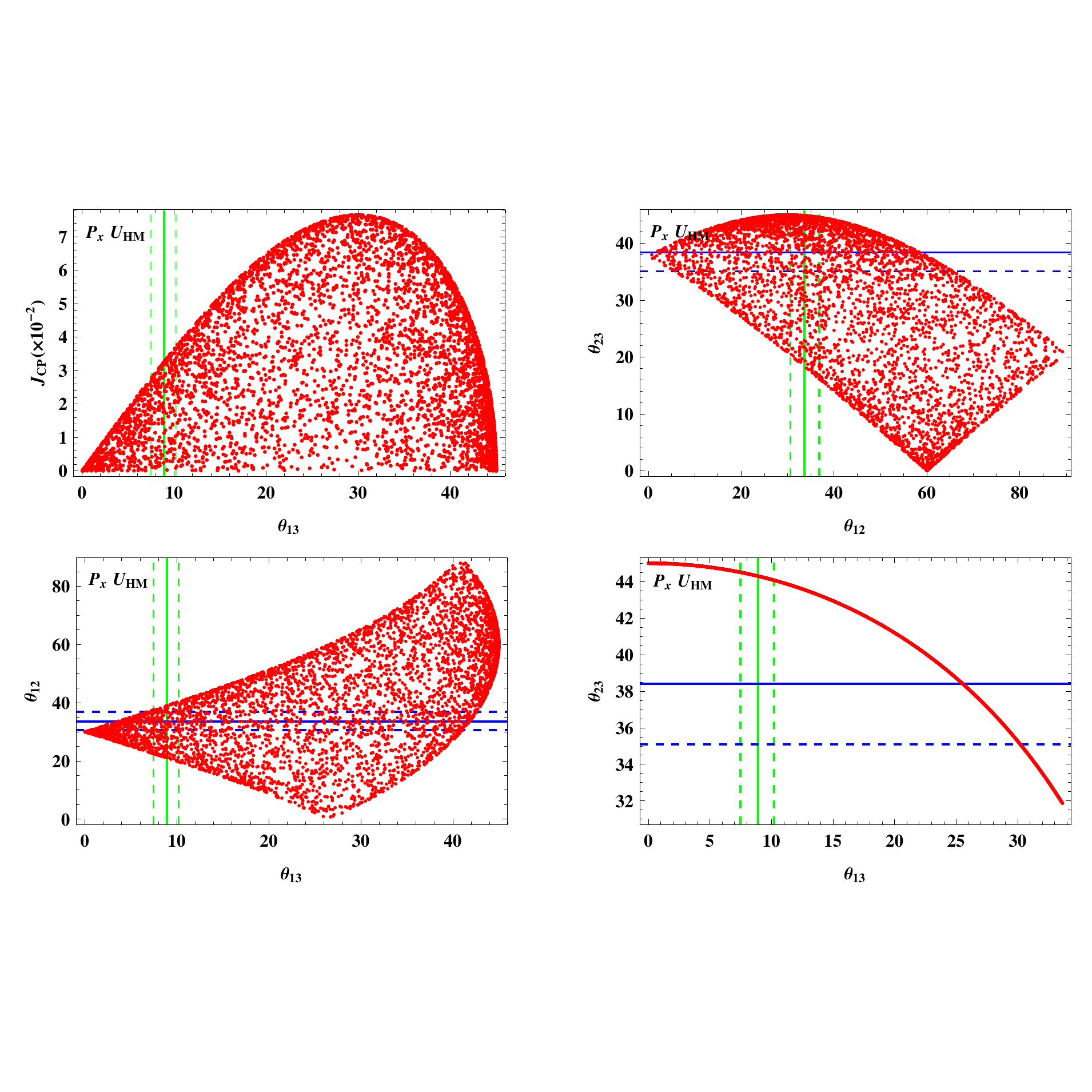}
\caption{The scatter plots for $\theta_{12}$, $\theta_{23}$, $\theta_{13}$ and $J_{\rm CP}$ with the $P_x\cdot U_{\rm{HM}}$ ansatz. Caption is the same as displayed in Fig. \ref{FigPxUtbm}.} \label{FigPxUhm}
\end{figure}

\clearpage
\subsubsection{$P_z\cdot U_{\rm{HM}}$}
\begin{figure}[t]
\centering
\includegraphics[bb=0 0 550 550, height=8.5cm, width=8.5cm, angle=0]{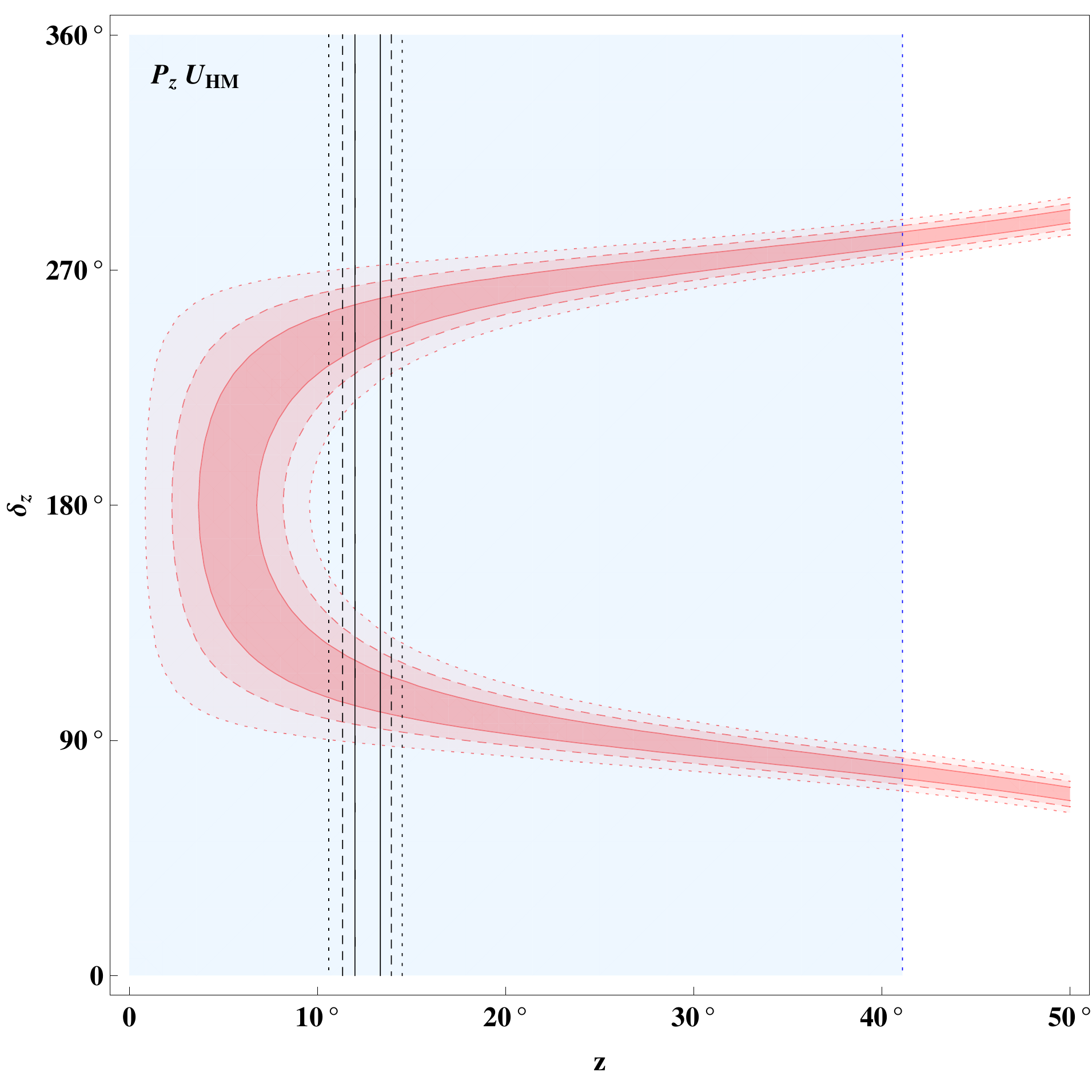}
\caption{The solutions corresponding to the measured $\theta_{12}$, $\theta_{23}$, and $\theta_{13}$ for $P_z\cdot U_{\rm{HM}}$ in the parameter space of $z - \delta_z$. Caption is the same as displayed in Fig. \ref{CPxUtbm}.} \label{CPzUhm}
\end{figure}

\clearpage
\begin{figure}[t]
\centering
\includegraphics[bb=0 110 550 550, height=12.5cm, width=15.5cm, angle=0]{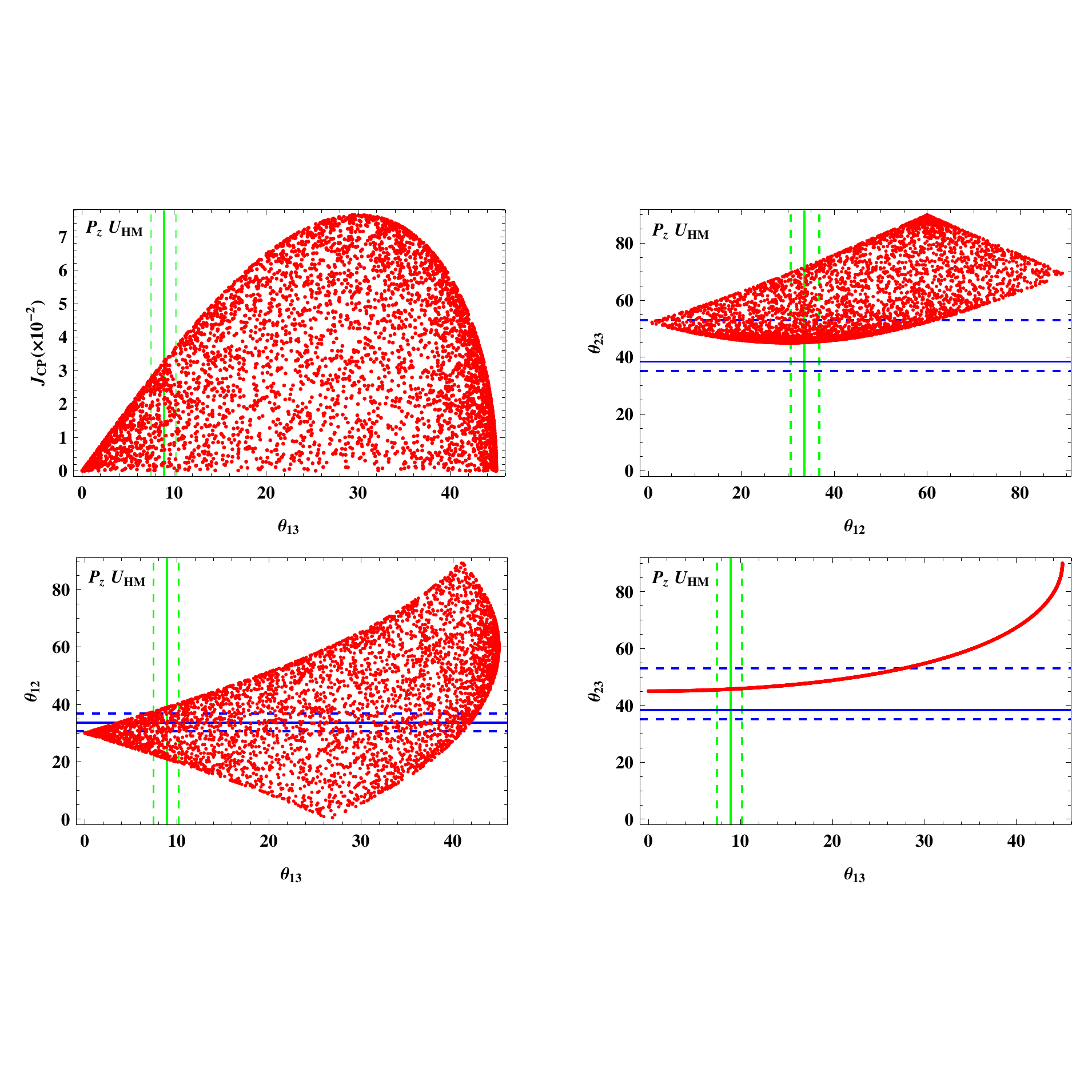}
\caption{The scatter plots for $\theta_{12}$, $\theta_{23}$, $\theta_{13}$ and $J_{\rm CP}$ with the $P_z\cdot U_{\rm{HM}}$ ansatz. Caption is the same as displayed in Fig. \ref{FigPxUtbm}.} \label{FigPzUhm}
\end{figure}

\clearpage
\subsubsection{$U_{\rm{HM}}\cdot P_y$}
\begin{figure}[t]
\centering
\includegraphics[bb=0 0 550 550, height=8.5cm, width=8.5cm, angle=0]{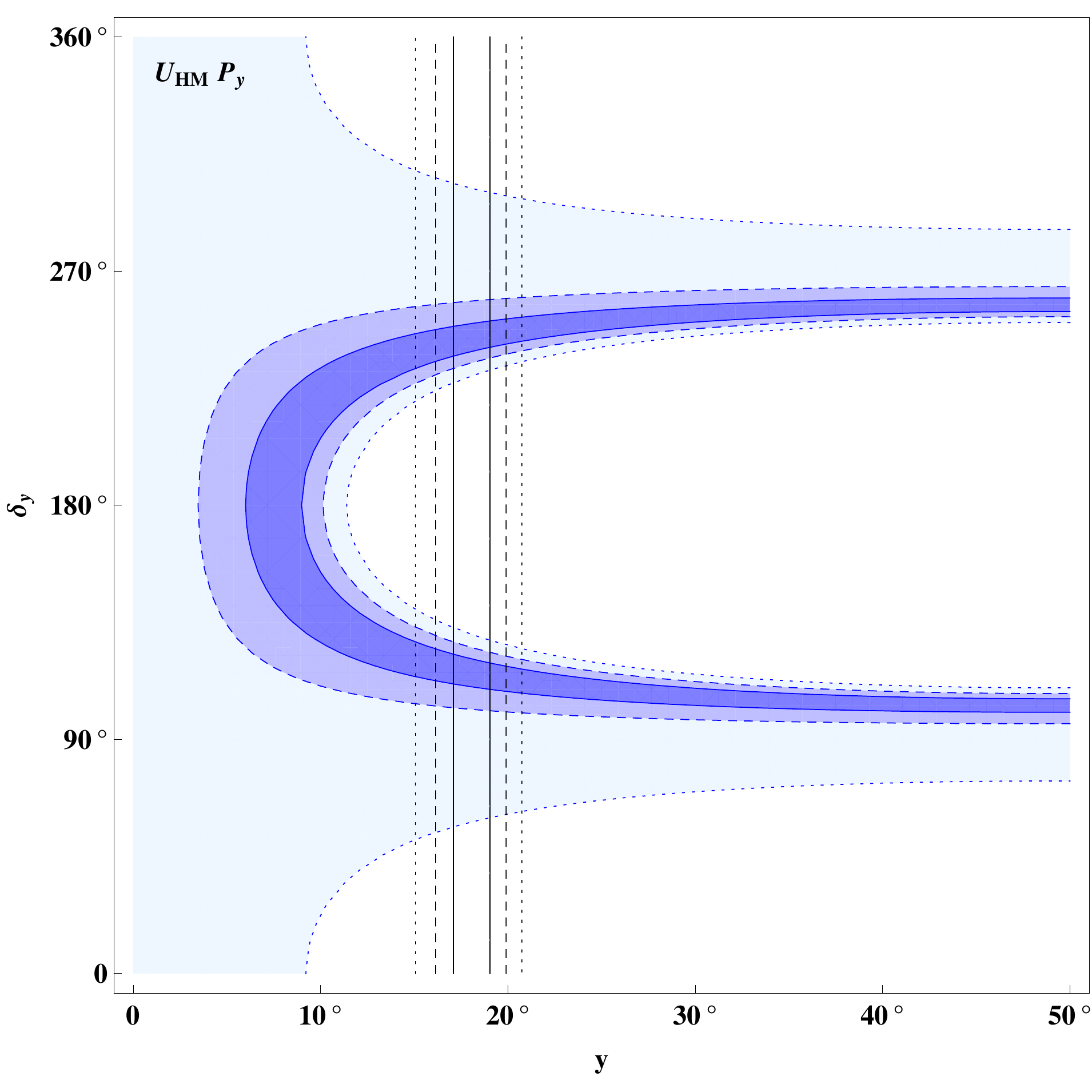}
\caption{The solutions corresponding to the measured $\theta_{12}$, $\theta_{23}$, and $\theta_{13}$ for $U_{\rm{HM}}\cdot P_y$ in the parameter space of $y - \delta_y$. Caption is the same as displayed in Fig. \ref{CPxUtbm}.} \label{CUhmPy}
\end{figure}

\clearpage
\subsubsection{$U_{\rm{HM}}\cdot P_z$}
\begin{figure}[t]
\centering
\includegraphics[bb=0 0 550 550, height=8.5cm, width=8.5cm, angle=0]{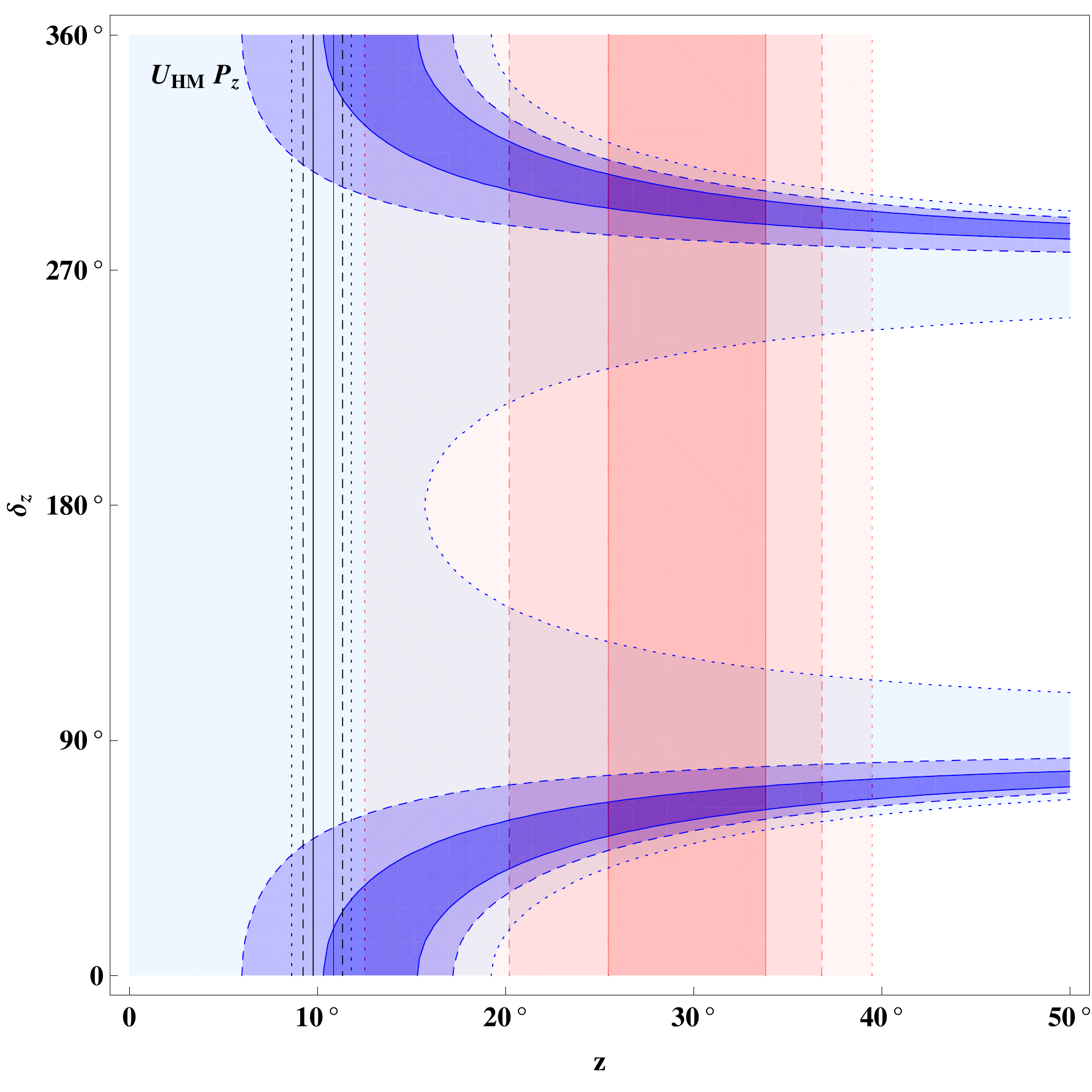}
\caption{The solutions corresponding to the measured $\theta_{12}$, $\theta_{23}$, and $\theta_{13}$ for $U_{\rm{HM}}\cdot P_z$ in the parameter space of $z - \delta_z$. Caption is the same as displayed in Fig. \ref{CPxUtbm}.} \label{CUhmPz}
\end{figure}

\clearpage
\subsection{Tetra-Maximal Mixing}
\subsubsection{$P_x\cdot U_{\rm{TMM}}$}
\begin{figure}[t]
\centering
\includegraphics[bb=0 0 550 550, height=8.5cm, width=8.5cm, angle=0]{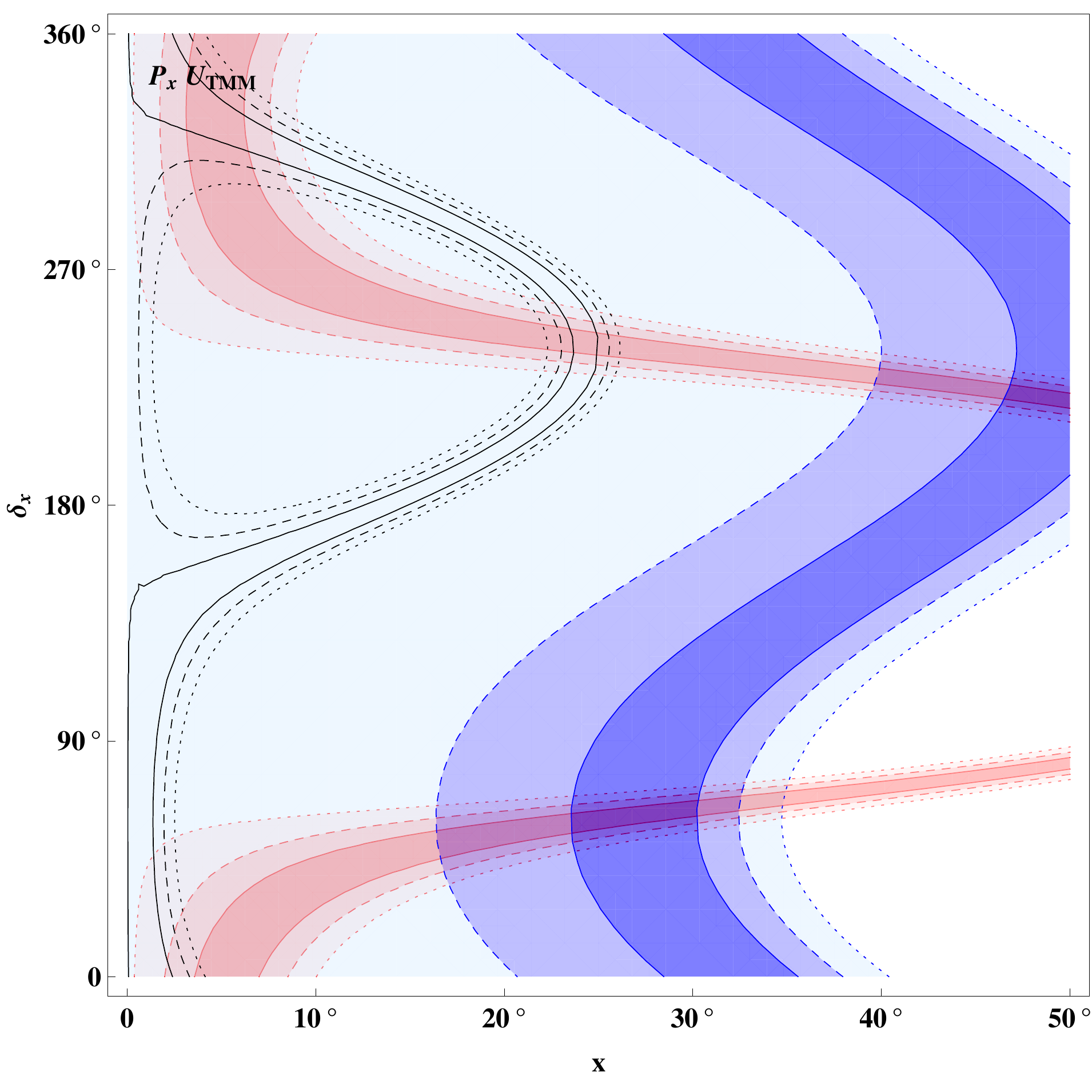}
\caption{The solutions corresponding to the measured $\theta_{12}$, $\theta_{23}$, and $\theta_{13}$ for $P_x\cdot U_{\rm{TMM}}$ in the parameter space of $x - \delta_x$. Caption is the same as displayed in Fig. \ref{CPxUtbm}.} \label{CPxUtmm}
\end{figure}

\clearpage
\begin{figure}[t]
\centering
\includegraphics[bb=0 110 550 550, height=12.5cm, width=15.5cm, angle=0]{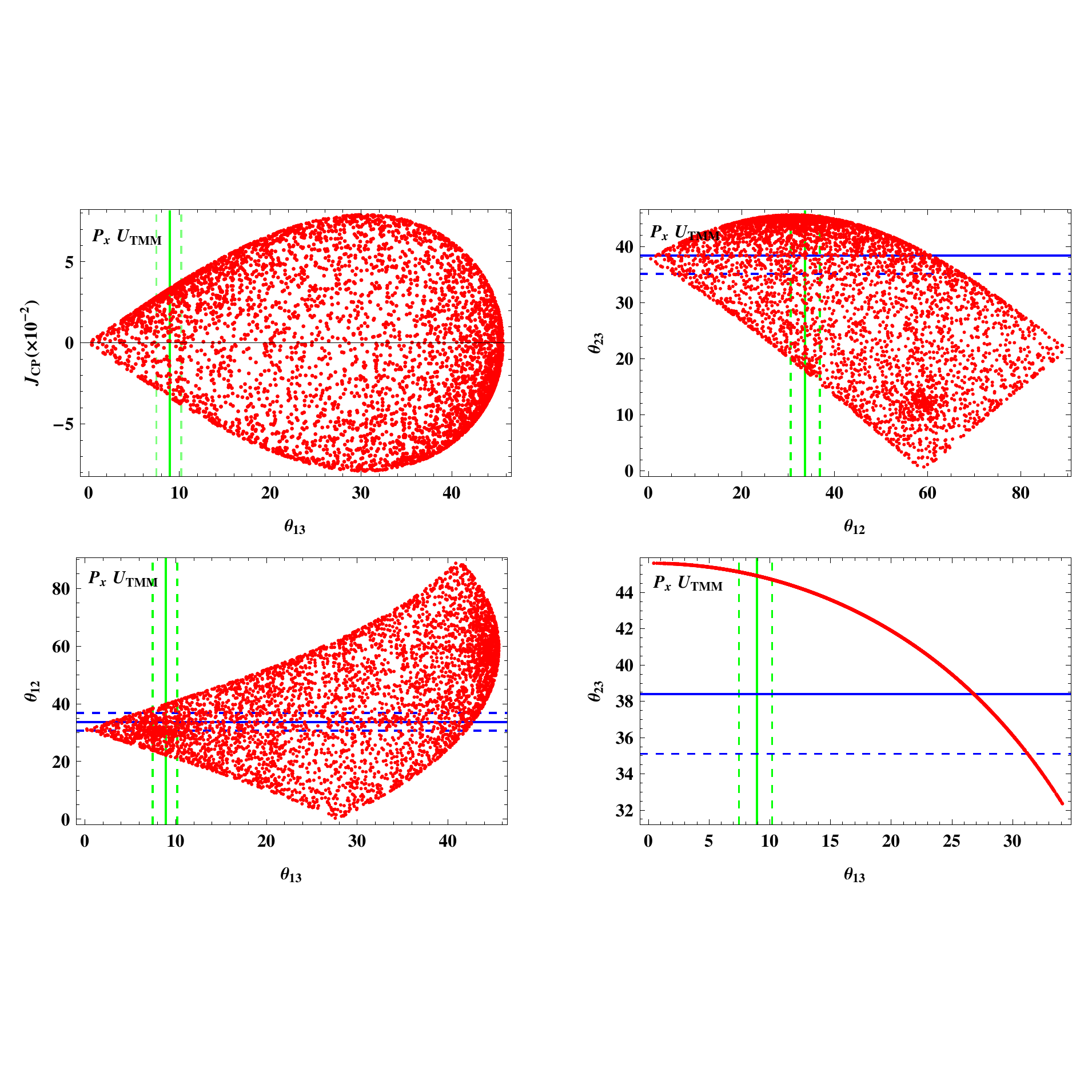}
\caption{The scatter plots for $\theta_{12}$, $\theta_{23}$, $\theta_{13}$ and $J_{\rm CP}$ with the $P_x\cdot U_{\rm{TMM}}$ ansatz. Caption is the same as displayed in Fig. \ref{FigPxUtbm}.} \label{FigPxUtmm}
\end{figure}

\clearpage
\subsubsection{$P_z\cdot U_{\rm{TMM}}$}
\begin{figure}[t]
\centering
\includegraphics[bb=0 0 550 550, height=8.5cm, width=8.5cm, angle=0]{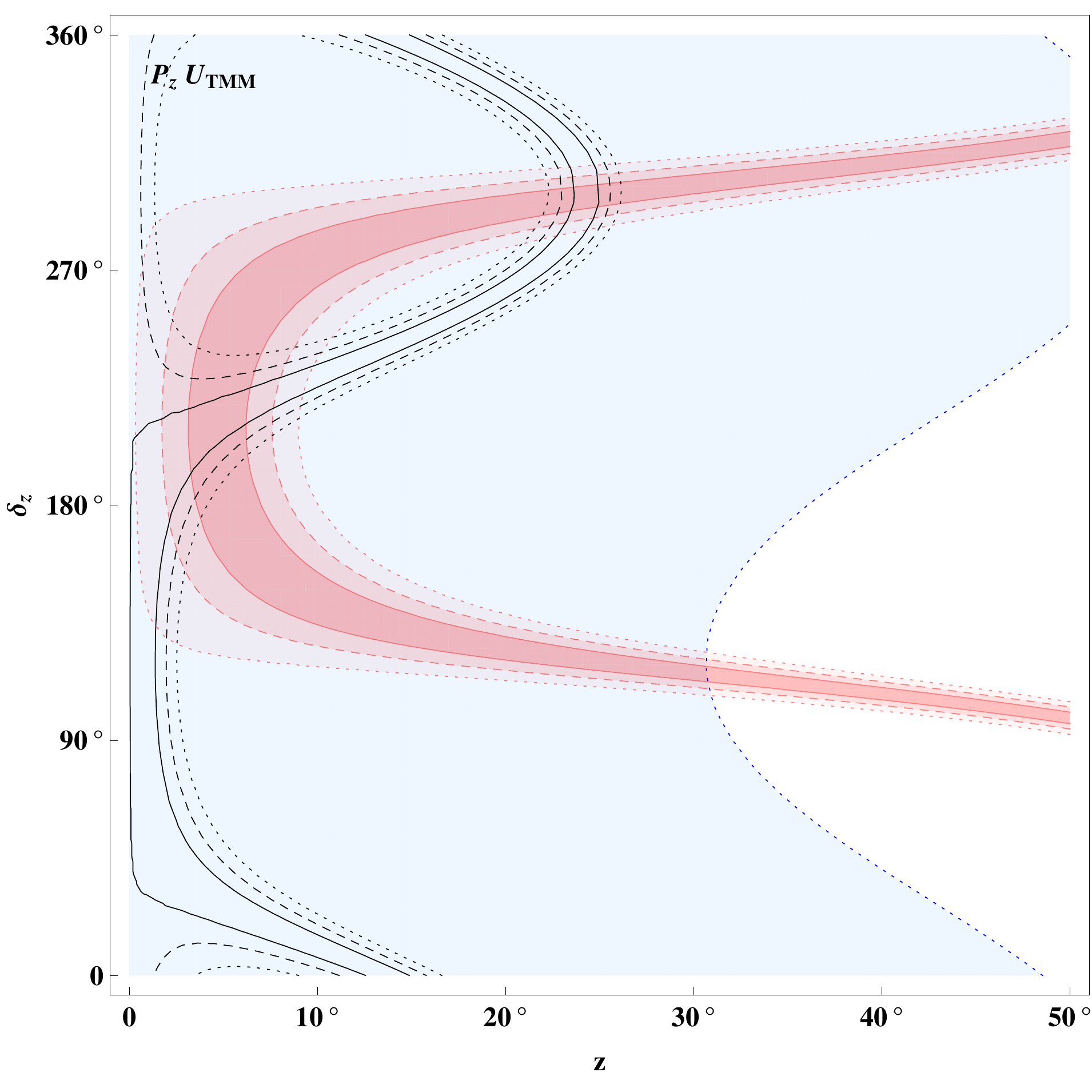}
\caption{The solutions corresponding to the measured $\theta_{12}$, $\theta_{23}$, and $\theta_{13}$ for $P_z\cdot U_{\rm{TMM}}$ in the parameter space of $z - \delta_z$. Caption is the same as displayed in Fig. \ref{CPxUtbm}.} \label{CPzUtmm}
\end{figure}

\clearpage
\begin{figure}[t]
\centering
\includegraphics[bb=0 110 550 550, height=12.5cm, width=15.5cm, angle=0]{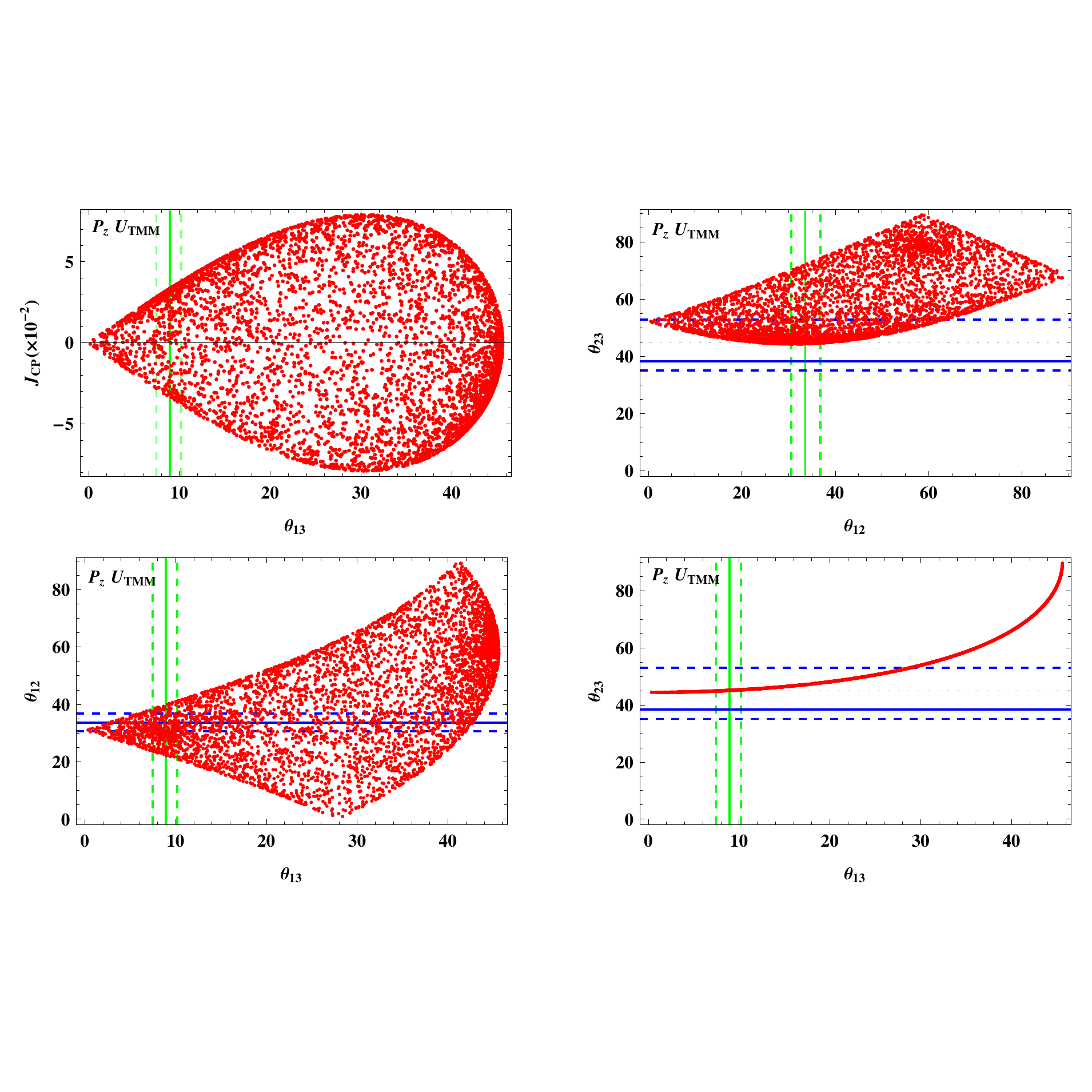}
\caption{The scatter plots for $\theta_{12}$, $\theta_{23}$, $\theta_{13}$ and $J_{\rm CP}$ with the $P_z\cdot U_{\rm{TMM}}$ ansatz. Caption is the same as displayed in Fig. \ref{FigPxUtbm}. $\theta_{23}=45^\circ$ labeled as horizontal dotted lines in upper-right and lower-right panels.} \label{FigPzUtmm}
\end{figure}

\clearpage
\subsubsection{$U_{\rm{TMM}}\cdot P_y$}
\begin{figure}[t]
\centering
\includegraphics[bb=0 0 550 550, height=8.5cm, width=8.5cm, angle=0]{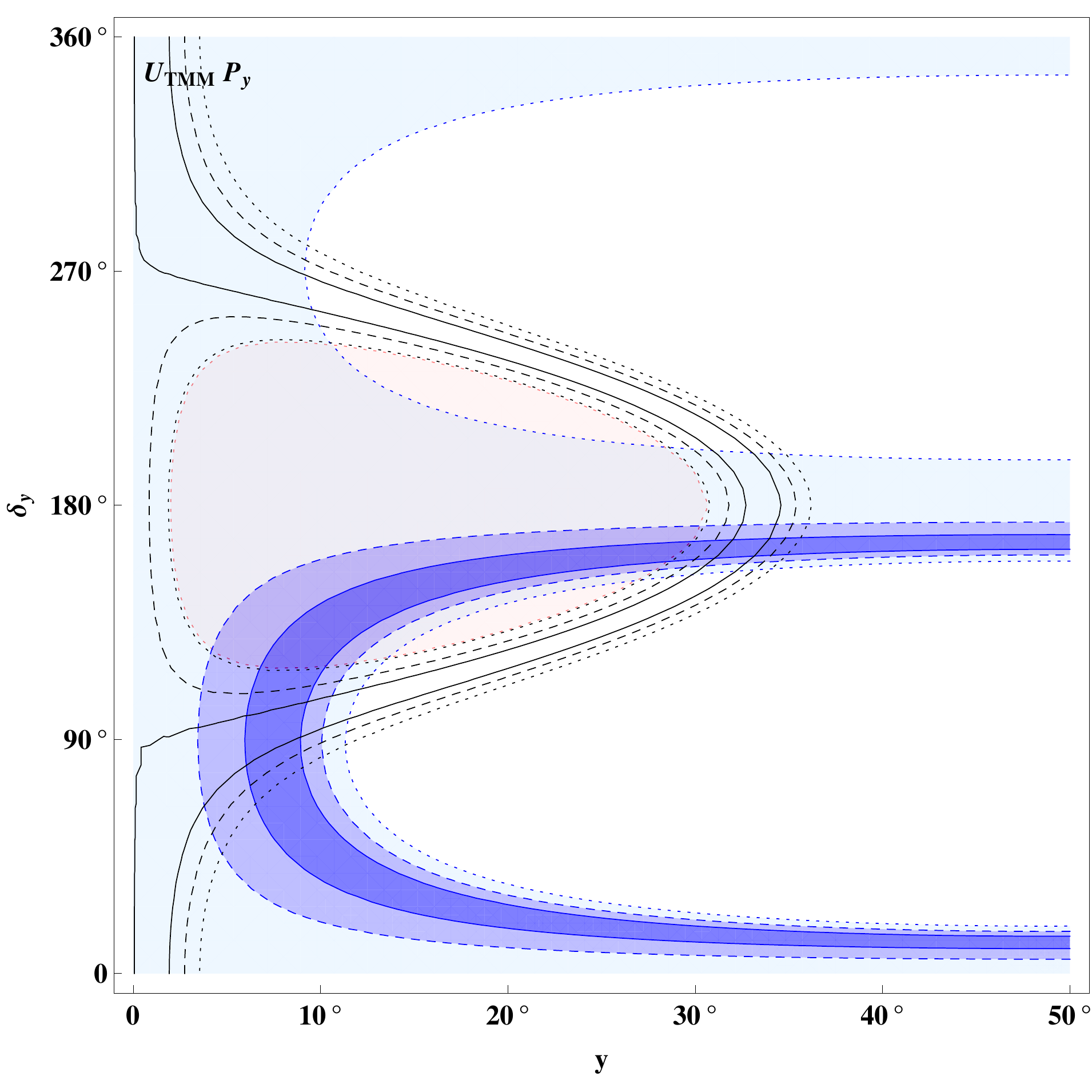}
\caption{The solutions corresponding to the measured $\theta_{12}$, $\theta_{23}$, and $\theta_{13}$ for $U_{\rm{TMM}}\cdot P_y$ in the parameter space of $y - \delta_y$. Caption is the same as displayed in Fig. \ref{CPxUtbm}.} \label{CUtmmPy}
\end{figure}

\clearpage
\subsubsection{$U_{\rm{TMM}}\cdot P_z$}
\begin{figure}[t]
\centering
\includegraphics[bb=0 0 550 550, height=8.5cm, width=8.5cm, angle=0]{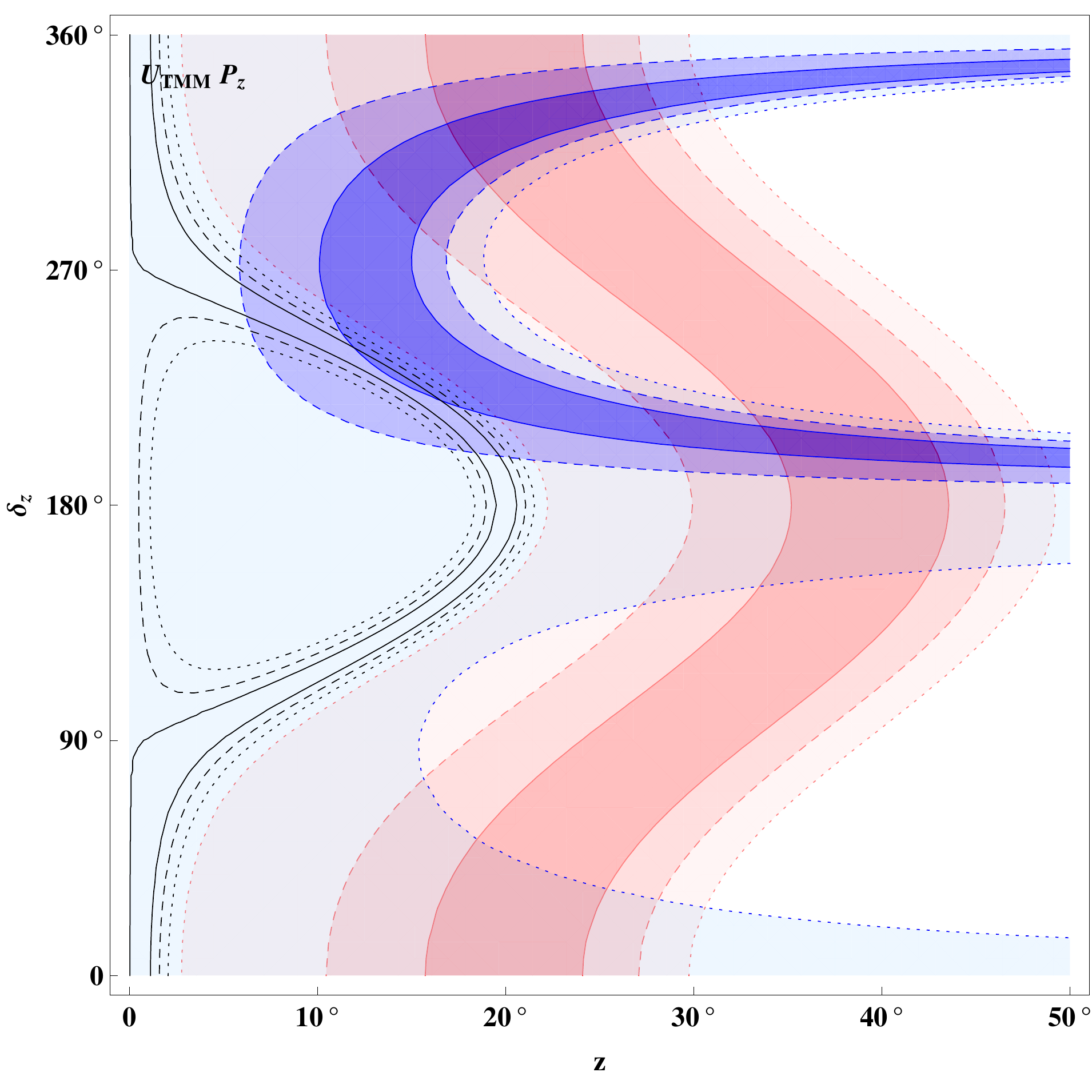}
\caption{The solutions corresponding to the measured $\theta_{12}$, $\theta_{23}$, and $\theta_{13}$ for $U_{\rm{TMM}}\cdot P_z$ in the parameter space of $z - \delta_z$. Caption is the same as displayed in Fig. \ref{CPxUtbm}.} \label{CUtmmPz}
\end{figure}

\clearpage
\subsection{Toorop-Feruglio-Hagedorn Mixing-1}
\subsubsection{$P_x\cdot U_{\rm{TFH1}}$}
\begin{figure}[t]
\centering
\includegraphics[bb=0 0 550 550, height=8.5cm, width=8.5cm, angle=0]{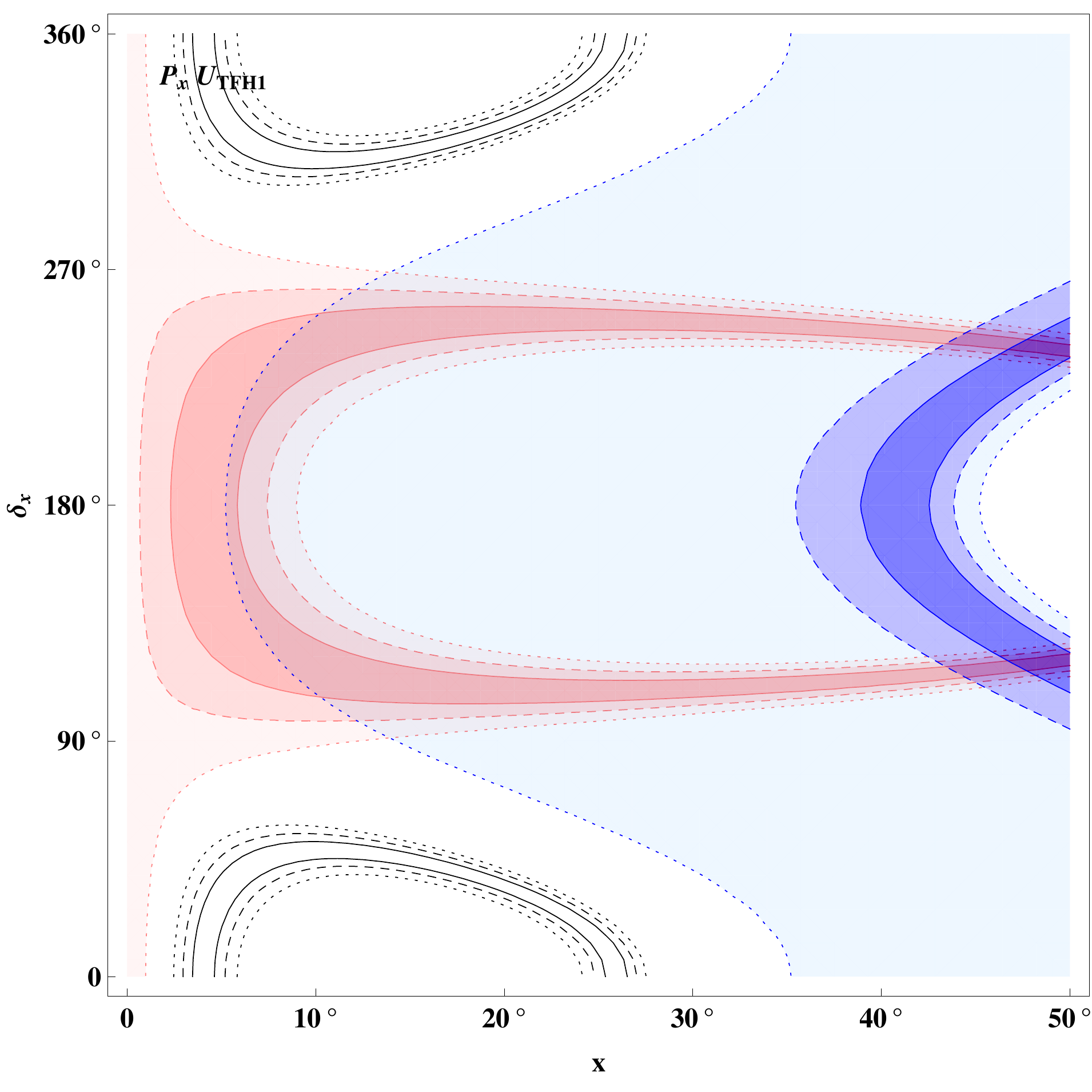}
\caption{The solutions corresponding to the measured $\theta_{12}$, $\theta_{23}$, and $\theta_{13}$ for $P_x\cdot U_{\rm{TFH1}}$ in the parameter space of $x - \delta_x$. Caption is the same as displayed in Fig. \ref{CPxUtbm}.} \label{CPxUtfh1}
\end{figure}

\clearpage
\subsubsection{$P_z\cdot U_{\rm{TFH1}}$}
\begin{figure}[t]
\centering
\includegraphics[bb=0 0 550 550, height=8.5cm, width=8.5cm, angle=0]{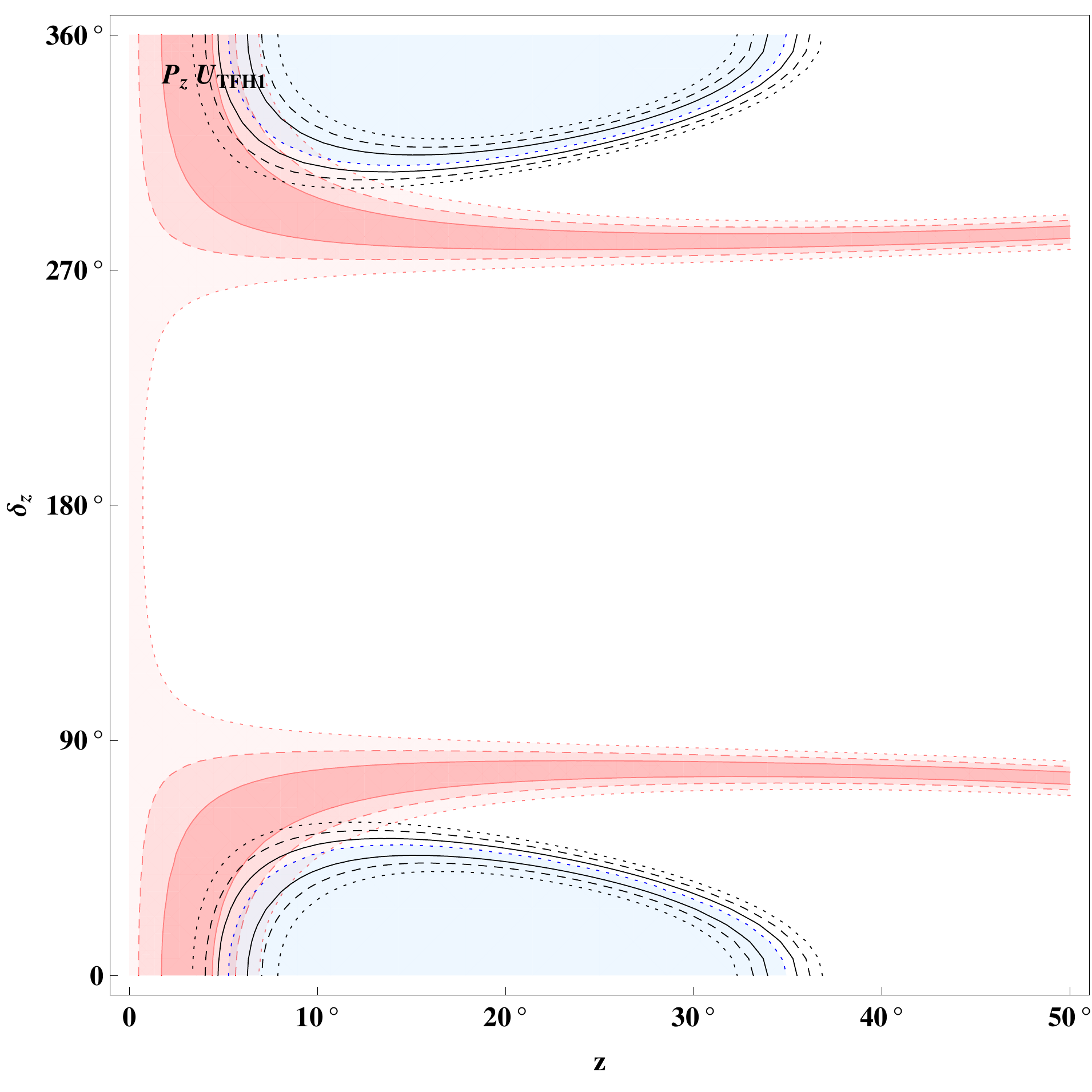}
\caption{The solutions corresponding to the measured $\theta_{12}$, $\theta_{23}$, and $\theta_{13}$ for $P_z\cdot U_{\rm{TFH1}}$ in the parameter space of $z - \delta_z$. Caption is the same as displayed in Fig. \ref{CPxUtbm}.} \label{CPzUtfh1}
\end{figure}

\clearpage
\begin{figure}[t]
\centering
\includegraphics[bb=0 110 550 550, height=12.5cm, width=15.5cm, angle=0]{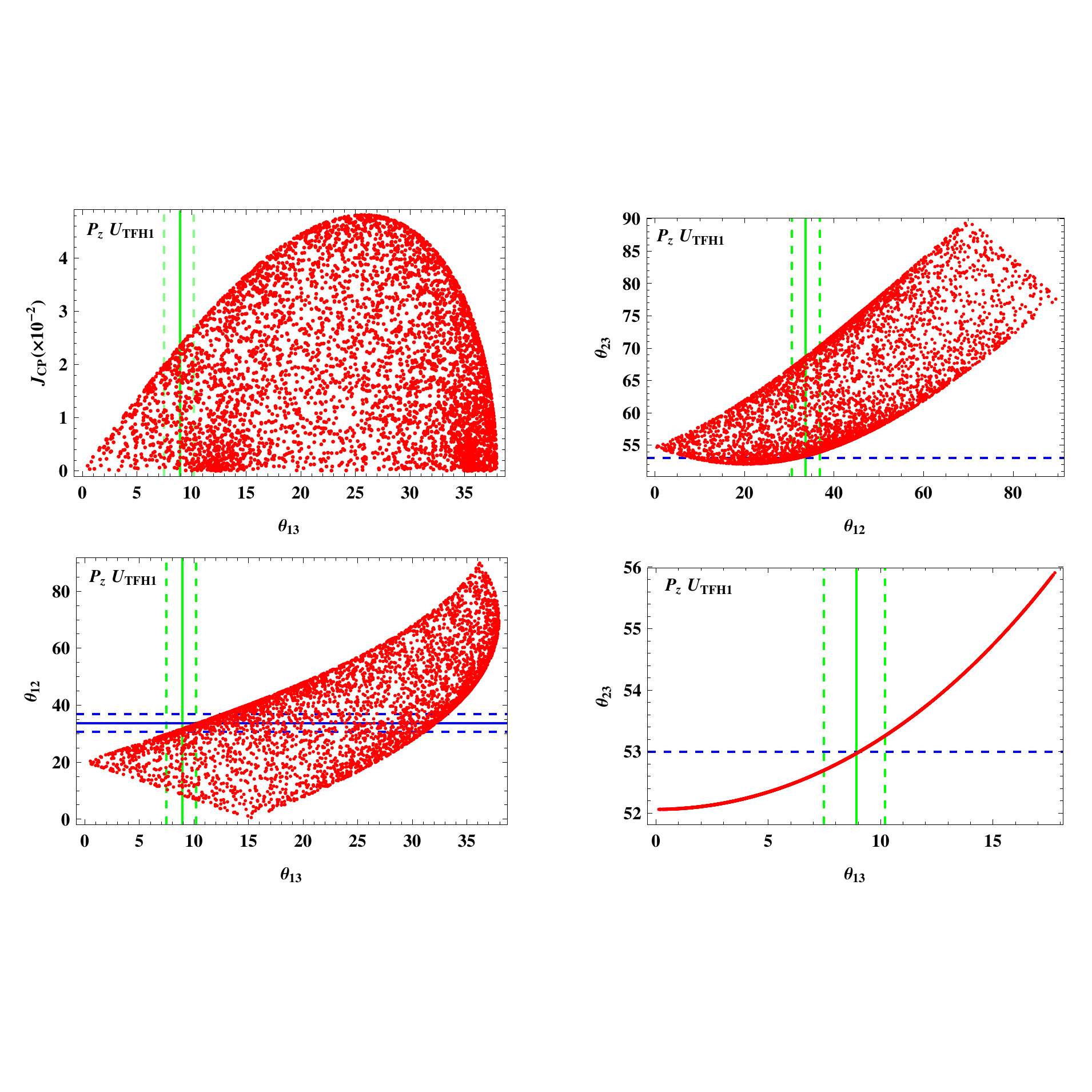}
\caption{The scatter plots for $\theta_{12}$, $\theta_{23}$, $\theta_{13}$ and $J_{\rm CP}$ with the $P_z\cdot U_{\rm{TFH1}}$ ansatz. Caption is the same as displayed in Fig. \ref{FigPxUtbm}.} \label{FigPzUtfh1}
\end{figure}

\clearpage
\subsubsection{$U_{\rm{TFH1}}\cdot P_y$}
\begin{figure}[t]
\centering
\includegraphics[bb=0 0 550 550, height=8.5cm, width=8.5cm, angle=0]{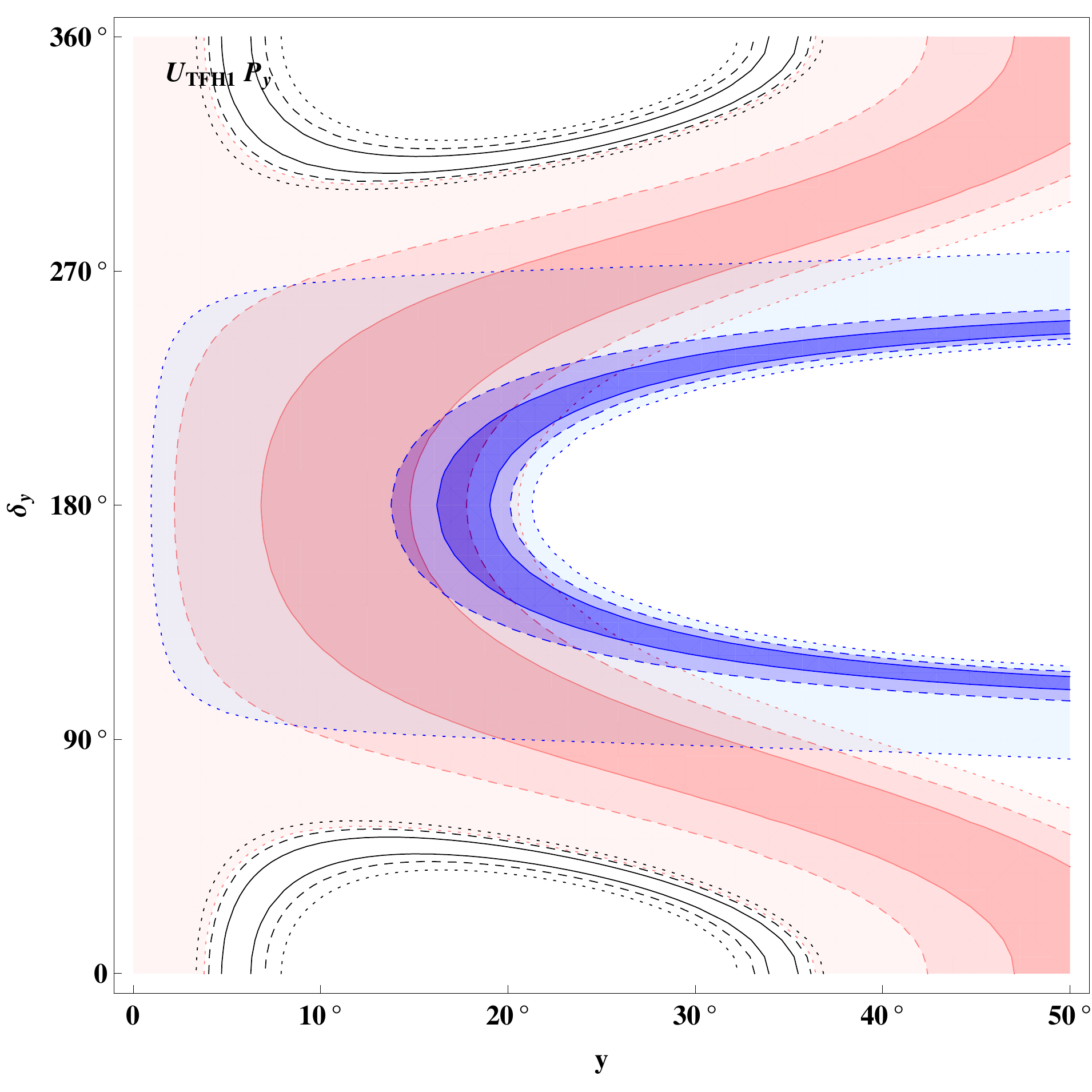}
\caption{The solutions corresponding to the measured $\theta_{12}$, $\theta_{23}$, and $\theta_{13}$ for $U_{\rm{TFH1}}\cdot P_y$ in the parameter space of $y - \delta_y$. Caption is the same  as displayed in Fig. \ref{CPxUtbm}.} \label{CUtfh1Py}
\end{figure}

\clearpage
\subsubsection{$U_{\rm{TFH1}}\cdot P_z$}
\begin{figure}[t]
\centering
\includegraphics[bb=0 0 550 550, height=8.5cm, width=8.5cm, angle=0]{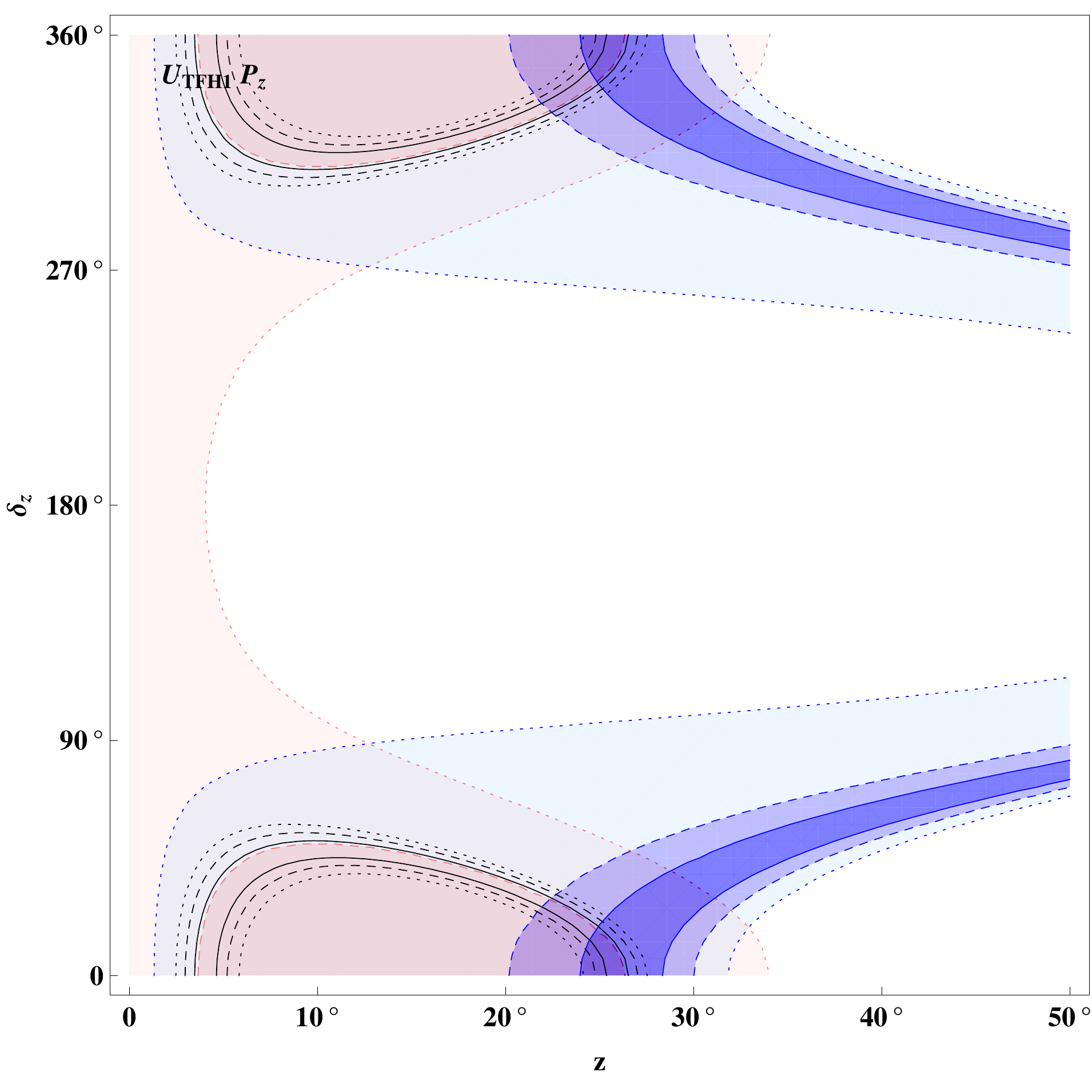}
\caption{The solutions corresponding to the measured $\theta_{12}$, $\theta_{23}$, and $\theta_{13}$ for $U_{\rm{TFH1}}\cdot P_z$ in the parameter space of $z - \delta_z$. Caption is the same as displayed in Fig. \ref{CPxUtbm}.} \label{CUtfh1Pz}
\end{figure}

\clearpage
\begin{figure}[t]
\centering
\includegraphics[bb=0 110 550 550, height=12.5cm, width=15.5cm, angle=0]{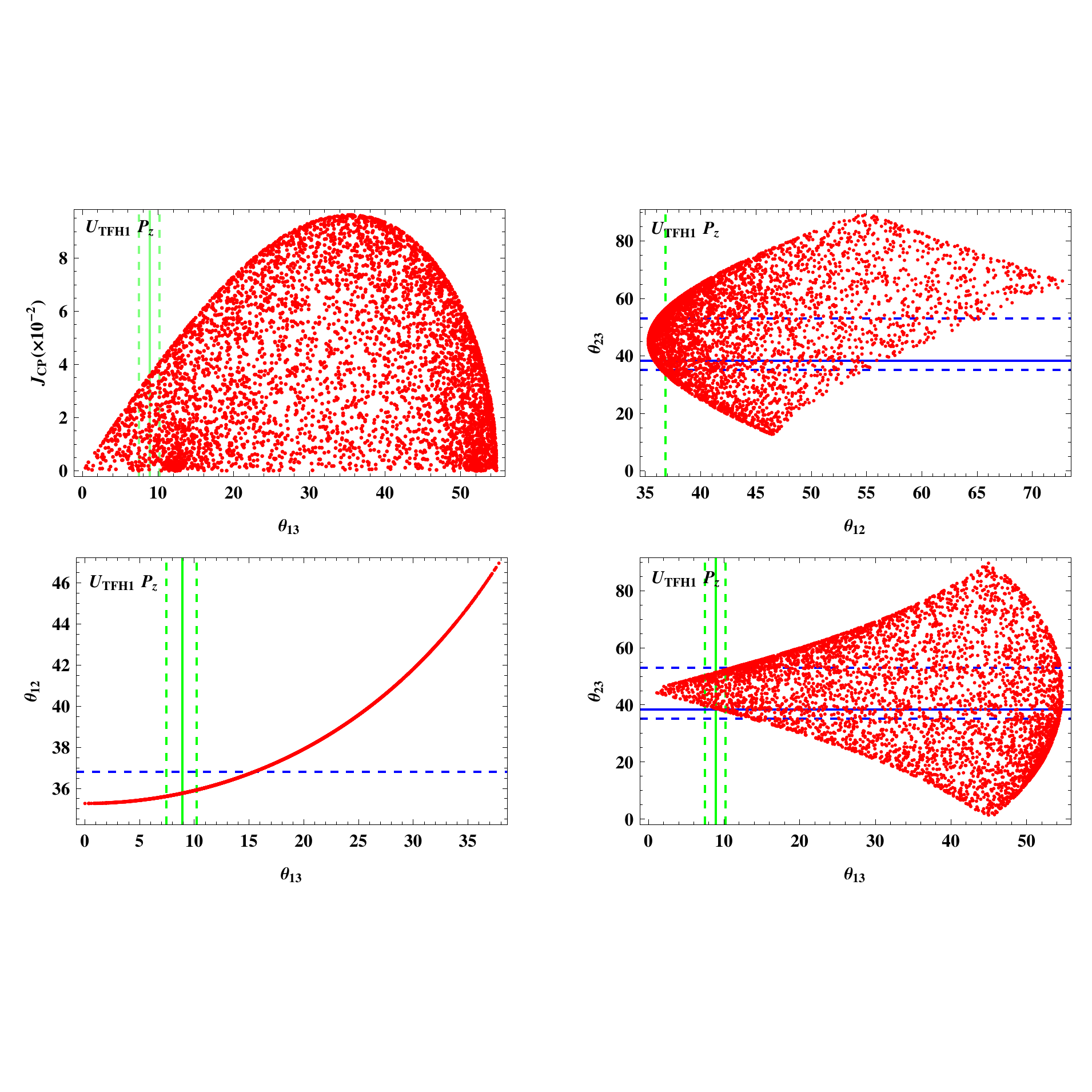}
\caption{The scatter plots for $\theta_{12}$, $\theta_{23}$, $\theta_{13}$ and $J_{\rm CP}$ with the $U_{\rm{TFH1}}\cdot P_z$ ansatz. Caption is the same as displayed in Fig. \ref{FigPxUtbm}.} \label{FigUtfh1Pz}
\end{figure}

\clearpage
\subsection{Toorop-Feruglio-Hagedorn Mixing-2}
\subsubsection{$P_x\cdot U_{\rm{TFH2}}$}
\begin{figure}[t]
\centering
\includegraphics[bb=0 0 550 550, height=8.5cm, width=8.5cm, angle=0]{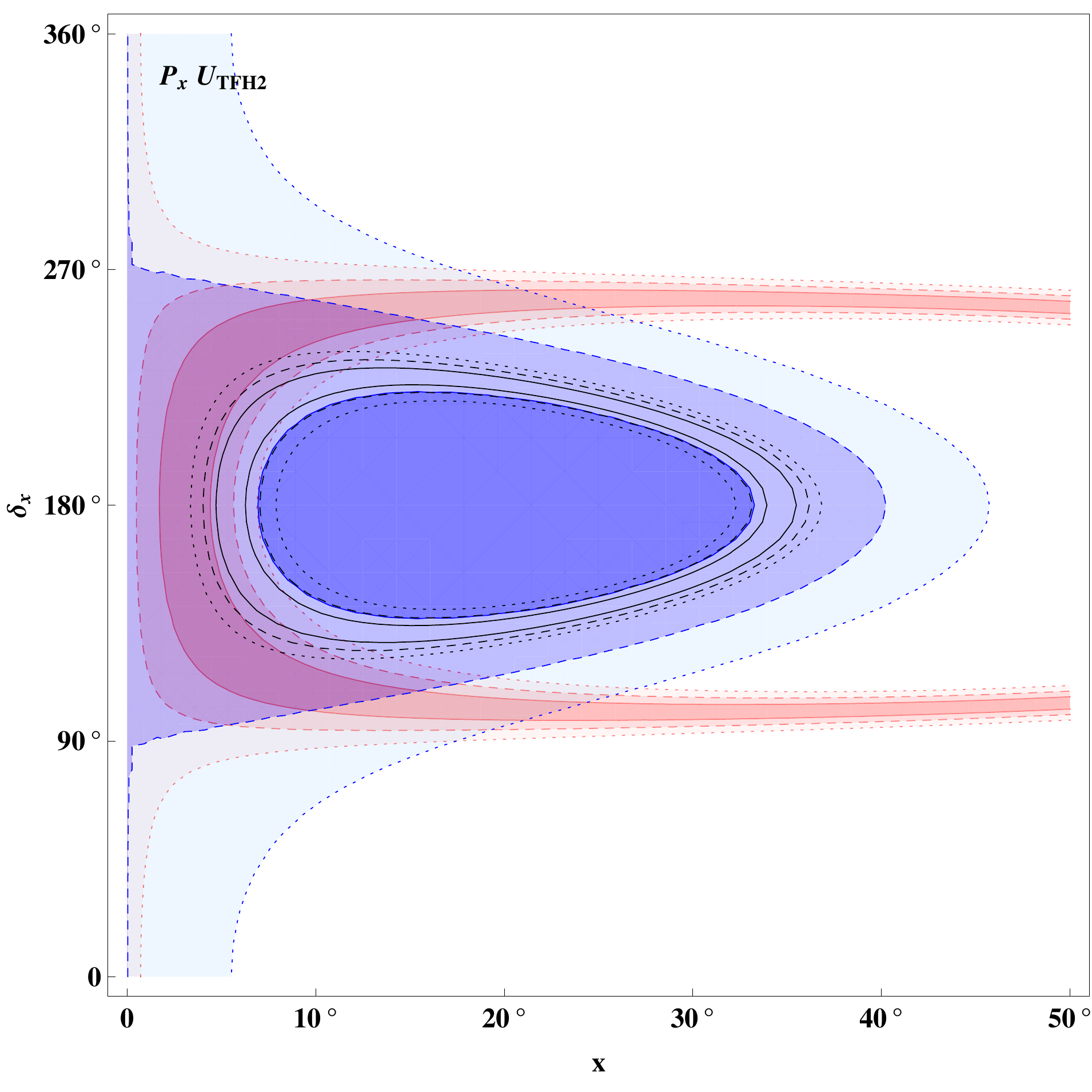}
\caption{The solutions corresponding to the measured $\theta_{12}$, $\theta_{23}$, and $\theta_{13}$ for $P_x\cdot U_{\rm{TFH2}}$ in the parameter space of $x - \delta_x$. Caption is the same as displayed in Fig. \ref{CPxUtbm}.} \label{CPxUtfh2}
\end{figure}

\clearpage
\begin{figure}[t]
\centering
\includegraphics[bb=0 110 550 550, height=12.5cm, width=15.5cm, angle=0]{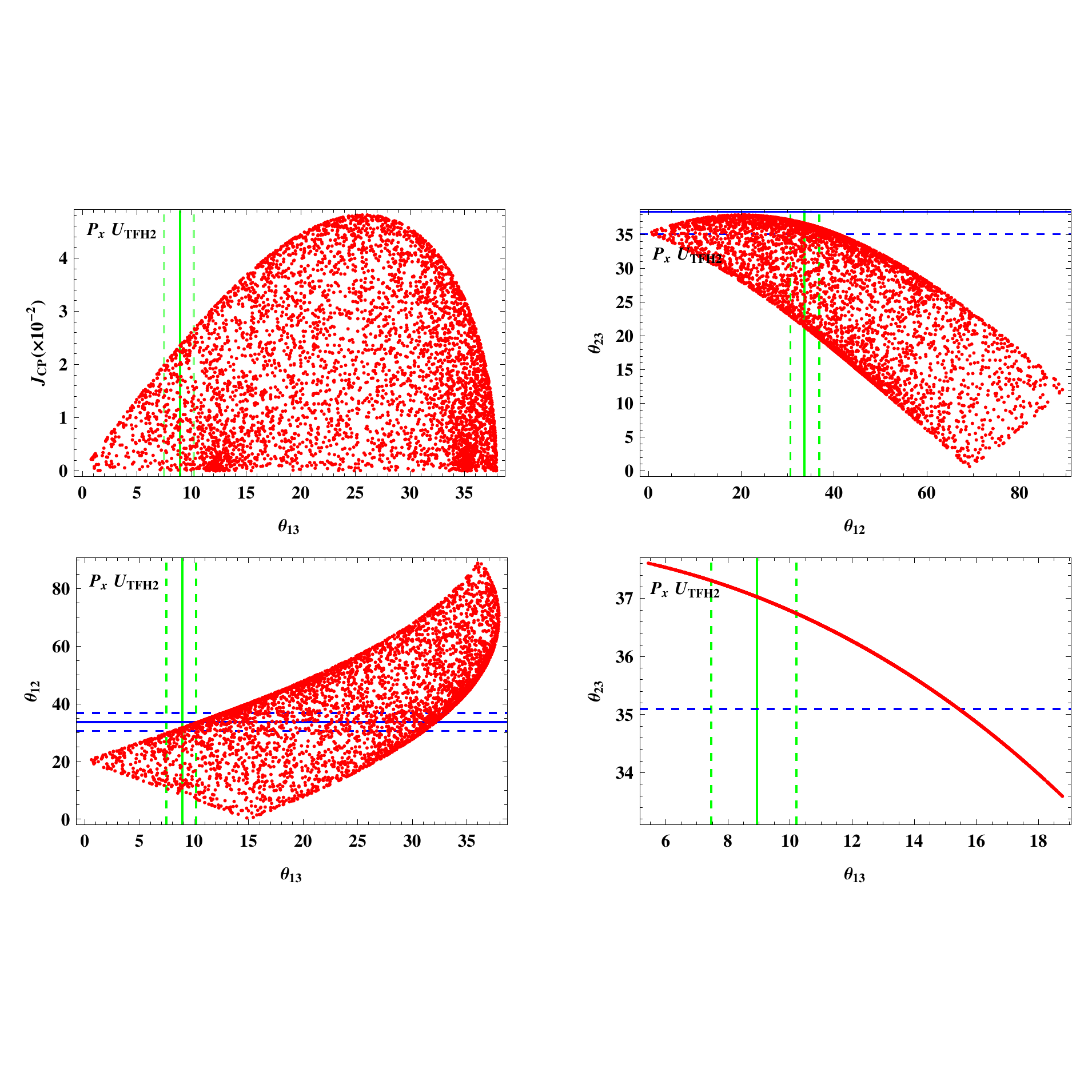}
\caption{The scatter plots for $\theta_{12}$, $\theta_{23}$, $\theta_{13}$ and $J_{\rm CP}$ with the $P_x\cdot U_{\rm{TFH2}}$ ansatz. Caption is the same as displayed in Fig. \ref{FigPxUtbm}.} \label{FigPxUtfh2}
\end{figure}

\clearpage
\subsubsection{$P_z\cdot U_{\rm{TFH2}}$}
\begin{figure}[t]
\centering
\includegraphics[bb=0 0 550 550, height=8.5cm, width=8.5cm, angle=0]{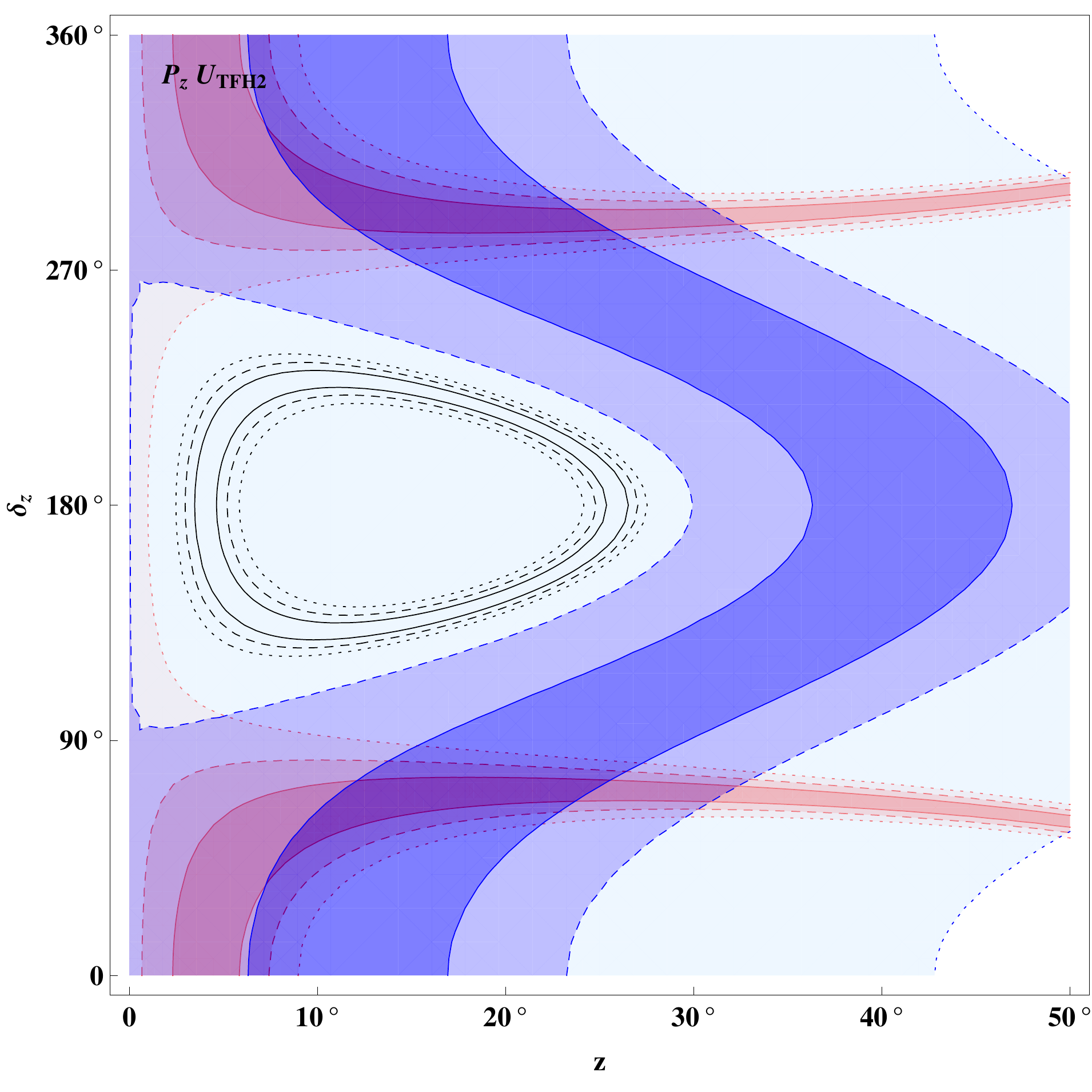}
\caption{The solutions corresponding to the measured $\theta_{12}$, $\theta_{23}$, and $\theta_{13}$ for $P_z\cdot U_{\rm{TFH2}}$ in the parameter space of $z - \delta_z$. Caption is the same as displayed in Fig. \ref{CPxUtbm}.} \label{CPzUtfh2}
\end{figure}

\clearpage
\subsubsection{$U_{\rm{TFH2}}\cdot P_y$}
\begin{figure}[t]
\centering
\includegraphics[bb=0 0 550 550, height=8.5cm, width=8.5cm, angle=0]{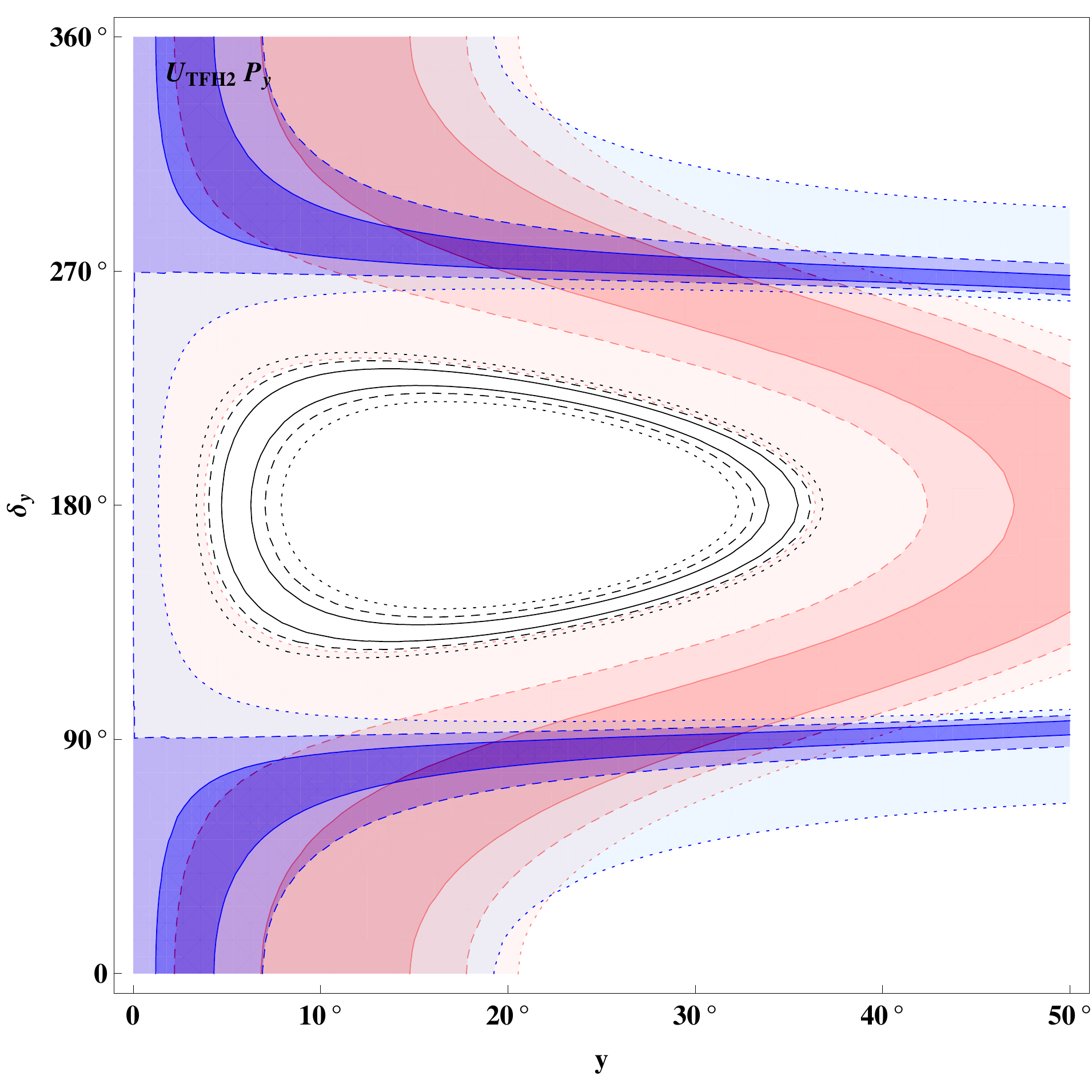}
\caption{The solutions corresponding to the measured $\theta_{12}$, $\theta_{23}$, and $\theta_{13}$ for $U_{\rm{TFH2}}\cdot P_y$ in the parameter space of $y - \delta_y$. Caption is the same as displayed in Fig. \ref{CPxUtbm}.} \label{CUtfh2Py}
\end{figure}

\clearpage
\subsubsection{$U_{\rm{TFH2}}\cdot P_z$}
\begin{figure}[t]
\centering
\includegraphics[bb=0 0 550 550, height=8.5cm, width=8.5cm, angle=0]{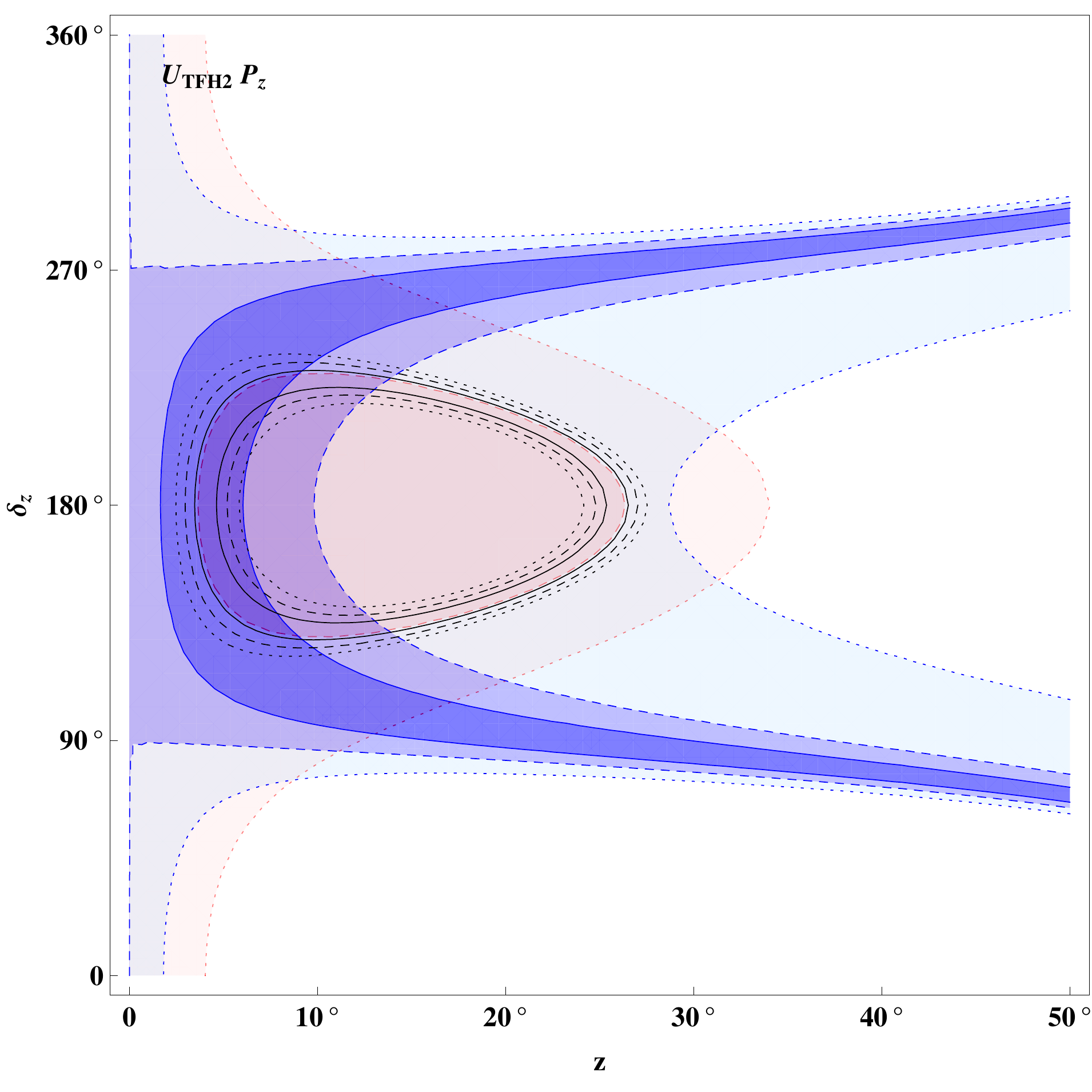}
\caption{The solutions corresponding to the measured $\theta_{12}$, $\theta_{23}$, and $\theta_{13}$ for $U_{\rm{GRM2}}\cdot P_z$ in the parameter space of $z - \delta_z$. Caption the same as displayed in Fig. \ref{CPxUtbm}.} \label{CUtfh2Pz}
\end{figure}

\clearpage
\begin{figure}[t]
\centering
\includegraphics[bb=0 110 550 550, height=12.5cm, width=15.5cm, angle=0]{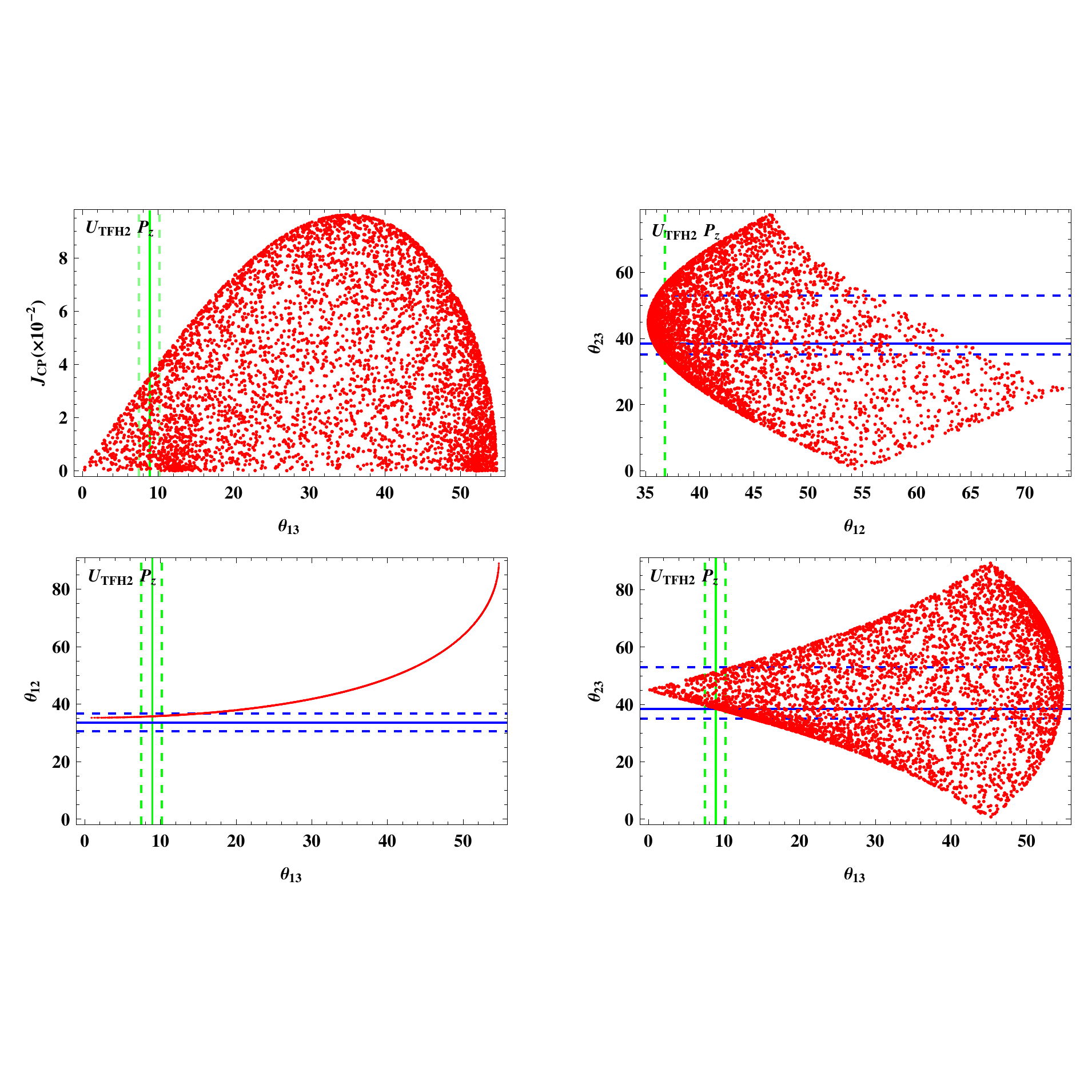}
\caption{The scatter plots for $\theta_{12}$, $\theta_{23}$, $\theta_{13}$ and $J_{\rm CP}$ with the $U_{\rm{TFH2}}\cdot P_z$ ansatz. Caption is the same as displayed in Fig. \ref{FigPxUtbm}.} \label{FigUtfh2Pz}
\end{figure}


\begin{thebibliography}{99}

\bibitem{Fritzsch}
H. Fritzsch, Phys. Lett. {\bf B73} (1978) 317; Nucl. Phys. {\bf B155} (1979) 189.

\bibitem{Lam}
C. S. Lam, Phys. Rev. {\bf D83} (2011) 113002; arXiv:1105.4622.

\bibitem{T2K}
K. Abe et al. (T2K Collaboration), Phys. Rev. Lett. 107 (2011) 041801,
arXiv:1106.2822.

\bibitem{DoubleChooz1}
Y. Abe et al. (Double-Chooz Collaboration), Phys. Rev. Lett. 108 (2012) 131801, arXiv:1112.6353.

\bibitem{DoubleChooz2}
Y. Abe et al. (Double-Chooz Collaboration), arXiv:1207.6632.

\bibitem{DayaBay1}
F. P. An et al. (Daya Bay Collaboration), Phys. Rev. Lett. 108 (2012) 171803, arXiv:1203.1669.

\bibitem{DayaBay2}
D. Dwyer, "Improved	Measurement	of Electron-antineutrino	 Disappearance at Daya Bay", presented at XXV International Conference on Neutrino Physics and Astrophysics, 3-9 June 2012, Kyoto, Japan.

\bibitem{DayaBay3}
L. Zhang, "Observation of Electron-Antineutrino Disappearance at Daya Bay", presented at International Symposium on Neutrino Physics and Beyond, 23-26 September 2012, Shenzhen, China.

\bibitem{RENO}
Soo-Bong Kim et al. (RENO), Phys. Rev. Lett. 108 (2012) 191802, arXiv:1204.0626.


\bibitem{PDG}
J. Beringer et al. (Particle Data Group), Phys. Rev. {\bf D86} (2012) 010001.

\bibitem{XZhBook}
Z.Z. Xing and S. Zhou, Neutrinos in Particle Physics, Astronomy and Cosmology (Zhejiang University Press and Springer-Verlag, 2011).

\bibitem{PMNS1}
B. Pontecorvo, Zh. Eksp. Theor. Fiz. {\bf 33} (1957) 549;
\textit{ibidem} {\bf 34} (1958) 247.

\bibitem{PMNS2}
Z. Maki, M. Nakagawa and S. Sakata, Prog. Theor. Phys. {\bf 28}
(1962) 870.

\bibitem{Jarlskog1}
C. Jarlskog, Phys. Rev. Lett. 55 (1985) 1039.

\bibitem{Jarlskog2}
D. D. Wu, Phys. Rev. {\bf D33} (1986) 860.


\bibitem{KamLAND}
A. Gando et al. (KamLAND Collaboration), Phys. Rev. {\bf D83} (2011) 052002, arXiv:1009.4771.



\bibitem{MINOS}
P. Adamson et al. (MINOS Collaboration), Phys. Rev. Lett. 107 (2011) 181802, arXiv:1108.0015.





\bibitem{GlobalFit}
G.L. Fogli, E. Lisi, A. Marrone, D. Montanino, A. Palazzo, A.M. Rotunno, Phys. Rev. {\bf D86} (2012) 013012, arXiv:1205.5254.

\bibitem{tribimaximal}
P.F. Harrison, D.H. Perkins and W.G. Scott, Phys. Lett. {\bf B530}
(2002) 167; Z.Z. Xing, Phys. Lett. {\bf B533} (2002) 85; P.F.
Harrison and W.G. Scott, Phys. Lett. {\bf B535} (2002) 163; X.G. He
and A. Zee, Phys. Lett. {\bf B560} (2003) 87; I. Stancu and D.V.
Ahluwalia, Phys. Lett. {\bf B460} (1999) 431.

\bibitem{democratic}
H. Fritzsch, Z. Z. Xing, Phys. Lett. {\bf B372} (1996) 265, hep-ph/9509389.

\bibitem{bimaximal}
V. D. Barger, S. Pakvasa, T. J. Weiler, and K. Whisnant, Phys. Lett. {\bf B437} (1998) 107, hep-ph/9806387.


\bibitem{GoldenRatioMixing1}
Y. Kajiyama, M. Raidal, and A. Strumia, Phys. Rev. {\bf D76} (2007) 117301, arXiv: 0705.4559.

\bibitem{GoldenRatioMixing2}
W. Rodejohann, Phys. Lett. {\bf B671} (2009) 267, arXiv: 0810.5239.

\bibitem{HexagonalMixing}
C. H. Albright, A. Dueck, and W. Rodejohann, Eur. Phys. J. {\bf C70} (2010) 1099, arXiv: 1004.2798.

\bibitem{Tetramaximal}
Z.Z. Xing, Phys. Rev. {\bf D78} (2008) 011301, arXiv: 0805.0416.

\bibitem{TFH1}
R. de Adelhart Toorop, F. Feruglio, and C. Hagedorn, Phys. Lett. {\bf B703} (2011) 447, arXiv: 1107.3468.

\bibitem{TFH2}
R. de Adelhart Toorop, F. Feruglio, and C. Hagedorn, arXiv: 1112.1340.

\bibitem{TFH3}
G. J. Ding, arXiv: 1201.3279.

\bibitem{mutau}
R.N. Mohapatra and A.Yu. Smirnov, Ann. Rev. Nucl. Part. Sci. {\bf
56} (2006) 569.







\end{thebibliography}
\end{document}